\documentclass[a4paper,11pt]{article}
\pdfoutput=1 

\usepackage{jheppub}
\usepackage[nice]{nicefrac}
\usepackage{booktabs}
\usepackage{pifont}
\newcommand{\cmark}{\ding{51}}%
\newcommand{\xmark}{\ding{55}}%

\RequirePackage{amsmath}
\RequirePackage{xspace}

\newcommand{\pt}{\ensuremath{p_{\mathrm{T}}}\xspace}
\newcommand{\GeV}{\ensuremath{\,\text{Ge\hspace{-.08em}V}}\xspace}

\DeclareMathOperator{\arcosh}{arcosh}

\usepackage[T1]{fontenc} 

\title{\boldmath Neural Embedding: Learning the Embedding of the Manifold of Physics Data}

\author[a,b,1]{Sang Eon Park,\note{Corresponding author.}}
\author[a,b]{Philip Harris}
\author[c,b]{Bryan Ostdiek}

\affiliation[a]{Massachusetts Institute of Technology,Cambridge, MA 02139, USA}
\affiliation[b]{The NSF AI Institute for Artificial Intelligence and Fundamental Interactions}
\affiliation[c]{Department of Physics, Harvard University, Cambridge MA 02138}

\emailAdd{sangeon@mit.edu}
\emailAdd{pcharris@mit.edu}
\emailAdd{bostdiek@gmail.com}

\abstract{
In this paper, we present a method of embedding physics data manifolds with metric structure into lower dimensional spaces with simpler metrics, such as Euclidean and Hyperbolic spaces. We then demonstrate that it can be a powerful step in the data analysis pipeline for many applications. Using progressively more realistic simulated collisions at the Large Hadron Collider, we show that this embedding approach learns the underlying latent structure. With the notion of volume in Euclidean spaces, we provide for the first time a viable solution to quantifying the true search capability of model agnostic search algorithms in collider physics (i.e. anomaly detection). 
Finally, we discuss how the ideas presented in this paper can be employed to solve many practical challenges that require the extraction of physically meaningful representations from  information in complex high dimensional datasets. 

}

\begin{document} 
\maketitle
\flushbottom

\section{Introduction}
\label{sec:intro}
Despite being high dimensional, physics datasets are highly structured since physical laws strictly govern the data generating process. Although the data is complicated, it is not hard to imagine that physics data can exist within low-dimensional manifolds inside a high-dimensional ambient space. 

%
%
%

There is a growing recent interest in endowing the space of collider events with a metric structure calculated directly in the space of its inputs. Metrics based on optimal transport, such as energy mover’s distance (EMD)~\cite{Komiske_2019} and Hellinger distance~\cite{Cai:2021hnn}, allow us to compare raw inputs directly and quantify the global structural difference between any pair of collider events. Since the advent of these studies, a broad range of use cases has been emerging for these metrics. These include event tagging, anomaly tagging\cite{Romao:2020ojy, Tsan:2021brw, Fraser:2021lxm}, and measurements of Quantum Chromo Dynamical (QCD) properties. 

However, the input dimension is usually very large for collider data; thus, the induced manifold of the metric lives in a very high dimensional space, making it challenging to work with directly. With just 50 particles and 3 features per particle, the induced manifold lives in $\mathbb{R}^{150}$, a prohibitively large dimensional space subject to the curse of dimensionality. 

After decades of searching at the LHC, no new physics beyond the discovery of the Higgs boson has been observed despite a large variety of targeted searches for new physics models. In light of this, we are starting to consider that maybe we are not looking in the right area of the collider data, and we should go beyond our existing new physics models. The shift from targeted searches to model agnostic searches is happening rapidly, and a diverse and rich variety of model agnostic search methods has been proposed by the community, based on a wide variety of different underlying principles \cite{Metodiev_2017,Collins_2018,Collins_2019,Nachman_2020, Heimel:2018mkt,Farina:2018fyg,Cerri_2019,Kuusela_2012,roy2020robust,Heimel_2019,roy2020robust,Blance_2019,Hajer_2020,dagnolo2020learning,D_Agnolo_2019,Romao:2020ocr, Fanelli:2022xwl, Dillon:2022tmm, Alvi:2022fkk, Bradshaw:2022qev, Ngairangbam:2021yma, Chekanov:2021pus, Mikuni:2021nwn, Aguilar-Saavedra:2021utu, Ostdiek:2021bem, Kasieczka:2021tew, Caron:2021wmq, Dorigo:2021iyy, Atkinson:2021nlt, Finke:2021sdf}. 

However, ways to evaluate and quantify the performance of these algorithms are less studied. 
We can’t systematically study different anomaly detection methods without a good method to quantify and study how each algorithm performs.  
Consequently, there is a strong need to come up with ways to quantify each method, especially to understand how far in the search space the algorithm can reach and how wide of a net is cast in the space of total possible physics events. 

This paper introduces a flexible framework for embedding the manifold of collider events in lower-dimensional spaces. This framework  allows physicists to get the most out of metric space properties of collider events and demonstrate that it can be used to quantify different anomaly detection algorithms for model agnostic searches.  Moreover, we show this embedding space captures core physical features and self assembles events into physically meaningful categories.

We primarily focus on learning embedding functions into lower-dimensional spaces with the goal of approximating the given original metric on the space of collider events. We will show that low distortion and robust embedding can be achieved in very low dimensions, down to two dimensions. Different choices of space where we can embed the physics event are also explored. We discuss the advantage of learning the embedding by training the embedding function to approximate the metric distance in the original space over out-of-the-box manifold learning methods such as t-SNE\cite{tsne} and UMAP\cite{umap}. 

The strength of the proposed method is presented with emphasis on quantifying anomaly detection algorithms. The embedding is a useful method of anomaly detection itself, but more importantly, it can address the bigger problem of quantifying the effectiveness of each technique. Using the notion of volume in the embedded space, we propose the volume-adjusted ROC curve, which in two dimensions becomes the area-adjusted ROC curve that tries to measure the “volume” of the total search space encompassed by an algorithm. We then quantify the performance of two different anomaly detection algorithms on a fixed dataset. We additionally show that low distortion embedding is useful for many different aspects of physics analysis by presenting a visualization and exploration of what is learned by the embedding.

Outside the realm of anomaly detection, we demonstrate that  with embedding, we can tackle some critical problems.  Mapping complicated metrics to simpler metrics gives access to a powerful algorithmic toolkit that allows us to do approximation, online analysis, data compression, and classification. Furthermore, mapping the original space to a lower-dimensional space makes many tasks, such as visualization, much easier. Embedding can also be used for data compression by  compressing the information about jets down to a few physically motivated numbers corresponding to the dimension of the embedded space and the physics motivated metrics learned in the embedding process. Using the embedding algorithm, we build on recent ideas in machine learning to develop effective strategies to abstract away non-essential information, leading to informed decision making and effective understanding of the core physics processes. 
We stress that there are a lot of compelling use cases, not limited to those mentioned and explored in this paper.

Embedding is particularly computationally tractable and scales better than the pairwise computation of the distance between events. Since embedding is embarrassingly parallel and can be calculated relatively cheaply through a single forward pass of the neural network, embedding can be computed in real-time, leading to further possibilities for low latency event classification and  online decision-making within a trigger system. 

Different embedding techniques have led to successes in many fields, such as dealing with biological sequences and phylogenetic analysis \cite{corso2021neural}, graph and  network analysis \cite{narayanan2017graph2vec, rozemberczki2018fast, gong2021smr, ahmed2018learning}, natural language processing \cite{glove, frogner2019learning, akbik-etal-2018-contextual,bartusiak2019wordnet2vec}, computer vision \cite{sanakoyeu2019divide,garcia2017visual}, amongst others. This technique is quickly gaining popularity and falls within a large space of machine learning aimed at effectively, scientifically motivated data representation. 

For this paper, we  demonstrate the embedding of hadronically decaying final states consisting of resonances that subsequently decay to intermediate resonances resolved as jets with  as many as 4-prongs over a large variety of masses. With this diverse set of events, we apply embedding and show that even with two dimensions, we can capture the core physical features. Furthermore, we   developed  a simplified “toy jet” generator to create complex objects under strict kinematic restrictions and a minimal set of parameters.
We use this toy dataset as a testing ground for our method to check whether the embedding learns the proper latent structure and self-organizes the space into distinct features. 


Embedding into a lower-dimensional space can also be seen as an alternate way of building a simpler space to perform physics measurements, searches, and classification, such as in \cite{dillon2021}. We bypass the need to do latent variable modeling by directly building the space through embedding with optimal transport distances on the original space, yielding a new handle in how to organize and classify data. 

Embedding has the potential to solve big problems within  physics analyses. We see this paper as the first of many future papers that could transform the way we approach physics data analyses altogether.

\section{Neural Embedding}
\label{sec:background}
In this paper, we develop a neural embedding to take collider events and embed them into a manifold governed by physically motivated principles. The key to performing this analysis is show that this embedding is robust across a variety of datasets.  In the following section, we will outline the core idea of the neural embedding, and motivate the choice and design of the simulated collider datasets. Our goal is to progressively build more complicated datasets towards an intuitive understanding for how this approach can be applied on fully realistic collider data.

\subsection{Problem Setting}

\begin{figure}[htbp!]
\centering
\includegraphics[width=.80\linewidth]{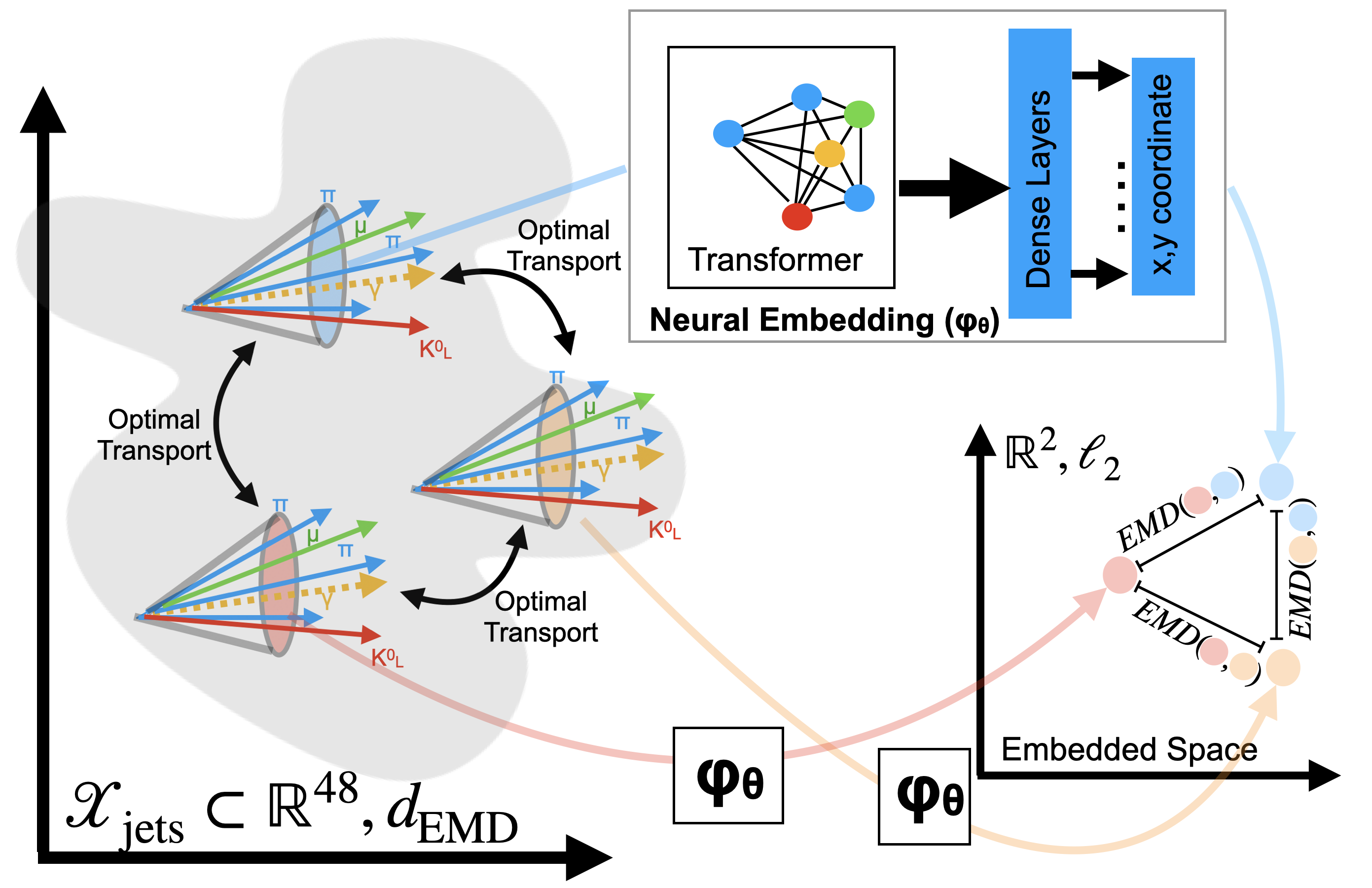}
\caption{ Diagrammatic representation of the distance preserving embedding. Grey region represents the data manifold, three  different types of jets represent three points on the manifold which gets mapped to $(\mathbb{R}^2, l_2)$ by the learned embedding. The energy mover's distance in the original space is preserved in the embedded space. }
\label{fig:diagram}
\end{figure}

Suppose we have a metric space $(\mathcal{X}, d)$ of collider events, where $\mathcal{X}$ denotes the space of collider events and $d$ is some metric defined on the space. For collider events, this space is often represented by a list of all the particles in the event with their subsequent features. With collisions comprising hundreds of particles in the final state, the resulting space is high dimensional and metrics on the space are computationally difficult.  Starting with the energy mover's distance (EMD)~\cite{Komiske_2019}, many metrics based on ideas from optimal transport have been proposed on the collider events where the space $\mathcal{X}$ is represented as a subset of high dimensional input space $\mathbb{R}^D$~\cite{Cai:2021hnn}. These metrics are capable of effectively interpreting collider events, but are often computationally intractable and hard to interpret within $\mathbb{R}^D$. 


Our goal in this paper is to simplify the interpretation of $\mathbb{R}^D$ by learning an embedding map to a low dimensional space that  preserves our metrics $d$ on the higher dimensional space. In other words, we aim to create a low dimensional space $\mathcal{Y}$ through $\phi: (\mathcal{X},d_{\mathcal{X}}) \rightarrow (\mathcal{Y}, d_\mathcal{Y}$). While a perfect embedding where the distance between any two events is perfectly preserved is not guaranteed, we aim to learn a mapping that satisfies a low distortion, namely it satisfies the relation
\begin{equation}
\begin{aligned}
\forall u,v \in \mathcal{X}, \: L\cdot d_{\mathcal{Y}}(\phi(u),\phi(v)) < d_\mathcal{X}(u,v) < C\cdot d_\mathcal{Y}(\phi(u),\phi(v)) 
\end{aligned}
\end{equation}
For some $0<L<1$ and $C \geq 1 $, where smaller $C$ indicates a smaller overall distortion in the space
and  the metric on the space $d_{\mathcal{Y}}$ is a simpler metric than the original metric $d_{\mathcal{X}}$. The constant $L$ is the inverse of the Lipschitz constant for the mapping $\phi$, and it guarantees that the metric distance doesn't blow up in the embedded space. 


Therefore our learning objective can be formulated as Eq.~\ref{eq:learningobjective}. For a family of functions parameterized by $\theta$, the goal is to minimize the empirical risk for the distortion for $N$ pairs $(u_i, v_i), i\in{1,\dots,N}$ from our training dataset, and we can do this with standard gradient descent algorithms on $\theta$ given as
\begin{equation}
\begin{aligned}
\label{eq:learningobjective}
\hat{\theta} = \underset{\theta}{\mathrm{arg\,min}} 
\frac{1}{N}\sum_{i=1}^{N}\frac{\left|d_\mathcal{Y}(\phi_{\theta}(u_i),\phi_{\theta}(v_i))-d_\mathcal{X}(u_i, v_i)\right|}{d_\mathcal{X}(u_i, v_i)}, \phi_\theta \in \mathcal{F}
\end{aligned}
\end{equation}

We denote this empirical risk minimization procedure as neural embedding~(NE). 

In this learning framework, we have to make two critical choices (1) which family of functions $\phi_{\theta} \in \mathcal{F}$ do we chose to approximate the embedding, and (2) which geometrical space $\mathcal{Y}$ do we choose to embed into.

The family of functions $\phi_\theta \in \mathcal{F}$ we choose in this paper is a family of deep neural networks. An appropriate choice of the neural network is made depending on the input format of the data. 

In this paper, we will look at two types of data. As a first demonstration of embedding, we embed the MNIST~\cite{mnist} handwritten digit images into a 2-dimensional Euclidean space.  As result, we choose convolutional neural networks (CNN) to handle image data. For the rest of the datasets, we use simulated collider events defined by a $\pt$-sorted sequence of final state particles. For the collider data, we use transformer networks with positional encoding. Since our input is a $\pt$-sorted sequence of particles where the flow of information between any particle is allowed, we believe this is a good choice and reflects the state of the art in data assimilation. 

The choice of metric space for $\mathcal{Y}$  is quite flexible. Previous studies, outside of physics, have considered Euclidean spaces, Hyperbolic spaces\cite{corso2021neural}, and Wasserstein spaces\cite{courty2017learning, frogner2019learning}. Here, we focus on a Euclidean space with a $l_2$-norm, $(\mathcal{Y}, d_\mathcal{Y}) = (\mathbb{R}^n, l_2)$, and the Hyberbolic sapce defined by the Poincaré ball $(\mathcal{Y}, d_\mathcal{Y}) = (\mathcal{B}^n, d_p)$. 
Lastly, to make the training tractable, we only train on the subset of available event pairs, by randomly sampling pairs from a total set of $10^{6}$ available events, and subsequently not considering all $10^{12}$ total possible pairings. 








\section{Datasets and Neural Network}
\label{sec:datasetsandnetwork}
Before we embark on the construction of the full NE, we would like to elaborate on the dataset construction used for these studies. Our ultimate goal with these studies is to demonstrate the broad applicability of this framework through the use of a variety of datasets including MNIST~\cite{mnist}. 

Furthermore, we will show the flexibility of the NE construction on progressively more complicated datasets leading towards a realistic dataset. Our goal with adding hierarchies of complexity is to show how NE is capable of transcending the obfuscation present from a more complicated dataset to extract the core physics features embedded within.  

To study NE, we utilize hadronically decaying particles at the LHC. This dataset consists of both new physics resonances with quarks in the final state or standard model production of quarks and gluons (QCD). Quarks, and gluons at the LHC will shower into many particles eventually leading to final state hadrons. These showers are then resolved at the LHC through jet clustering algorithms, yielding jets\cite{Cacciari:2008gp,Cacciari:2011ma}.  In this paper, we will focus on applying NE to a single jet. The large number of particles and complex topologies within a jet make them a difficult tool to study, and in many studies, it has been shown that jets benefit enormously from machine learning approaches.

Since jets are complicated objects, we created a series of hierarchical datasets whereby we progressively made each dataset more and more complicated. 
As a consequence, we developed several jet simulations that allow for the isolation of a fixed number of hidden parameters, so that we can effectively study how NE can extract the critical patterns hidden within the data.  In this section we present the two main simulations used to study NE on jets, the toy jet generator, and the realistic jet generator. 

\subsection{Toy Jet Generator}

In order to be able to progressively add levels of complexity in the data generation, a toy jet generator was constructed. The toy jet 
emulates a typical  parton shower, while also storing the individual latent variables, so that we can later extract them directly, and explicitly check what information is learned.

\subsubsection{Jet Generation}

The main goal of the toy jet generator is to isolate parameters of the parton shower so that we can see how the NE organizes the embedded space. In light of this, we constructed the toy jet generator such that the masses and momentum of each splitting can be fixed, and the angles of the subsequent splittings can be sampled from a fixed prior. For these studies, we fixed the momentum to be 400 GeV, while we allowed the masses to be sampled from a fixed distribution.  

We implemented two different versions of the toy jet generator, a simple version where the hard and soft splittings are distinctly different and a realistic version where the soft splittings approximate the matrix element of normal quark and gluon fragmentation. 

In both types of jet generators, jets are generated with specified fixed "prongs", a fixed total number of particles, a fixed momentum, and fixed mass distributions. The hard splittings are designed to mimic the decays of a resonance. Namely, they are sampled from an angular distribution in the rest frame of the mother particle of the shower. The number of prongs within a jet defines the number of hard splittings used in the shower. To reach the total number of particles required by the generation, we continue to shower the jet with soft splitting until we reach the final multiplicity. 

In all cases, we force the jet construction to be in a fixed coordinate system whereby the original parton direction is at 0,0, and the first splitting occurs along the x-axis. Subsequent splittings are then randomized in $\phi$ about the particle direction, with the angle of the two particle split, $\theta$, being defined by a sampling prior that varies depending on the jet generator type. The sampling prior for the soft and hard splittings is what defines the difference between the simplified and realistic toy jet generator. All other components of the generation remain the same.

For the hard splitting "signal" models, we force a decay chain of resonances characteristic of the top quark. In particular, we set the mass of the first "signal" parton always to be 172~\GeV, and with potential decay components having a resonance of 80~\GeV and 4~\GeV. As a result, when we simulate two prong signal jets, we take the jet mass to be 172~\GeV, with its decay components being massless quarks. For 3-prong jets, we have the 172~\GeV resonance decays to  a secondary resonance of 80 GeV that decays to quarks and a massless quark. For 4 prong jets, we have a final splitting with a mass of 4 GeV that then decays to two massless quarks. We do not explore jets beyond four prongs. However, we continue to decay the particles until we reach the particle multiplicity of the desired generator.  

In the following subsections, we present the difference between the two toy jet generators. The only difference is the splitting angle in the rest frame. However, this has a large impact on the resulting kinematics.

\subsection{Simple Jet Generator}
\label{subsec:simpletoyjet}

For the simple jet generator scenario, we want to enhance the ability to distinguish hard scatters from soft scatters. To that extent, we define an unphysical sampling prior that is distinctly different between the hard and the soft scatters. 
This is achieved by drawing $\theta_{branch}$, the splitting in the rest frame of the parton, from two distinct distributions.  The hard splitting angle $\theta_{branch}$ is drawn from a normal distribution $\mathcal{N}(\nicefrac{1}{2}, 0.1)$, about $\pi/2$ distribution with narrow variance. The soft splitting angle is drawn from the half-normal distribution with a wide variance of 0.1 radians. 

In addition to the angle in the rest frame, we also plot the splitting angle of the first splitting, we define as
\begin{equation}
\label{eq:zg}
    z_g = \frac{\max{p_{\mathrm{T,1}},p_{\mathrm{T,2}}}}{p_{\mathrm{T,1}}+p_{\mathrm{T,2}}}  
\end{equation}

\begin{figure}[htbp]
\centering
\includegraphics[width=.45\linewidth]{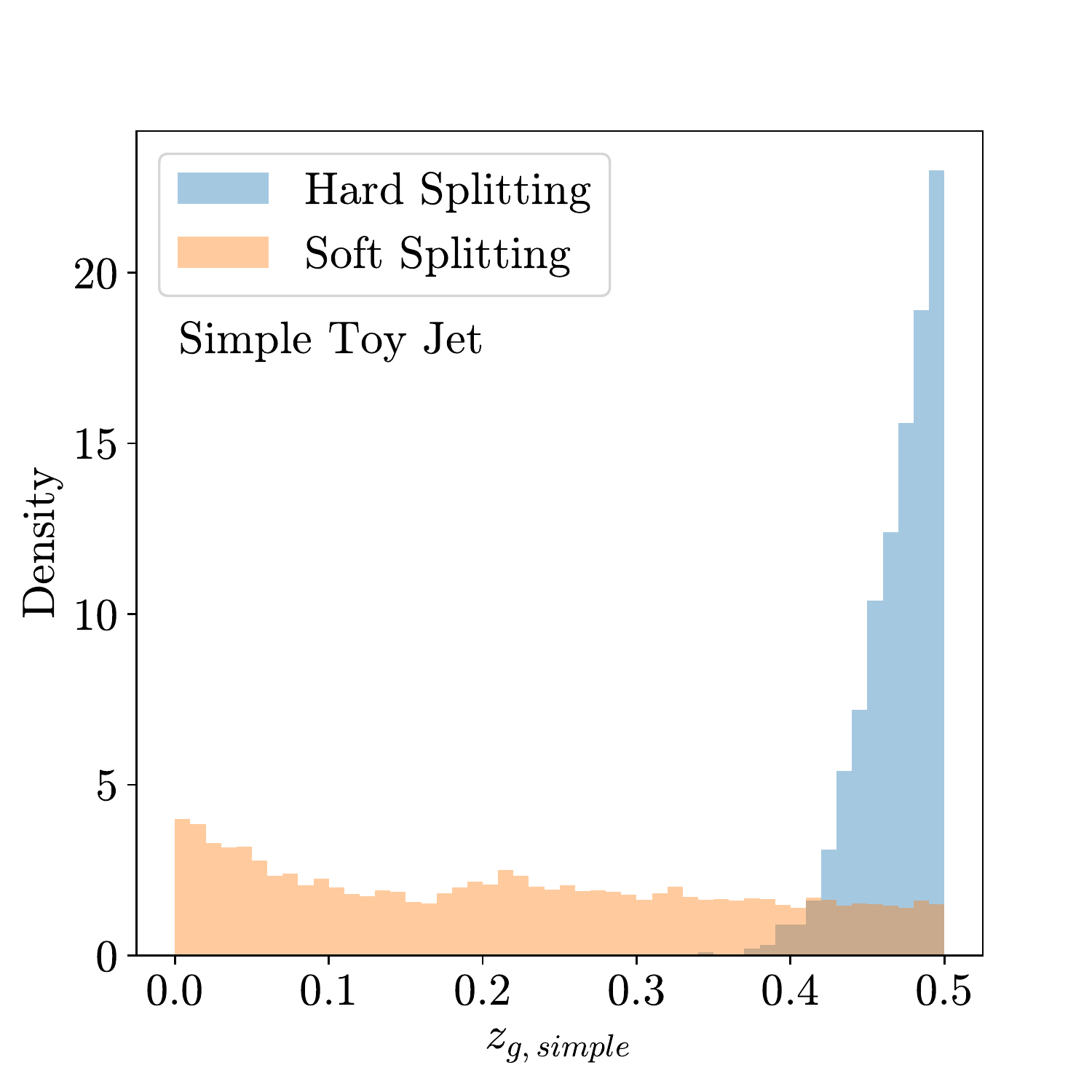}
\includegraphics[width=.45\linewidth]{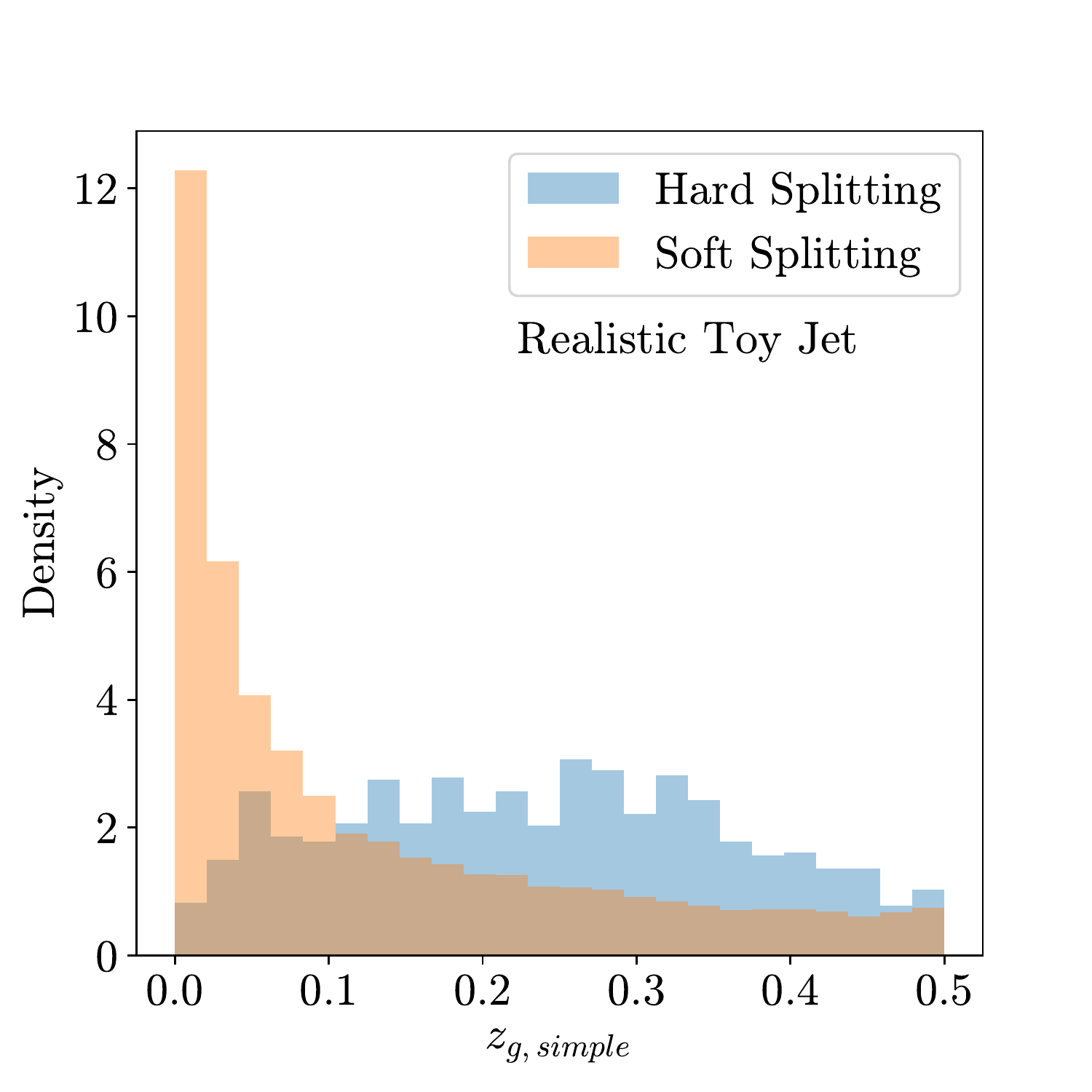}
\caption{ (Left) The distribution of parton momentum sharing variable $z_g$ for each parton splitting for hard and soft splitting, for simple toy jet generator. (Right) The distribution of parton momentum sharing variable $z_g$ for each parton splitting for hard and soft splitting, for realistic toy jet generator. 
}
\label{fig:momentumsharing}
\end{figure}

For the simple jet generator, the splitting angle of the first splitting is shown in Fig.~\ref{fig:splittingangle}.  In addition to the splitting angle we observe a distinct difference in the splitting fraction $z_{g}$ of the first splitting (Fig.~\ref{fig:momentumsharing}).
For this sample, we generated 2M simple toy jets with prong numbers varying from 1 to 4 prongs (QCD(1p), 2p, 3p, 4p). We use 200k jets of each type for validation and testing. Plots of  sample jets are shown in~\ref{subsec:simpletoyjet_examples}.

\subsection{Realistic Jet Generator}
\label{subsec:realistictoyjet}
\begin{figure}[htbp]
\centering
\includegraphics[width=.45\linewidth]{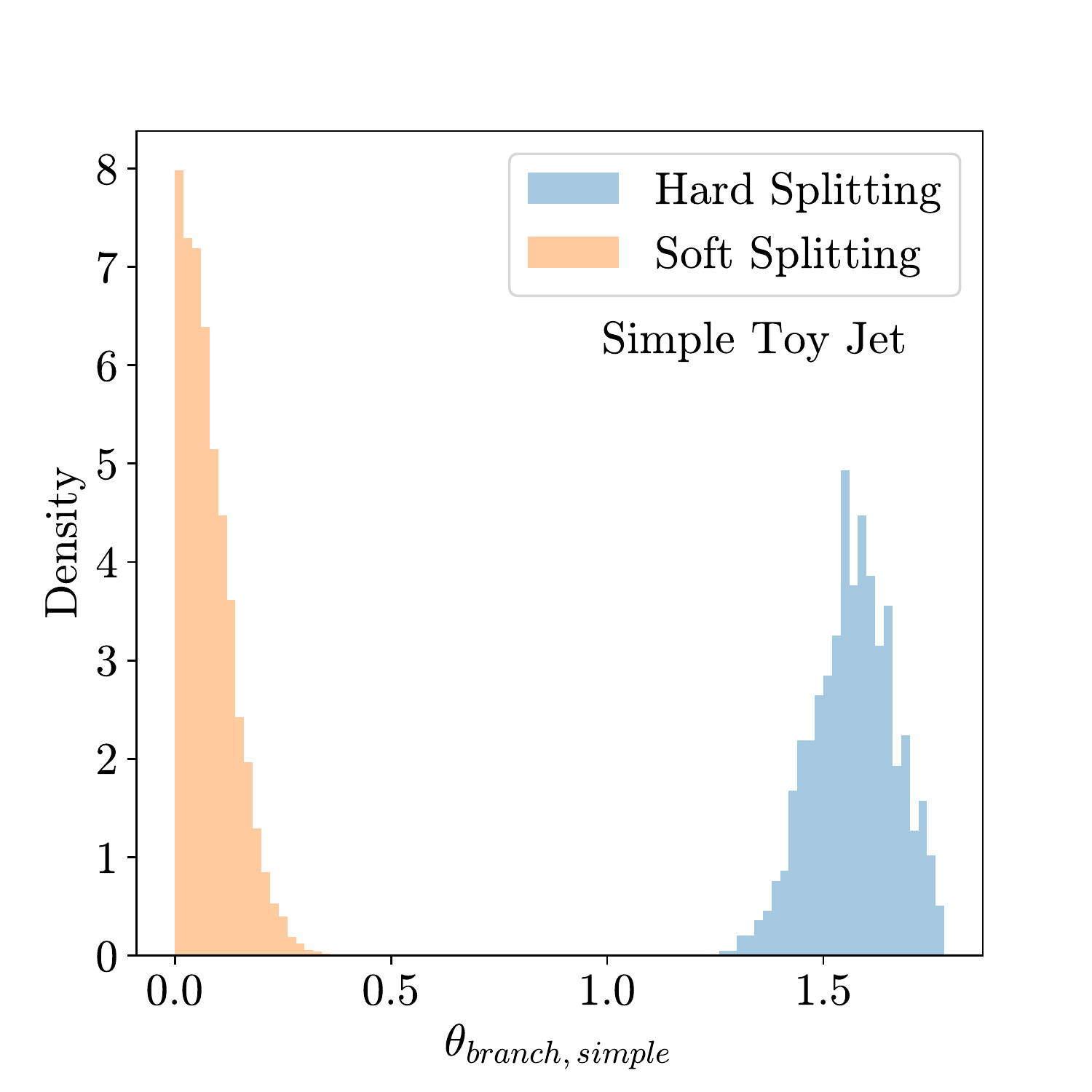}
\includegraphics[width=.45\linewidth]{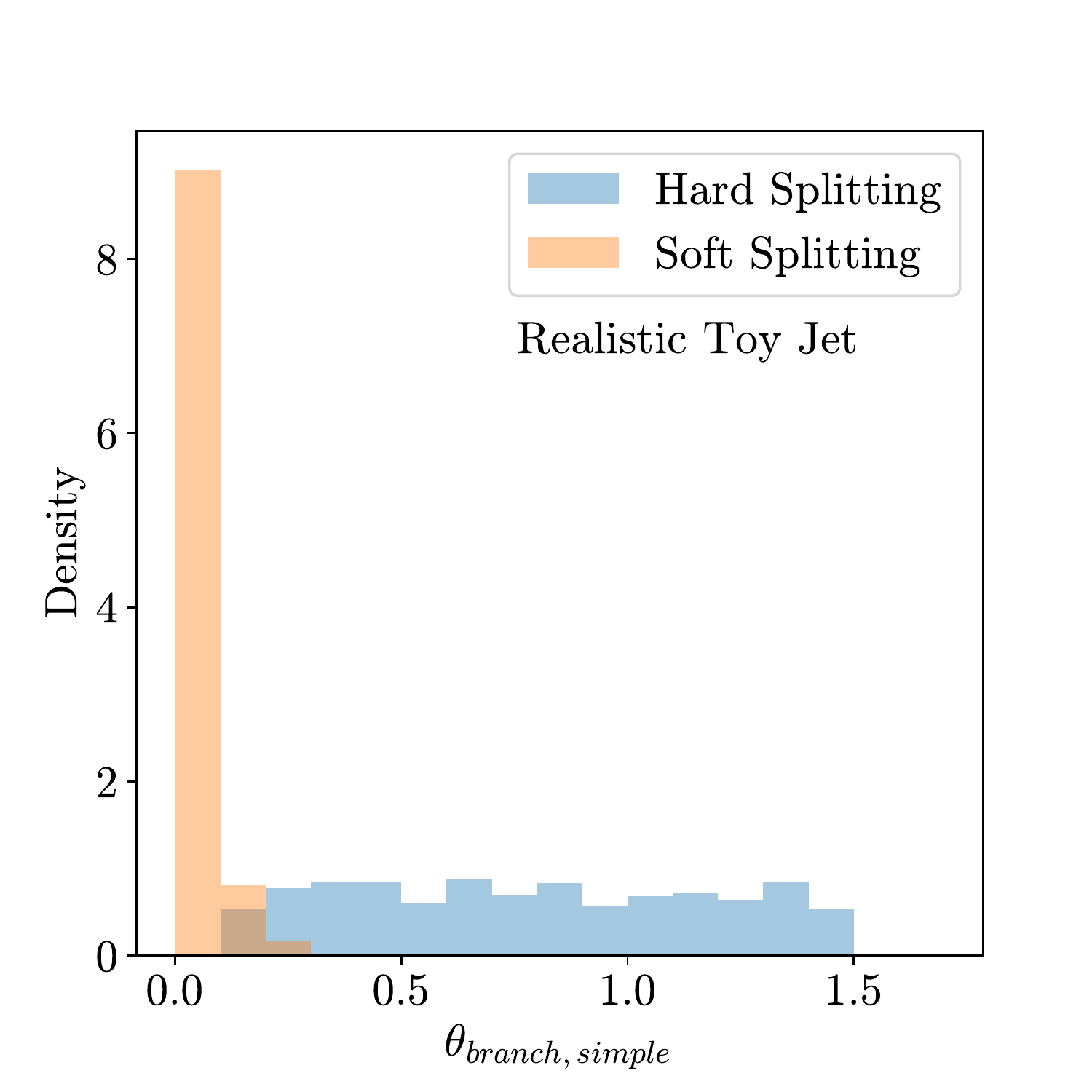}
\caption{ (Left) The distribution of splitting angle $\theta_{branch}$ for each parton splitting for hard and soft splitting, for simple toy jet generator. (Right) The distribution of parton momentum sharing variable $\theta_{branch}$ for each parton splitting for hard and soft splitting, for realistic toy jet generator. 
}
\label{fig:splittingangle}
\end{figure}

For the realistic jet generator, we follow a sampling prior characteristic of real physical decays. For the hard splitting, we sample $\theta_{branch}$ from a flat prior. The soft splittings are computed by sampling a probability distribution given by $p=\frac{1}{\theta z}$ where $\theta$ is the angle of the splitting and $z$ is the momentum fraction of the jet. This closely approximates a typical true parton shower. Fig.~\ref{fig:splittingangle} shows the splitting energy fraction $z_{g}$ for a hard and soft splitting. We observe a behavior similar to what is observed in previous studies in data\cite{CMS:2017qlm,ALargeIonColliderExperiment:2021mqf,ATLAS:2018zhf}. Similarly, Fig.~\ref{fig:splittingangle} shows the splitting for both hard and soft splittings, the bias towards small $\theta$ is very clear in this scenario.

For this study, we generate 1M realistic toy jets of each category (QCD(1p), 2p, 3p, 4p) and 200k for validation and testing. Plots of  sample jets are shown in~\ref{subsec:realistictoyjet_examples}. 






\subsection{Network Architectures}

The choice of a parametric family of functions $\phi_\theta \in \mathcal{F}$ used to approximate the embedding is important since we have to choose a family of deep neural networks with strong expressive power. We furthermore want the chosen neural network family to have properties (such as invariance and equivariance) that is appropriate for the data we have. 


As a first demonstration of how to apply  NE to a generic dataset, we perform training on the MNIST character recognition dataset~\cite{mnist}.
For these collections of digit images, we use CNNs to approximate the embedding. 

Next, we aim to demonstrate NE in a simplified physics-like environment by creating a ``toy jet'' generator that constructs jets with a fixed momentum, mass, and a fixed number of particles but a varying number of hard and soft splittings. For this dataset, we rely on transformer networks with multi-headed attention \cite{vaswani2017attention} applied to the $\pt$-sorted particle 4-vector dataset. 


Finally, we demonstrate the NE on a set of fully simulated jets with characteristic detector resolutions using Delphes\cite{delphes}. Since this dataset yields particles in a similar format to the ``toy jet'' dataset, we employ an identical network architecture to that of the ``toy jet'' generator.

The details of the neural network architecture are explained in detail in Appendix~\ref{sec:nndetails}.

\subsection{Summary}

The details of the studies are summarized in Table~\ref{tab:summary}. 

\begin{table}[htbp!]
\centering
\caption{Summary of datasets, network architecture and geometry of the embedded space presented in this paper. } \label{tab:summary}

\vspace{3pt}

\small
\begin{tabular}{cccc}
\toprule
\textbf{Section} & \textbf{Dataset} & \textbf{Architecture} & \textbf{Geometry} \\ \midrule
Section \ref{sec:mnist} & MNIST  \cite{mnist}     & CNN                        & Euclidean \\
Section \ref{sec:simpletoyjets} & Simple Toy Jets \ref{subsec:simpletoyjet}  & Transformer  & Euclidean  \\
Section \ref{sec:realistictoyjets} & Realistic Toy Jets \ref{subsec:realistictoyjet}  & Transformer  & Euclidean  \\
Section \ref{sec:SimulatedJetsEuclidean} & Simulated Jets \ref{sec:SimulatedJetsEuclidean}  & Transformer  & Euclidean  \\
Section \ref{sec:simulatedjethyperbolic} & Simulated Jets \ref{sec:SimulatedJetsEuclidean}  & Transformer  & Hyperbolic  \\
\bottomrule
\end{tabular}
\end{table}

\section{Experiment}
\label{sec:experiment}
In this section, we present the results of NE on the three progressively more realistic datasets ranging from a proof of concept using the benchmark MNIST dataset to the jet datasets that are progressively more realistic.


\subsection{MNIST}
\label{sec:mnist}
The MNIST dataset \cite{mnist} consists of images of characters, each presented in a square array of 28x28 pixels, or 784 total pixels. To perform the NE, we consider  one million pairs of MNIST images, including all ten digits. To define the distance between any image, we utilize the optimal transport calculated by POT package~\cite{flamary2021pot}. The embedding function is approximated by convolutional neural networks (CNNs) with 4 hidden layers with MLP layers attached at the end that output two numbers, yielding a two-dimensional embedding space. 

The embedding into a two-dimensional Euclidean space with a $l_2$-norm is achieved by learning the function:

\begin{equation}
\begin{aligned}
\label{eq:mnistembed}
\phi_{\theta, \mathrm{CNN}}: (\mathcal{X}_{\mathrm{digits}} \subset \mathbb{R}^{784}, \mathcal{W}_2 ) \rightarrow (\mathbb{R}^2, l_2)~.
\end{aligned}
\end{equation}

\begin{figure}[htbp!]
\centering
\includegraphics[width=.32\linewidth]{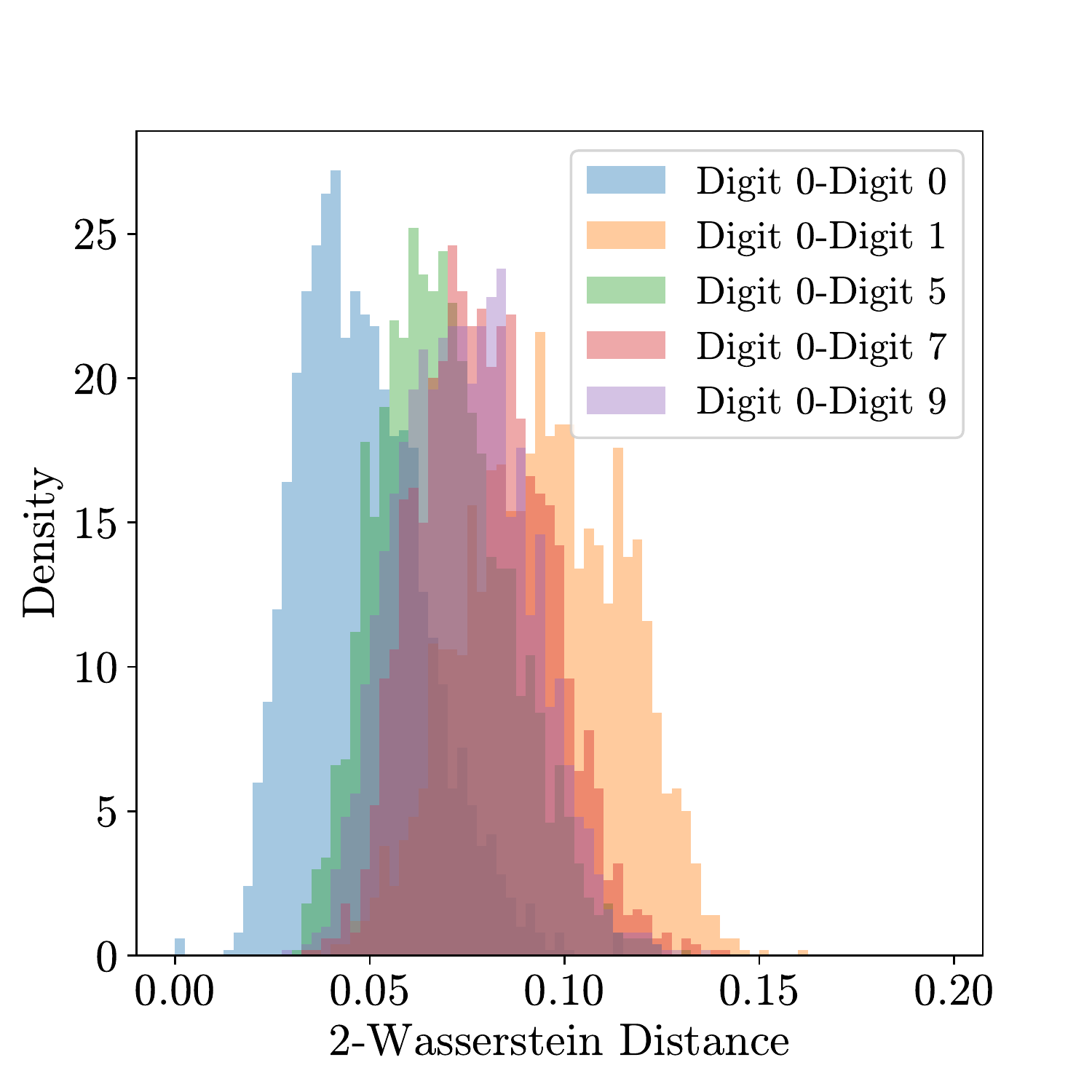}
\includegraphics[width=.32\linewidth]{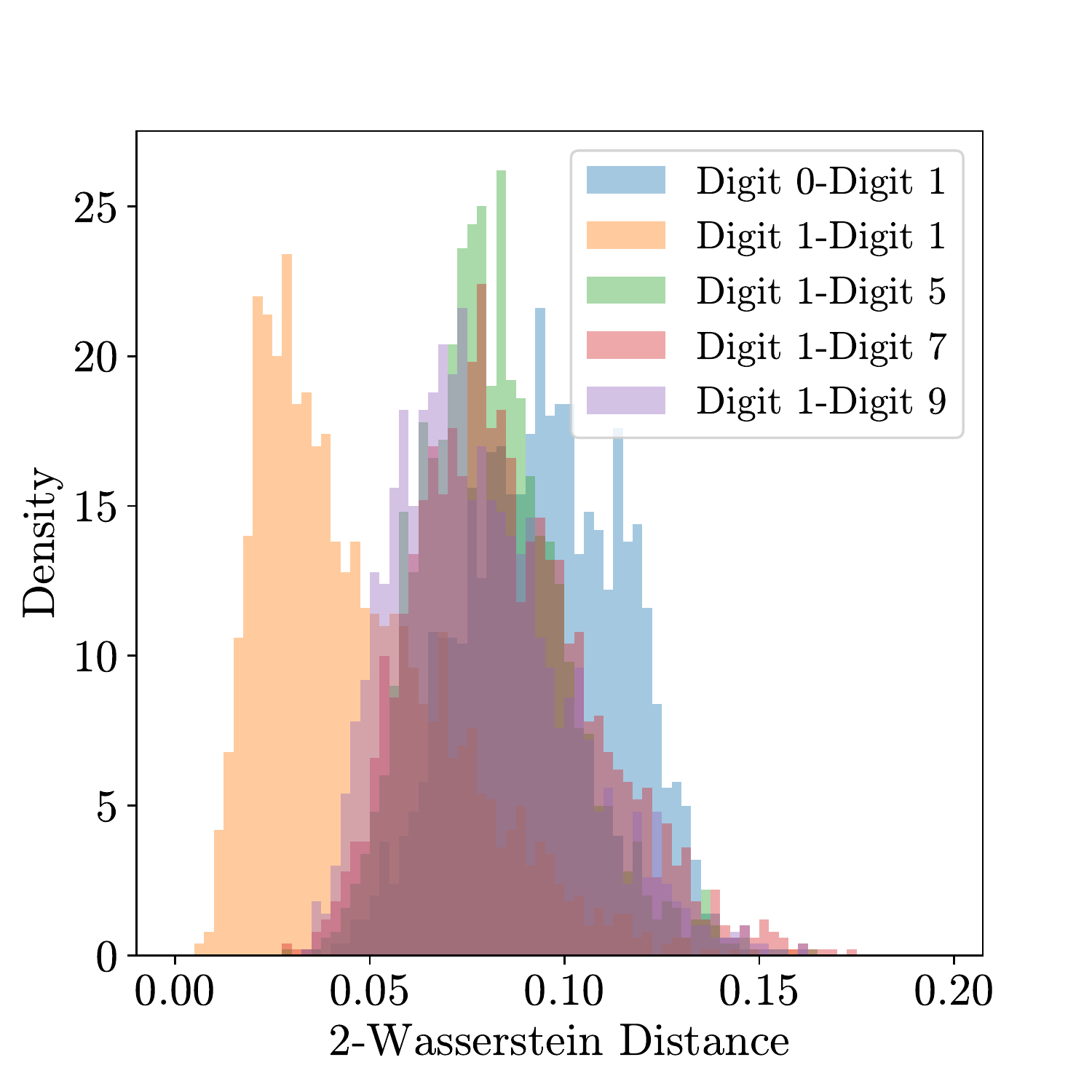}
\includegraphics[width=.32\linewidth]{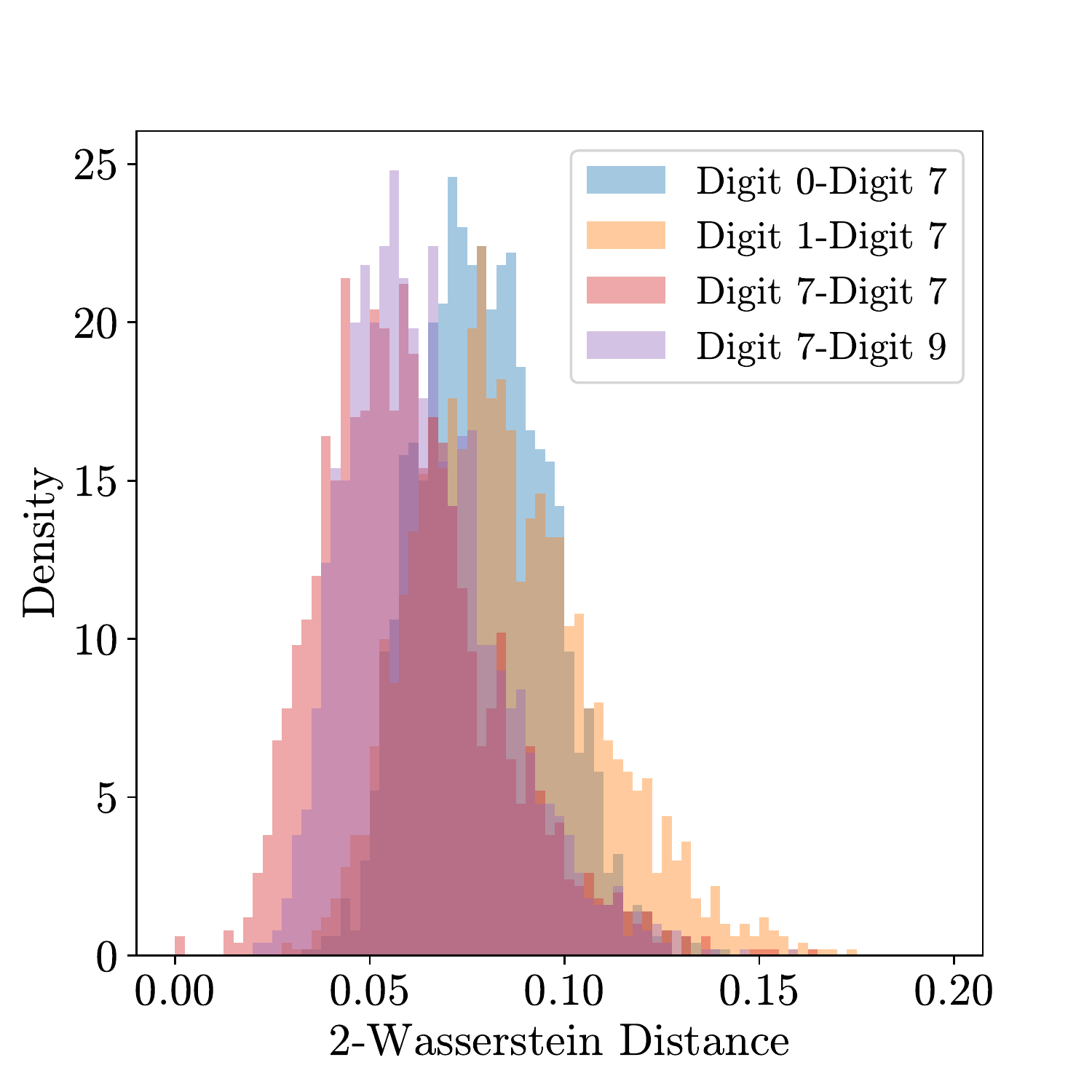}
\caption{ 2-Wasserstein distance $\mathcal{W}_2$ distribution for select digits. (Left) The optimal transport distance with respect to digit 0. (Middle) The optimal transport distance with respect to digit 1. (Right) The optimal transport distance with respect to digit 7. 
}
\label{fig:mnistotdist}
\end{figure}

The distributions of optimal transport distances between pairs of images for selected digits are shown in Fig.~\ref{fig:mnistotdist}. If we look at the histograms of the 2-Wasserstein distance distributions, we can see that the distance between 0 and 1 is very far, and digit 9 is about equidistant from both 0 and 1.

\begin{figure}[htbp!]
\centering
\includegraphics[width=.48\linewidth]{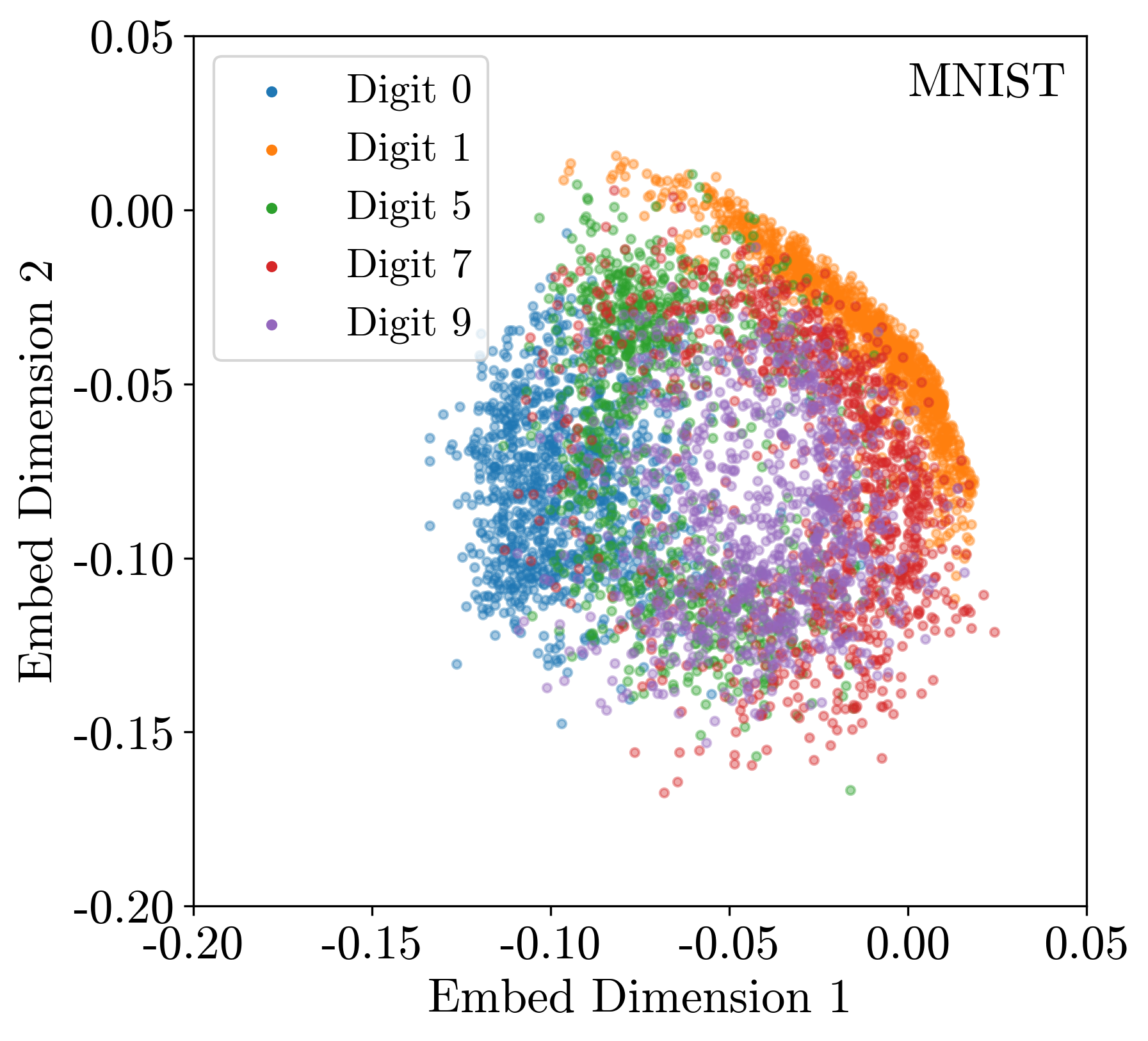}
\includegraphics[width=.48\linewidth]{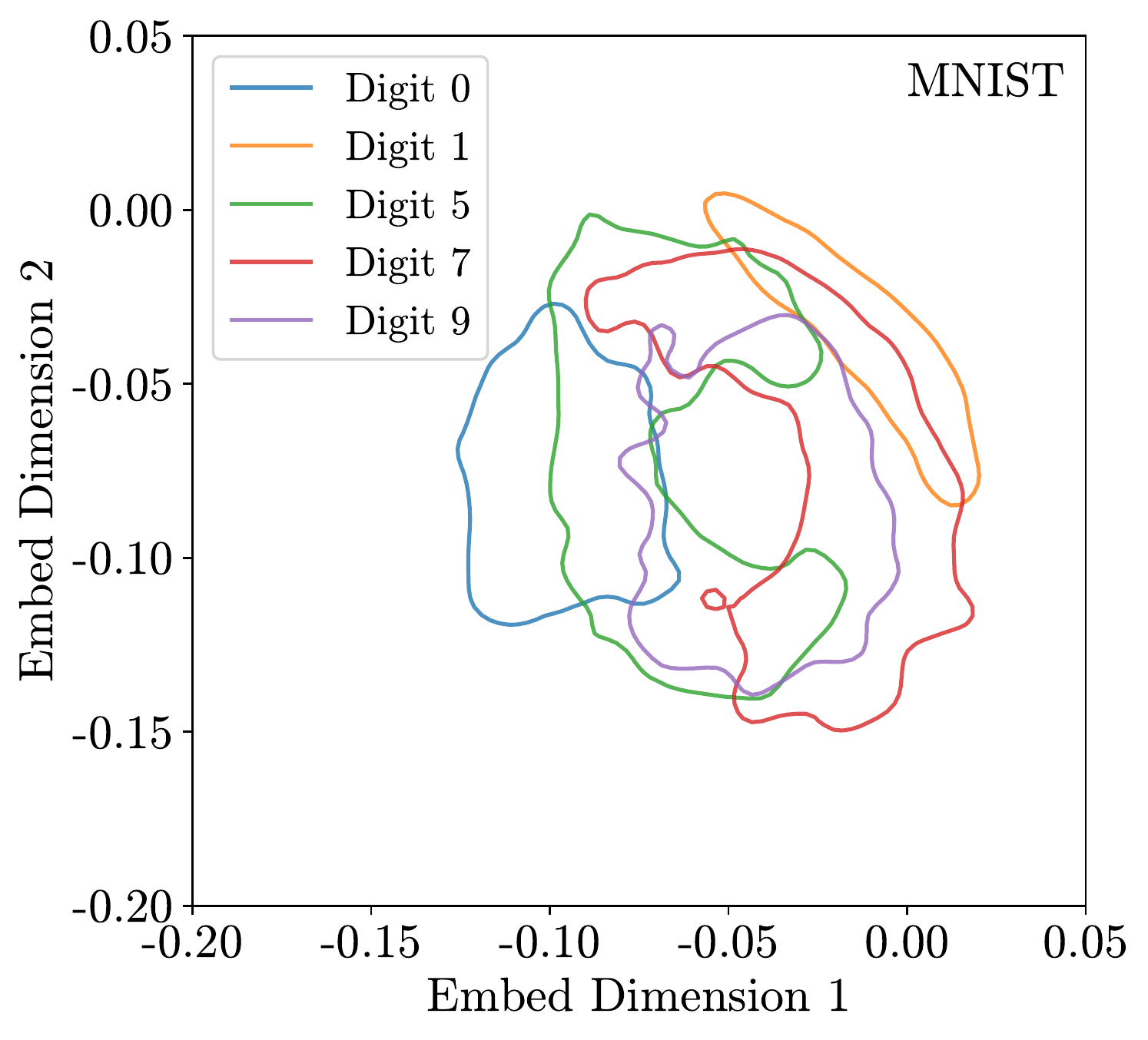}
\caption{ (Left) The scatterplot of embedding of select MNIST digits, 0,1,5,7,9. (Right) The same embedding smoothed with kernel density estimator, with contour lines corresponding to cdf value 0.8.  
}
\label{fig:mnistembedding}
\end{figure}

With the pairwise optimal transport distances, Fig.~\ref{fig:mnistembedding} shows the embedding into the Euclidean space with $l_2$-norm for five selected digits. We show both scatter plots of embedded digits and the contours of the cumulative distribution function (CDF). 
The contours are obtained by first applying kernel density estimation, then integrating from the maximum probability density function (PDF) value of the two-dimensional distribution by lowering the threshold of the PDF value until the desired enclosed probability mass is achieved (usually chosen to be 0.5 and 0.8). 

The result of the corresponding NE  yields a space with  the similarity between digits that we would naively expect from our knowledge of digits and from the optimal transport distance between the images we observed from Fig.~\ref{fig:mnistotdist}. In particular, we observe in the embedded space that the digit 0 and 1 form two distant clusters, while the cluster of digits 5,7, and 9 are located between those two clusters.

\subsection{Toy Jets}

To test the NE on a more complicated dataset, we consider varying sets of progressively more complicated datasets using the toy jet generator. With each dataset, we take the first $K$ highest transverse momentum constituents, each with $(\pt, \eta, \phi)$ information yielding a mapping from  $\mathbb{R}^{3K}$ to the lower dimensional space. For the toy jets generator and future particle based studies, we take $K=16$, and our embedding function thus becomes:


\begin{equation}
\begin{aligned}
\label{eq:jetembed}
\phi_{\theta, \mathrm{Transformer}}: (\mathcal{X}_{\mathrm{jets}} \subset \mathbb{R}^{48}, d_\mathrm{EMD} ) \rightarrow (\mathbb{R}^2, l_2)
\end{aligned}
\end{equation}

\subsubsection{Simple Toy Jets}
\label{sec:simpletoyjets}
For toy jets, we train on 1-prong(QCD) jets and resonant 2-prong and 3-prong jets with a fixed mass at $172\GeV$ and transverse momentum $400\GeV$, and we test on different  1-prong, 2-prong, 3-prong jets, with 4-prong generated jets added as well. We train the transformer model on 2M jets for each type of jet and validate on 2k jets each. The 4-prong jets are reserved just for prediction  to see if the embedding can be extrapolated to jets drawn from the toy jet model with different parameters compared to what was shown in the training.  

\begin{figure}[htbp]
\centering
\includegraphics[width=.99\linewidth]{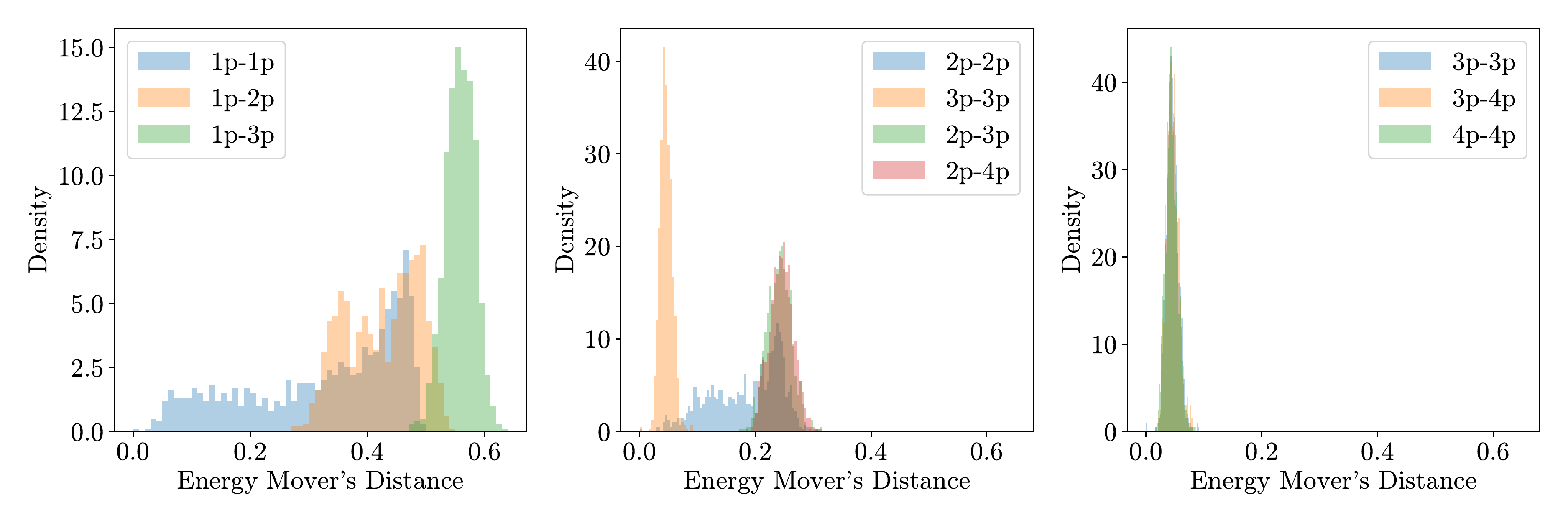}

\caption{ (Left) The distribution of energy mover's distance (EMD) between QCD jets and two-prong, three-prong, and four-prong jets. (Middle) The distribution of energy mover's distance between QCD jets and two-prong, three-prong, and four-prong jets. (Right) The distribution of the parton momentum sharing variable $z_g$ for each parton splitting for hard and soft splitting for the realistic toy jet generator. 
}
\label{fig:emdsimplejet}
\end{figure}

The distribution of pairwise energy mover's distance(EMD) for simple toy jets is shown in Fig.~\ref{fig:emdsimplejet}. 

From the observed EMD, we can start to infer the expected shape of the embedded space. Since the distances between 1-prong jets are wide, we  expect that they would not form a closely grouped cluster in the embedded space. Also, since EMD distribution between 1-prong and 3-prong is larger than between 1-prong and 2-prong, we can expect that the distance between clusters of 1-prong and 3-prong jets would be larger than 1-prong and 2-prong. 
Furthermore, we expect that 3-prong jets will form a small cluster since the EMD between 3-prong jets are small. We see that 3-prong and 4-prong jets have a similar distribution, and we can guess that 3- and 4-prong will form close clusters while 2-prong and 1-prong(QCD) jets will form separate distinct clusters. 

\begin{figure}[htbp]
\centering
\includegraphics[width=.45\linewidth]{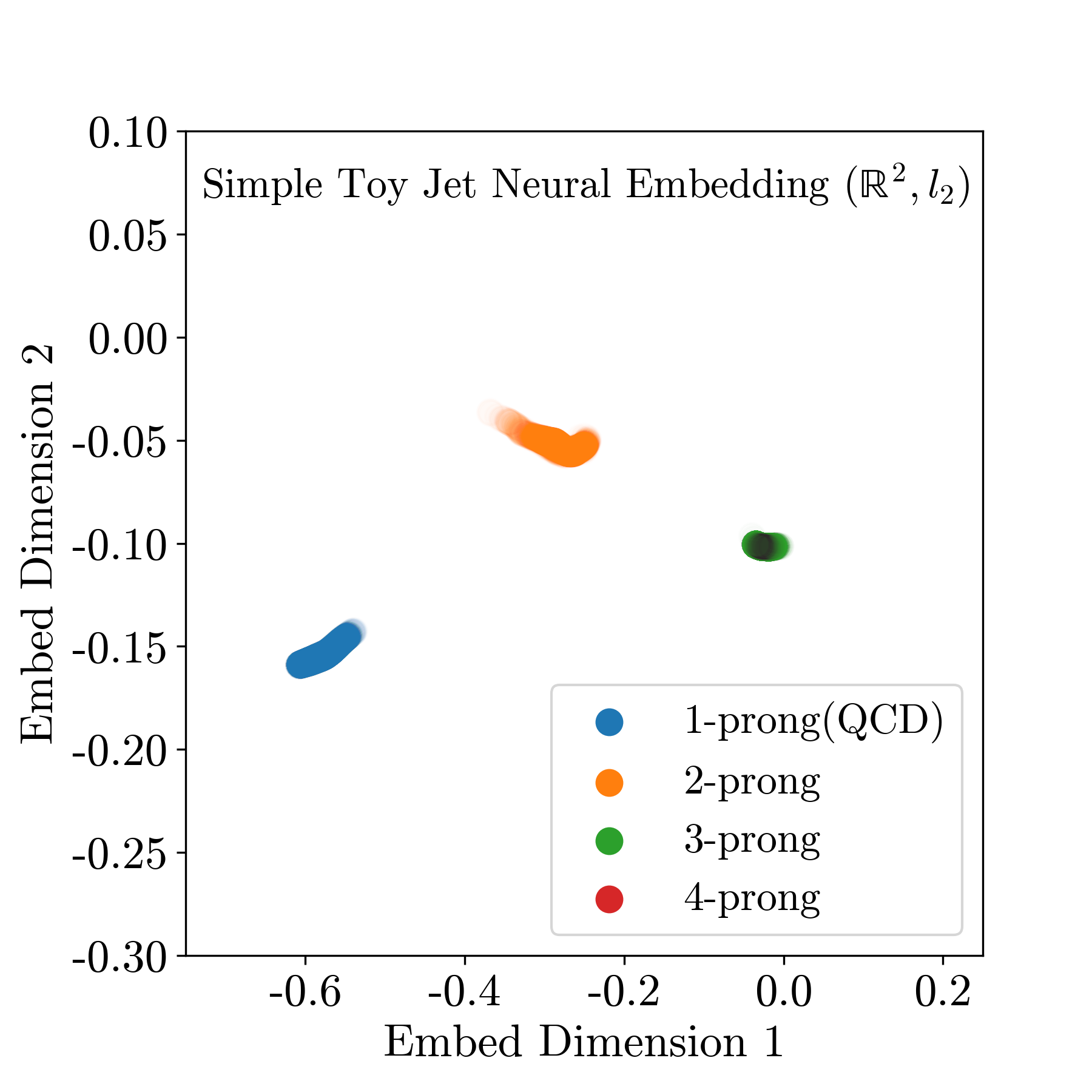}
\includegraphics[width=.45\linewidth]{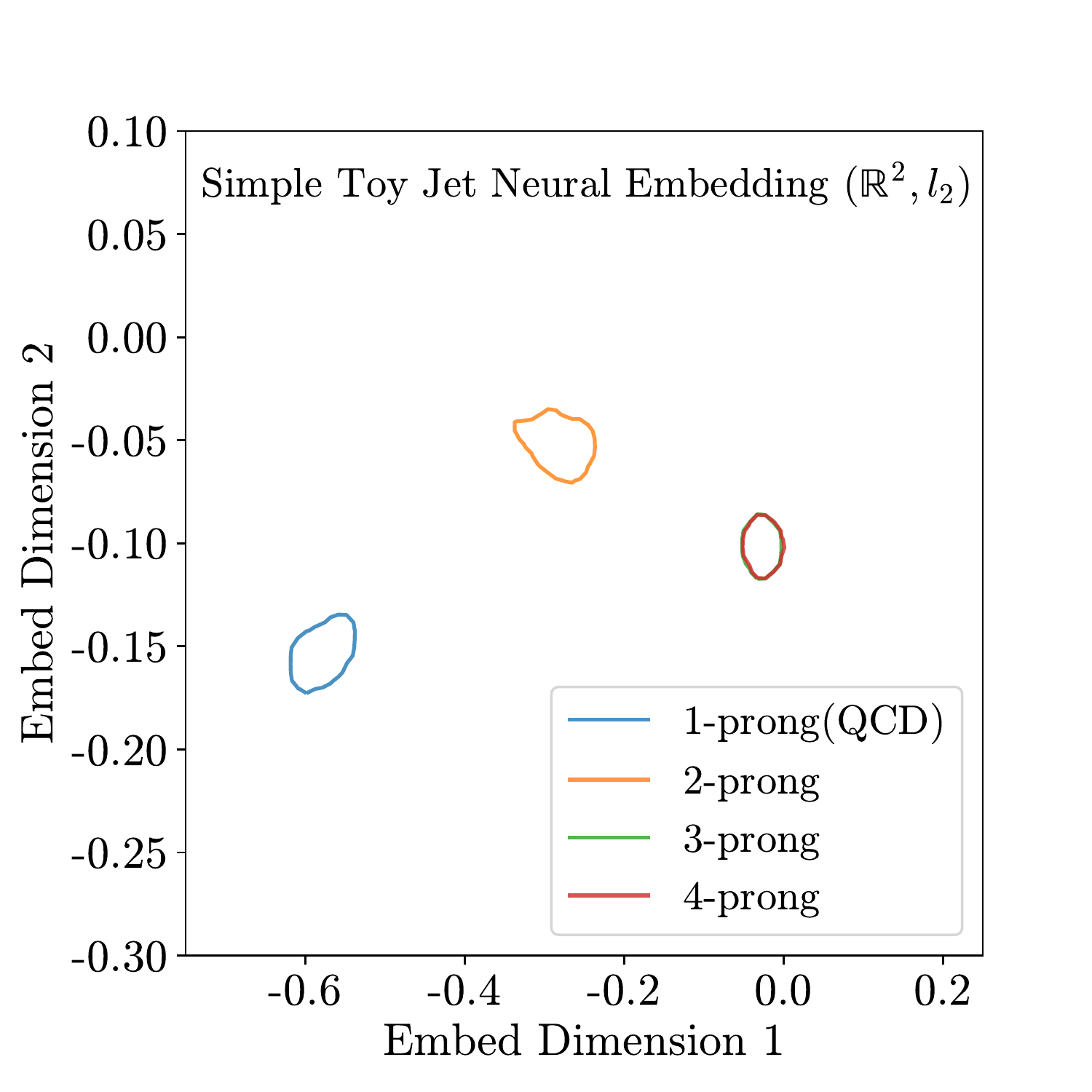}
\caption{ (Left) The embedding of realistic toy jets for 1-prong(QCD), 2-prong, 3-prong, and 4-prong jets. The same embedding smoothed with kernel density estimator, with contour lines corresponding to cdf value 0.5 and 0.8. 
}
\label{fig:simple_embeddingplot}
\end{figure}

The result of the NE is shown in Fig.~\ref{fig:simple_embeddingplot}. We observe that all the points form a small cluster according to their pronginess, with a small diffuse shape for 2-prong and 1-prong(QCD) jets, and 3-prong and 4-prong jets get mapped to an almost identical region. The embedding shows a  simple structure and we see that it reflects the raw EMD distribution in Fig.~\ref{fig:emdsimplejet} well.

\subsubsection{Realistic Toy Jets}
\label{sec:realistictoyjets} 
 
\begin{figure}[htbp]
\centering
\includegraphics[width=.242\linewidth]{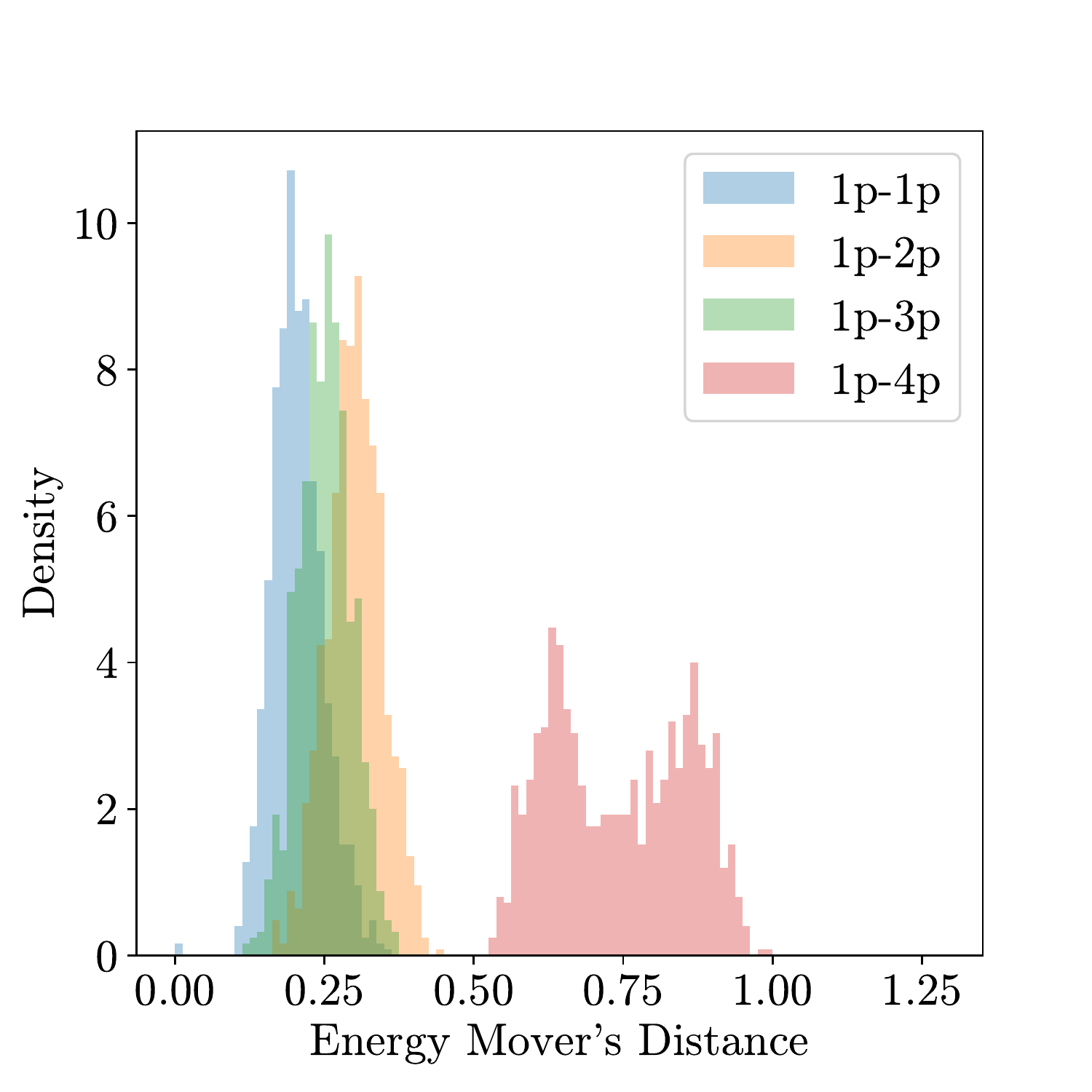}
\includegraphics[width=.242\linewidth]{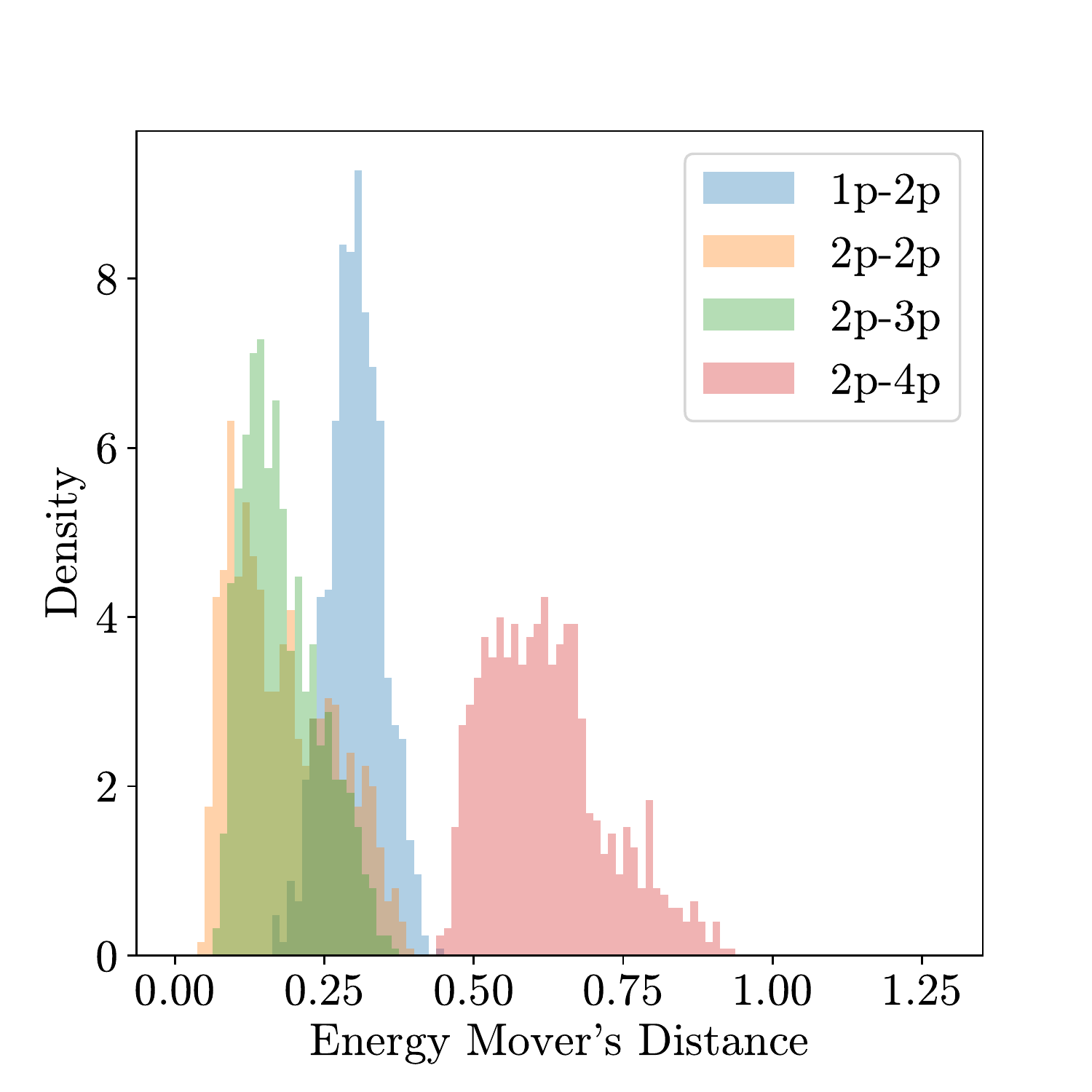}
\includegraphics[width=.242\linewidth]{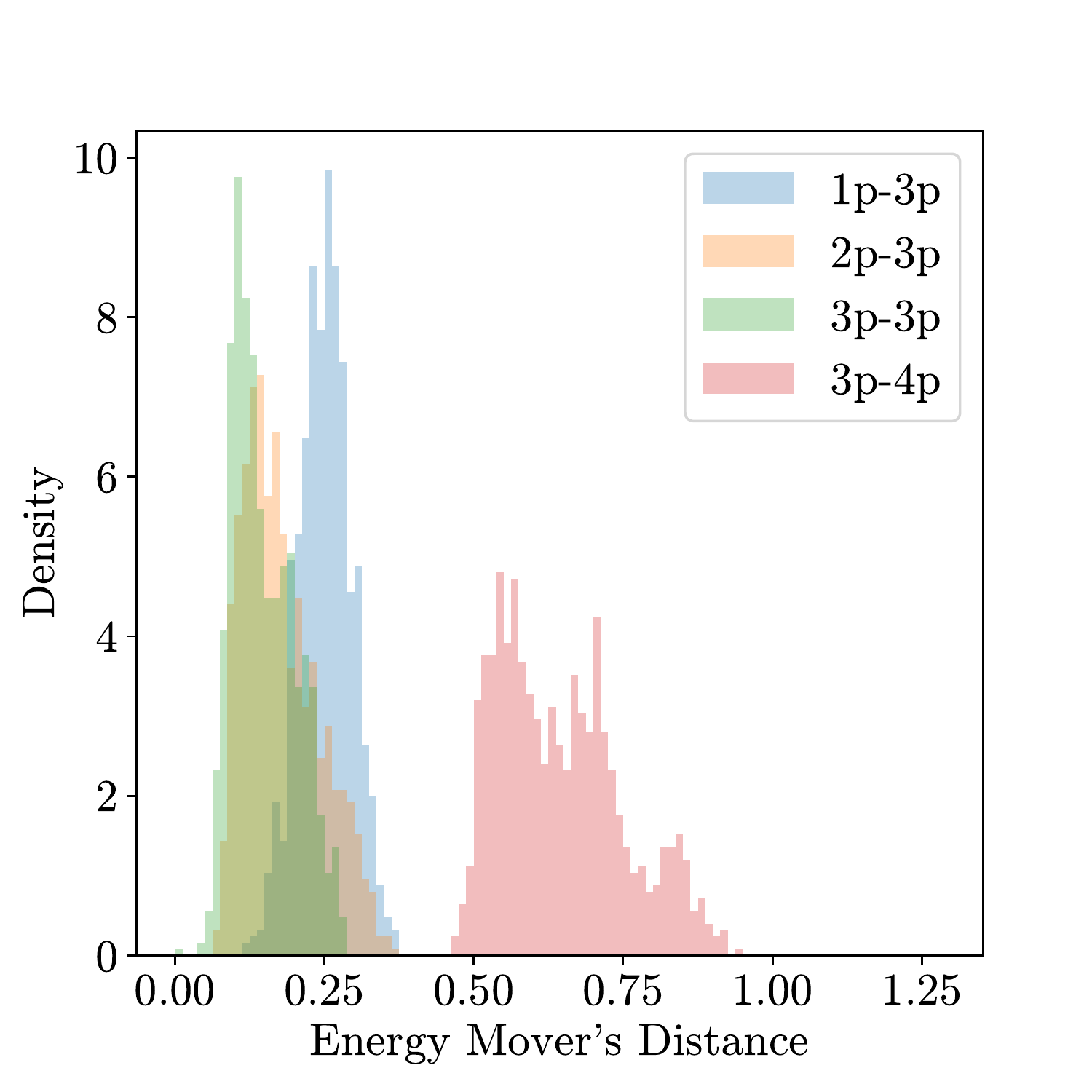}
\includegraphics[width=.242\linewidth]{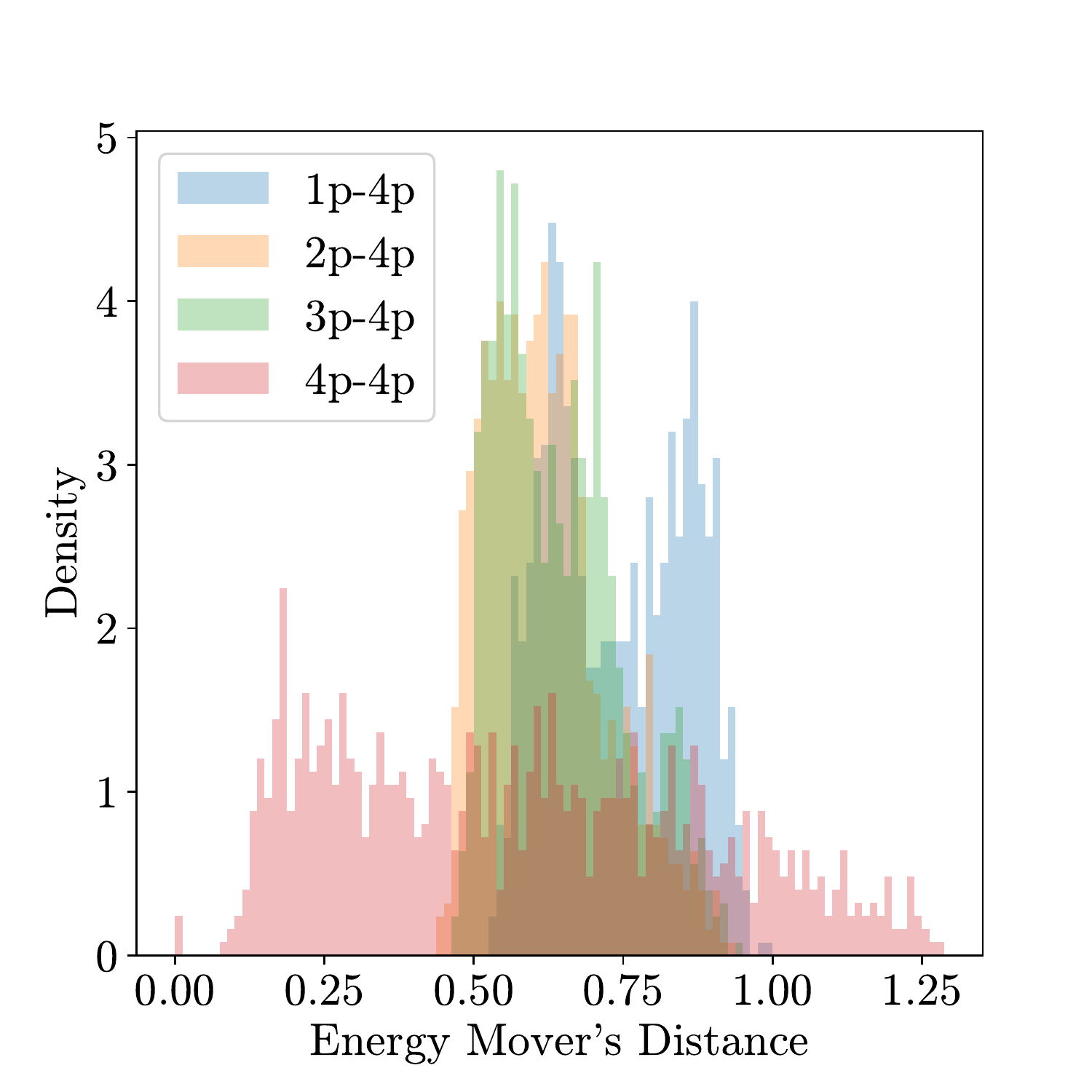}

\caption{(Left) The distribution of energy mover's distance between QCD jets and other jets. (Middle Left) The distribution of energy mover's distance between 2-prong jets and other jets. (Middle Right) The distribution of energy mover's distance between 3-prong jets and other jets. (Right) The distribution of energy mover's distance between four prong jets and other jets.}
\label{fig:emdrealisticjet}
\end{figure}

\begin{figure}[htbp]
\centering
\includegraphics[width=.45\linewidth]{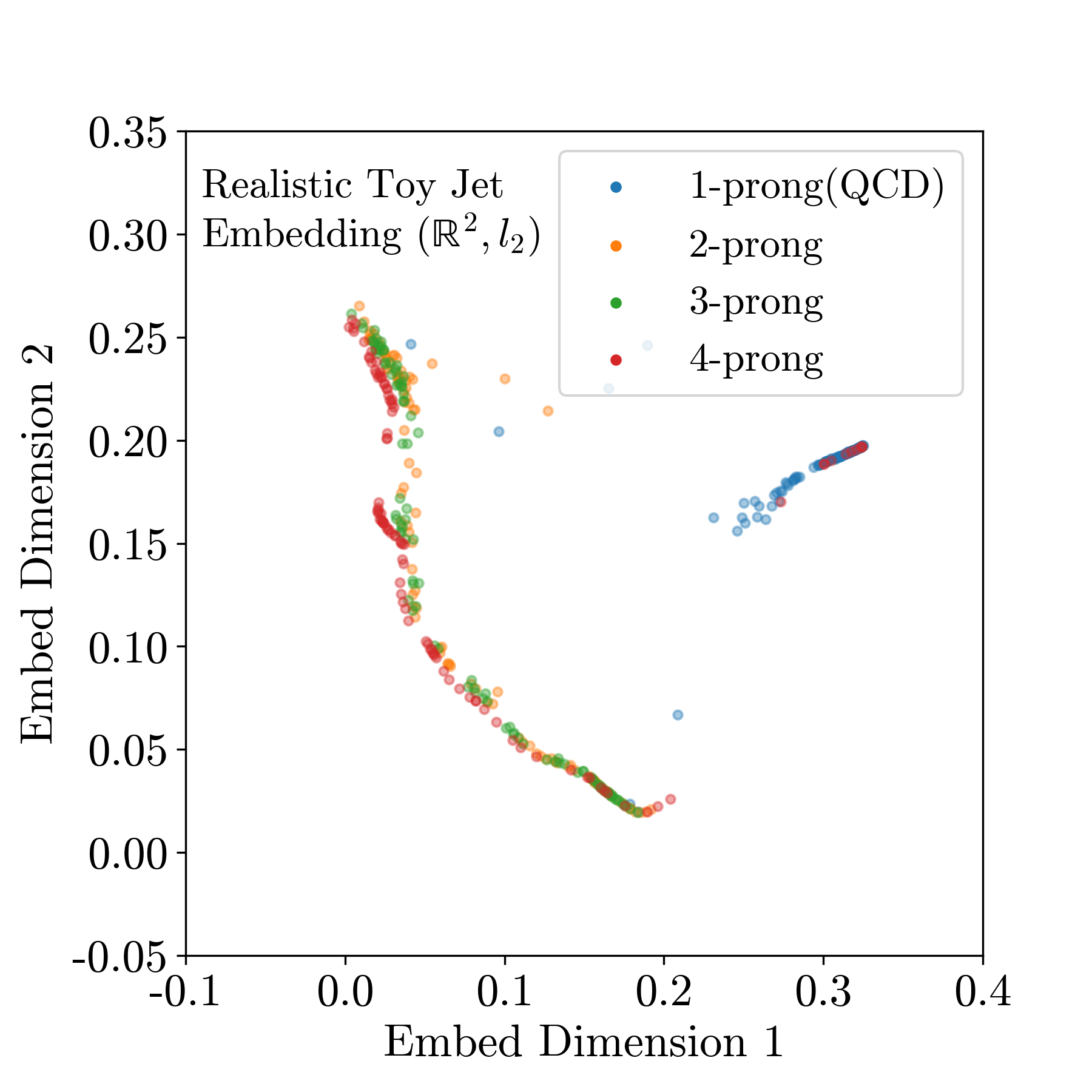}
\includegraphics[width=.45\linewidth]{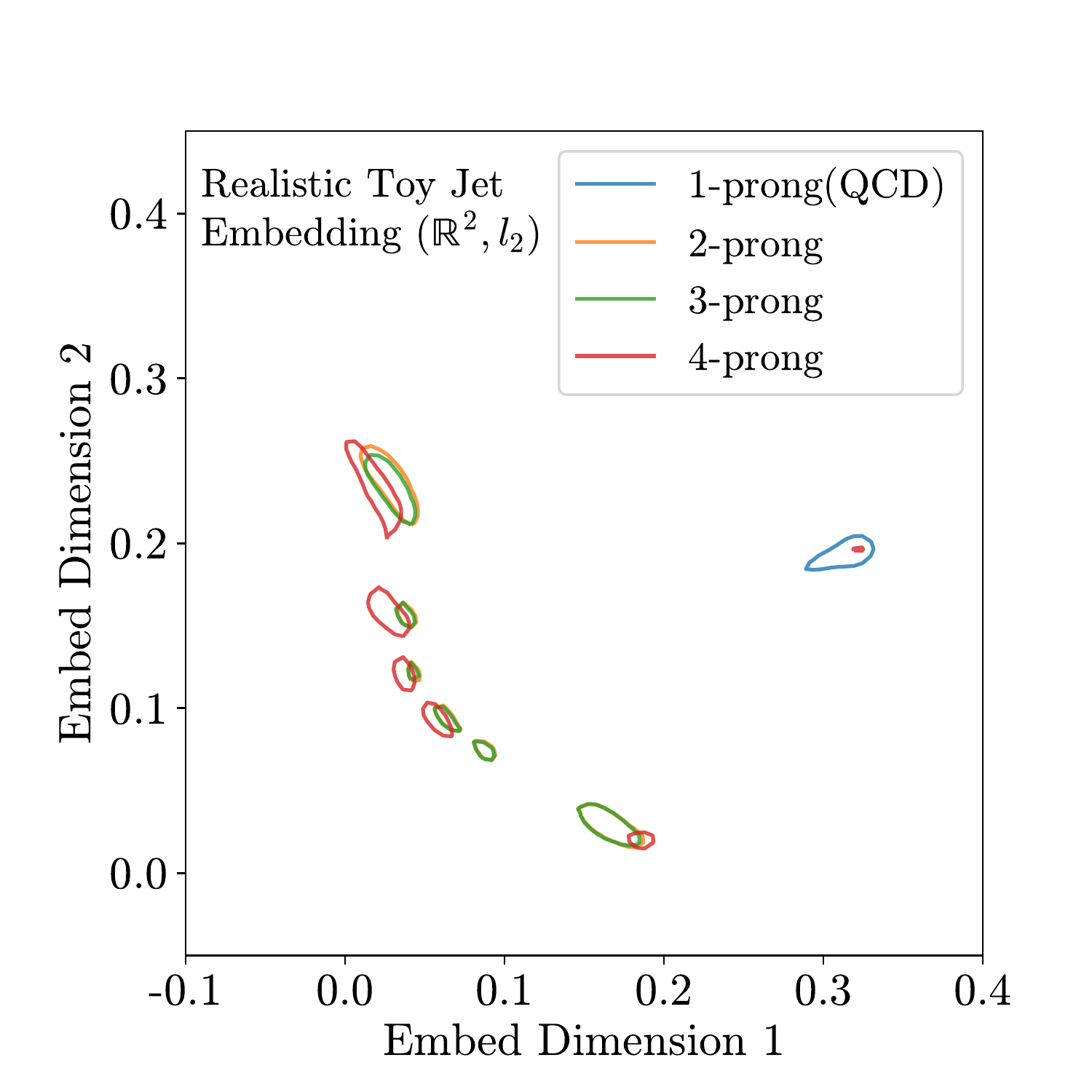}
\caption{ (Left) The embedding of realistic toy jets for 1-prong(QCD), 2-prong, 3-prong, and 4-prong jets. (Right) The same embedding smoothed with kernel density estimator, with contour lines corresponding to cdf value 0.5. 
}
\label{fig:realistic_embeddingplot}
\end{figure}

The distribution of pairwise energy mover's distance for realistic toy jets is shown in Fig.~\ref{fig:emdrealisticjet}. The NE is shown in Fig.~\ref{fig:realistic_embeddingplot}. We observe a similar trend to that of the simple toy jets. We see that 1-prong(QCD) jets form a cluster on their own, and 2-prong, 3-prong, and 4-prong forming clusters around 1-prong jets.

\begin{figure}[htbp]
\centering
\includegraphics[width=.99\linewidth]{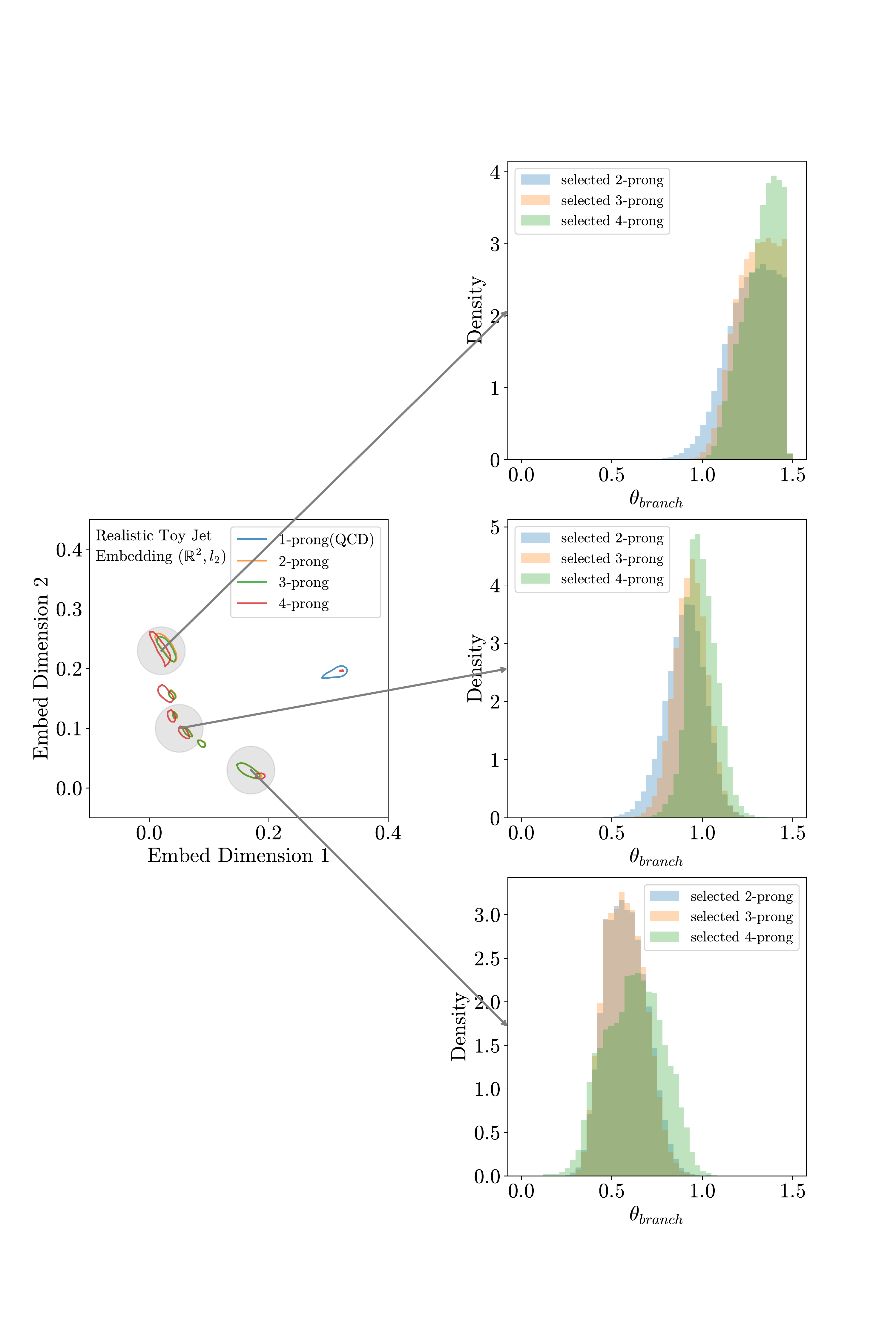}
\caption{ Selecting three different regions of the embedded space for realistic toy jets and plotting the first splitting angle $\theta_{branch}$ of the jets that fall into each of those regions.}
\label{fig:realisticwhatislearned}
\end{figure}

We can further investigate what is learned in these embeddings by choosing different regions of the embedded space and looking at the first splitting angle $\theta_{branch}$ in the rest frame of the jets. We see in Fig.~\ref{fig:realisticwhatislearned} that in the case of realistic toy jets, the first splitting angle is learned very well by the embedding, and it uses this feature to start organizing the dataset. 
 
\subsection{Simulated Jets}
\label{sec:SimulatedJetsEuclidean}
Finally, we consider a set of simulated true jets. For these events, we rely on events generated with MADGRAPH \cite{Alwall:2014hca}, showered with Pythia \cite{Skands:2014pea, Sjostrand:2014zea} and then smeared using Delphes \cite{delphes}. Jets were clustered with FastJet \cite{cacciari2006dispelling, cacciari2012fastjet}.  In this scenario, we generate events  from a wide variety of different topologies  consisting of QCD, 2-prong, and 3-prong jets of varying masses. From this dataset, we perform a single training on all of these topologies to construct the embedded space. In all cases, we train on a single jet.

\begin{table}[h]
\centering
\caption{Summary of simulated jet samples} \label{tab:samples}

\vspace{3pt}

\small
\begin{tabular}{ccccc}
\toprule
\textbf{Events} & \textbf{Jet Pronginess} & \textbf{Jet Mass} & \textbf{Used in Training} & \textbf{Test Dataset} 
\\ \midrule
$X \rightarrow Y Y^{\prime}$ & 2 & $25\GeV$ & \xmark & Extrapolation\\
$X \rightarrow Y Y^{\prime}$ & 2 & $80\GeV$ & \cmark & Interpolation\\
$X \rightarrow Y Y^{\prime}$ & 2 & $170\GeV$ & \xmark & Extrapolation\\
$X \rightarrow Y Y^{\prime}$ & 2 & $400\GeV$ & \cmark & Interpolation\\
$W^{\prime}\rightarrow B^{\prime} T$ & 3 & $25\GeV$ & \xmark & Extrapolation\\
$W^{\prime}\rightarrow B^{\prime} T$ & 3 & $80\GeV$ & \cmark & Interpolation\\
$W^{\prime}\rightarrow B^{\prime} T$ & 3 & $170\GeV$ & \xmark & Extrapolation\\
$W^{\prime}\rightarrow B^{\prime} T$ & 3 & $400\GeV$ & \cmark & Interpolation\\
$V_{kk} \rightarrow (VV)V$ & 4 & $170\GeV$ & \xmark & Extrapolation\\
$V_{kk} \rightarrow (VV)V$ & 4 & $400\GeV$ & \xmark & Extrapolation\\

\bottomrule
\end{tabular}
\end{table}

Table~\ref{tab:samples} summarizes the different samples utilized for the training and testing of the embedded space. To demonstrate the robustness of the construction, we eliminated a variety of mass points in the training, along with all 4-prong samples. However, we still use these samples in the testing of the space. 

The testing is done in two different datasets, the interpolation, and the extrapolation dataset. The extrapolation set is a collection of jets of types not shown in training. This dataset includes 4-prong jets and 2-prong and 3-prong jets with masses eliminated from the training. The interpolation set is a collection of jets of types shown in training that shows the interpolation capability of these methods, such as 2-prong and 3-prong jets with masses shown in training that were held out for testing and have no overlap with the training dataset. For all jet types, one million jets were used in training, 200k jets each for validation and testing, and 10k jets were used for presenting the results.

The QCD jets are constructed from events generated with MADGRAPH and showered with PYTHIA, with an HT range of $1500$ to $2000 \GeV$. A pre-selection on the jets is applied so that the $\pt$ of the jets are greater than $300 \GeV$.  The two prong jets are generated from $X \rightarrow Y Y^{\prime}$ process, with $Y$ and $Y^{\prime}$ masses 25,80, 170, 400 $\GeV$. We use masses $80, 400 \GeV$ for training and $25, 170 \GeV$ for testing in the interpolated dataset.  The 3-prong jets are generated from $W^{\prime}\rightarrow B^{\prime} T$ events, with both $W^{\prime}$ and Top quark mass varied, both decaying to a 3-prong. As with the 2-prong sample masses, 80, 400$\GeV$ are used for the training and 25 and 170$\GeV$ for testing.  Lastly, one million 3-prong jets are used for training, and 200k jets for validation and testing, just as in the 2-prong case. For the 4-prong jets, two mass points 170 and 400~\GeV are generated from triboson events $V_{kk} \rightarrow (VV)V$, where two bosons, both decaying 2-prong, get clustered in the same jet. These jets weren't shown in the training at all and constitute the extrapolation test dataset. For the respective labels, we only take jets with explicitly 2,3,or 4 prongs for the NE.

The OT-based distances between the jets are very sensitive to preprocessing. As a result, we apply a more complex pre-processing scheme. First, the jets are centered so that the jet $\eta$ and $\phi$ are centered to the origin. Then the jet constituents are rotated with respect to the origin so that two most energetic components are aligned along the y axis of the $(\Delta \eta = \eta_i - \eta_{\text{jet}}, \Delta \phi=\phi_{\text{jet}})$ coordinate system. Finally, the jets are flipped so that the maximum sum of energy of constituents is placed in the first quadrant in the  $(\Delta \eta, \Delta \phi)$ plane. Examples of such jets are shown in Appendix~\ref{sec:example_jets}.  

\begin{figure}[htbp!]
\centering
\includegraphics[width=.455\linewidth]{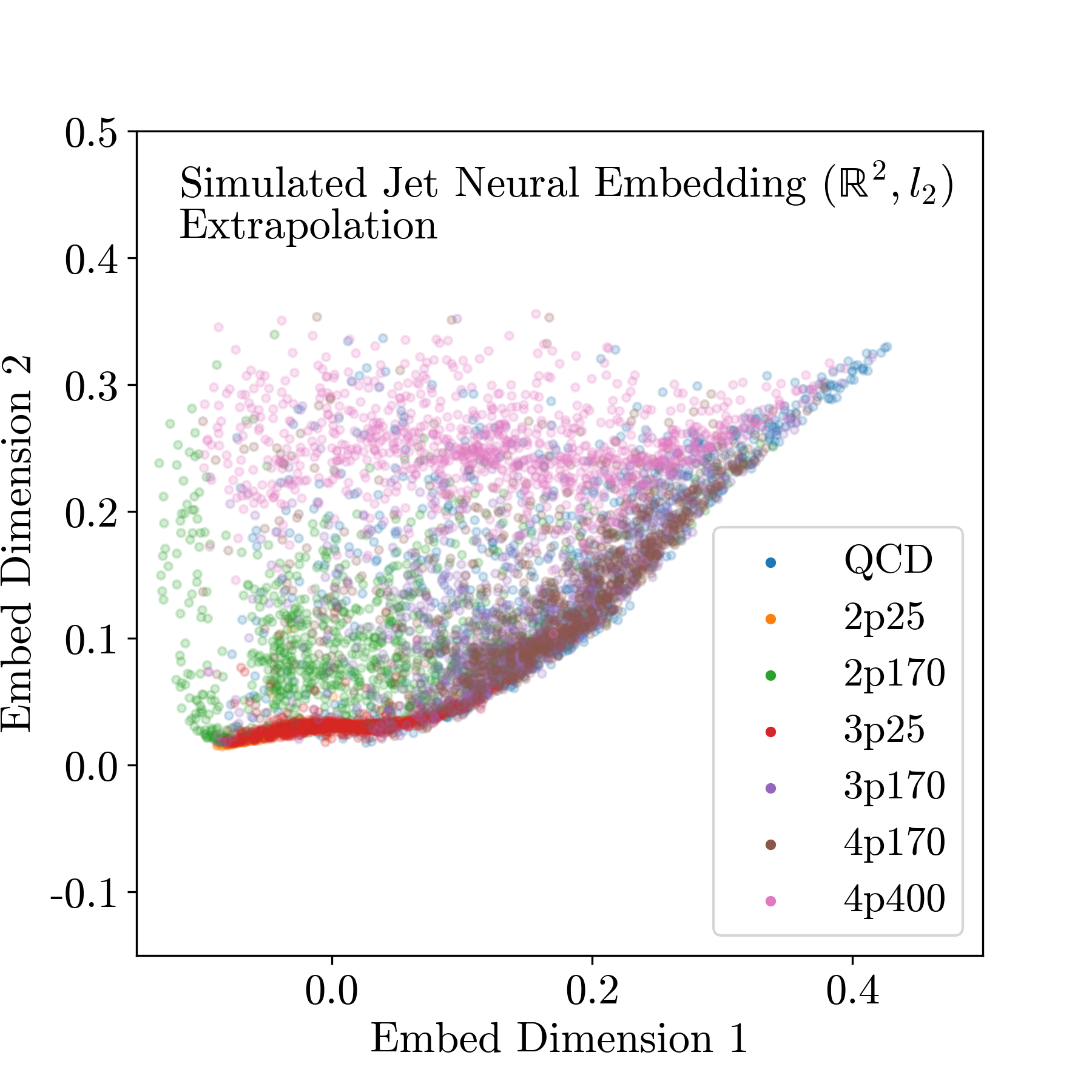}
\includegraphics[width=.43\linewidth]{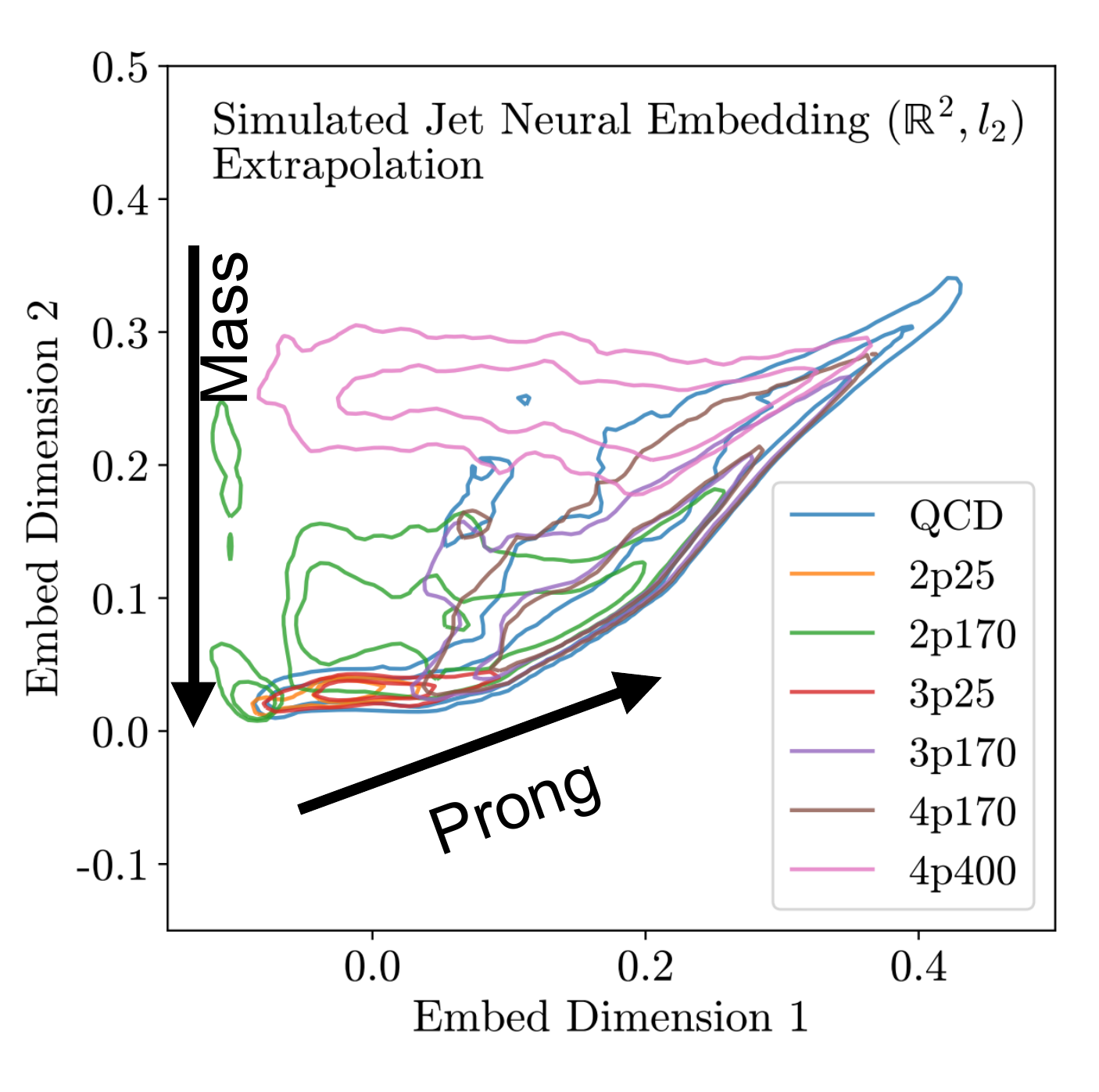}
\caption{ (Left) The embedding (Right) The same embedding smoothed with kernel density estimator, with contour lines corresponding to CDF value 0.5 and 0.8. Labels indicate the pronginess and the mass of the jet. For instance, 2p25 indicates 2-prong jets with mass 25$\GeV$, generated from $X\rightarrow Y Y^{\prime}$ model.}
\label{fig:embeddingplot_simulated_extrapolate}
\end{figure}

Fig. ~\ref{fig:simulatedjetEMDdist} shows the EMD for the jets, we observe that the energy mover's distance is more sensitive to varying the mass compared to varying the pronginess of the jets. Fig.~\ref{fig:embeddingplot_simulated_extrapolate} shows the result of the NE applied to the extrapolation dataset. We observe a strong grouping according to the jet masses with a general progression towards smaller masses as one goes to smaller values on the y-axis. We also observe a trend towards lower prongs as one moves closer to the origin in the embedded space. As a consequence of these trends, we find 2-prong and 3-prong jets with 25$\GeV$ mass get grouped in the bottom left corner, and 2-prong, 3-prong, and 4-prong jets with masses $170\GeV$ get grouped above the 25$\GeV$ mass group. Above the $170\GeV$ group, 4-prong jets with $400\GeV$ mass are placed.

\begin{figure}[htbp!]
\centering
\includegraphics[width=.97\linewidth]{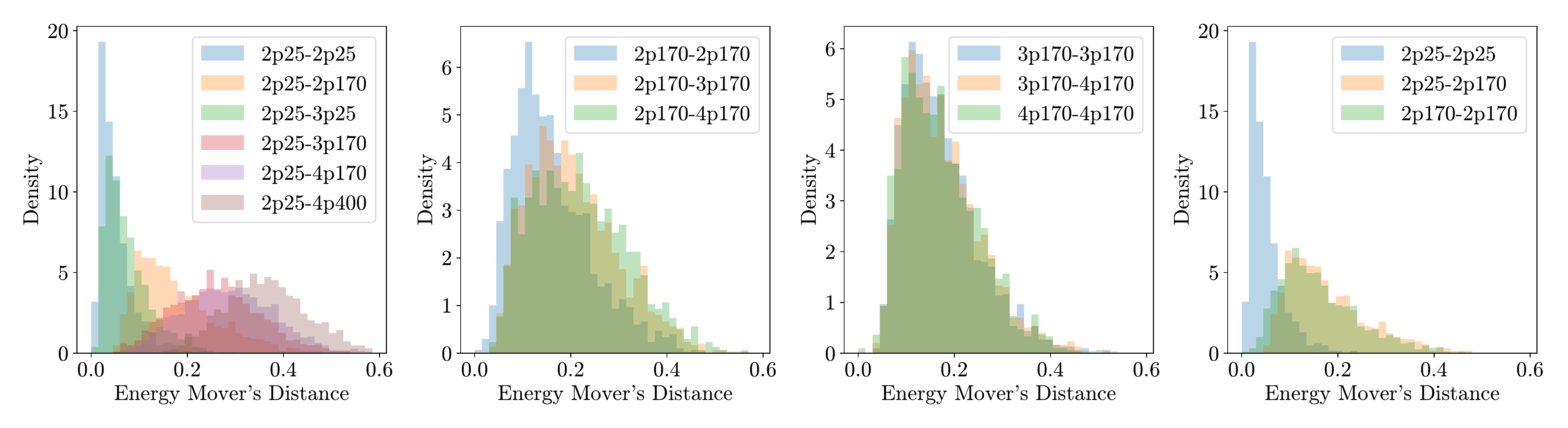}
\caption{ (Left) The distribution of energy mover's distance(EMD) between 2-prong jets with mass 25 \GeV jets and other jets. (Middle Left) The distribution of energy mover's distance between different jets with fixed mass of 170 \GeV. (Middle Right) The distribution of energy mover's distance between different jets with fixed mass of 170 \GeV.  (Right) The distribution of energy mover's distance between 2-prong jets with different masses}
\label{fig:simulatedjetEMDdist}
\end{figure}

\begin{figure}[htbp!]
\centering
\includegraphics[width=.45\linewidth]{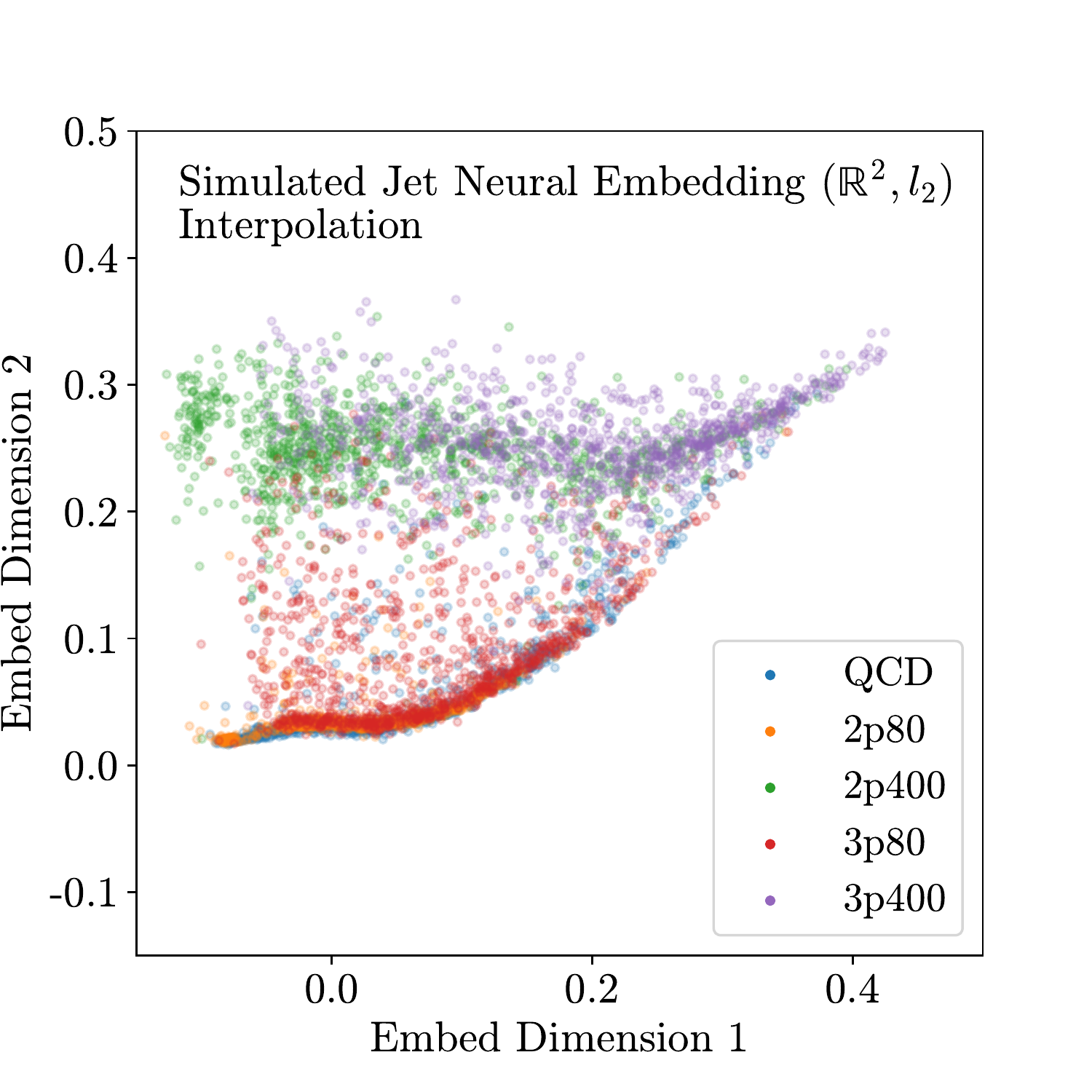}
\includegraphics[width=.45\linewidth]{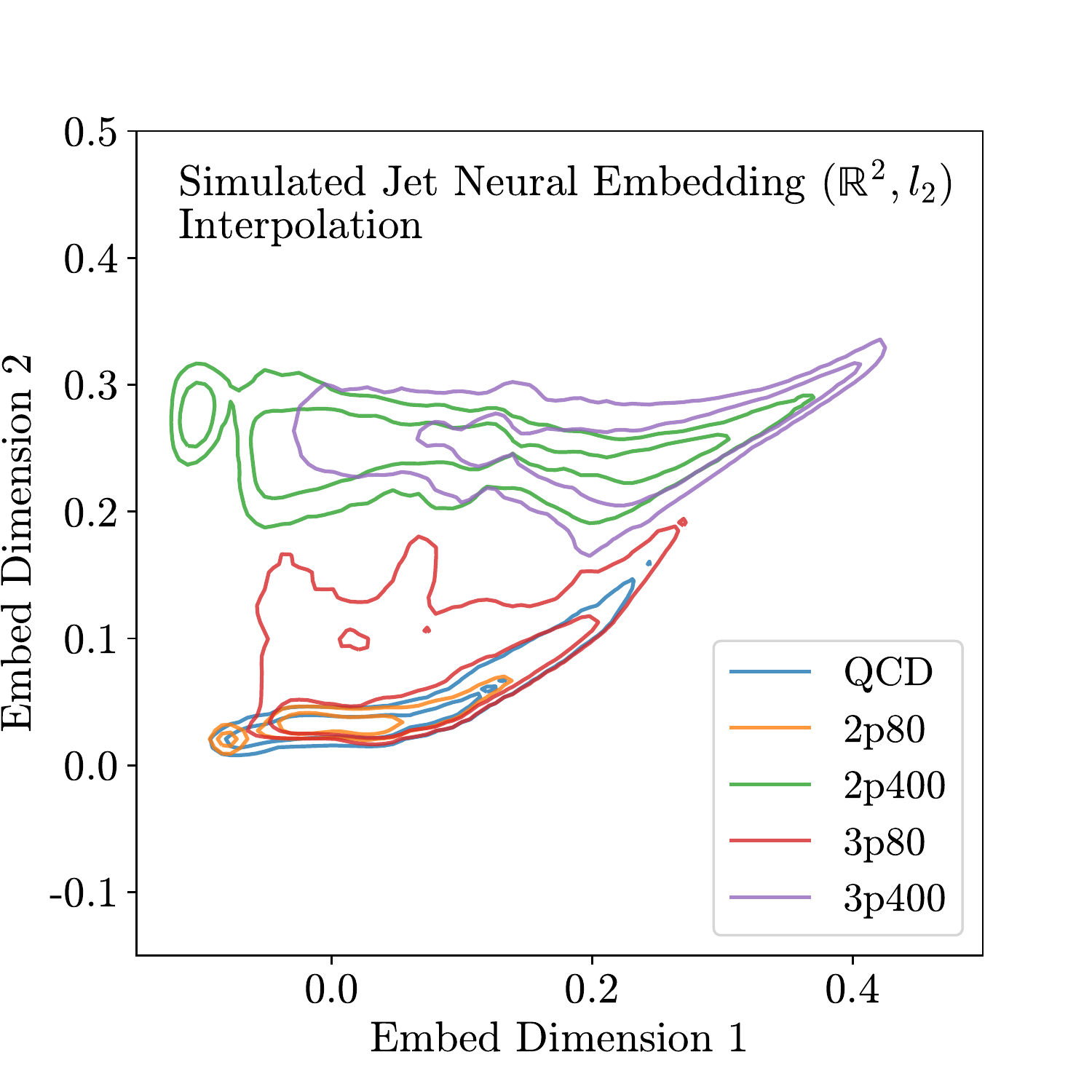}
\caption{ (Left) The embedding (Right) The same embedding smoothed with kernel density estimator, with contour lines corresponding to cdf value 0.5 and 0.8.  
}
\label{fig:embeddingplot_simulated_interpolate}
\end{figure}

Fig.~\ref{fig:embeddingplot_simulated_interpolate} shows the result of the NE on the interpolated datasets. With these datapoints, we again observe that there is an even stronger grouping based on the mass, QCD with all the 80$\GeV$ jets getting mapped to the bottom half of the space, and all the $400\GeV$ jets getting mapped to the top half of the space.

In order to further understand what is learned by the embedding functions, in Fig.~\ref{fig:embeddingplot_simulated_interpolate}, we look at different regions of the NE and plot physical observables. We choose different regions of the Euclidean space and plot the histograms of subjettiness variables $\tau_{21}, \tau_{32}, \tau_{43}$. We find a strong correlation among selected subjettiness variables even for QCD jets. In particular, the pronginess consistently goes down to lower values as one progresses towards the origin of the embedded space.  Already, we can see that with this embedded space, we can start to classify jets into distinct regions based on their features and their generated properties.

\begin{figure}[htbp!]
\centering
\includegraphics[width=.99\linewidth]{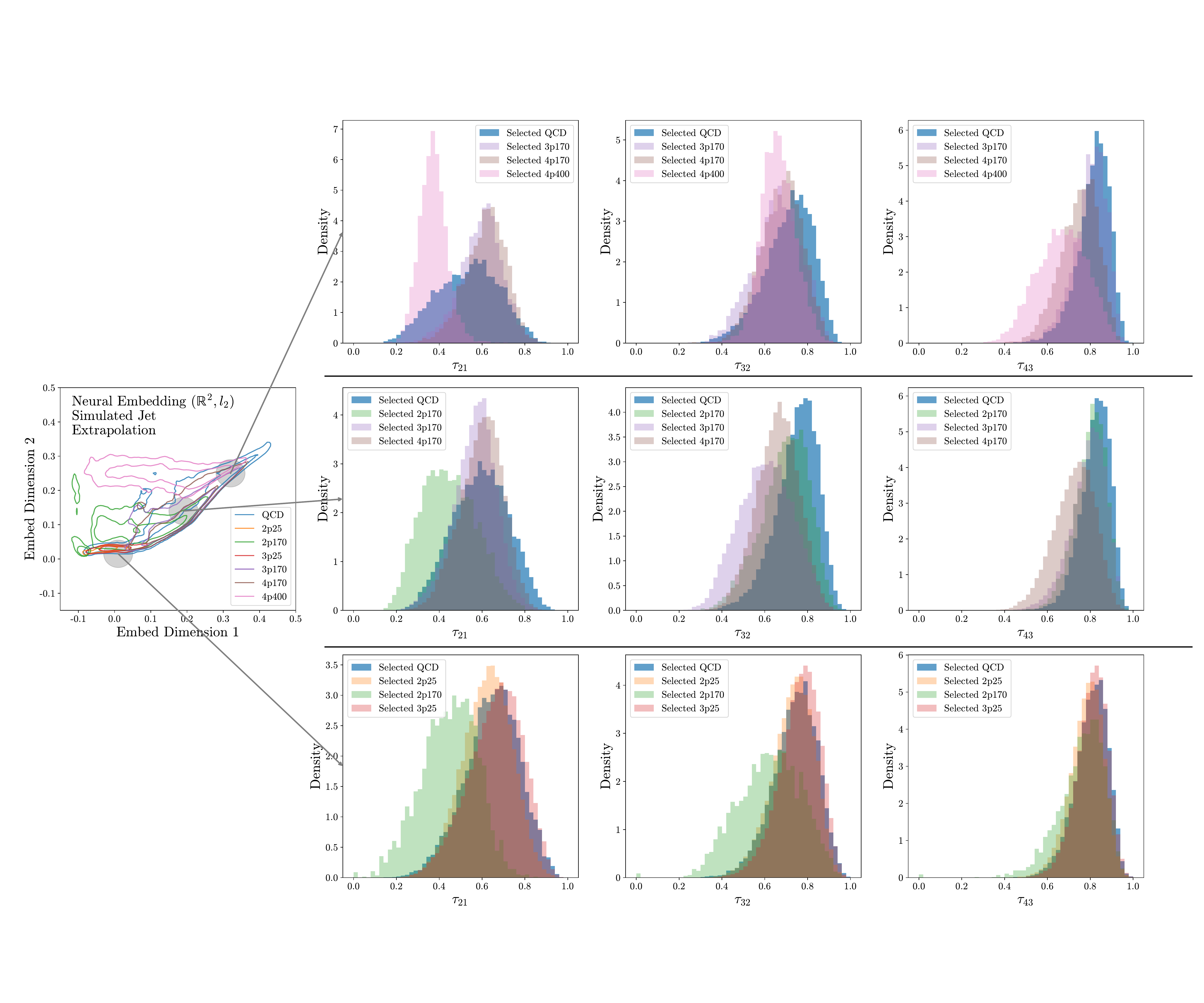}
\caption{ (Left) The embedding (Right) The same embedding smoothed with kernel density estimator, with contour lines corresponding to cdf value 0.5. 
}
\label{fig:embeddingplot_whatislearned_extrapolate}
\end{figure}


\subsection{Hyperbolic Embedding of Jets}
\label{sec:simulatedjethyperbolic}
In the above section, we performed a NE into a Euclidean space. In this section, we present results on embedding into non-Euclidean, Hyperbolic spaces. For this paper, we primarily study embedding into the two-dimensional Poincaré disks. It is well known that tree-like structure embeds well into the Poincaré disk since the distance gets stretched close to the boundary of Poincaré disks \cite{nickel2017poincare, corso2021neural, peng2021hyperbolic, klimovskaia2020poincare}. Thus it is interesting to view jets as tree structures and embed them into Poincaré disks. 

Hyperbolic space has a physical analog and is used to define the motion of high momentum objects within Minkowski space. The jet is a high momentum object that decays into a spray of high momentum particles. As a result, its decay products and the forces causing the decay undergo relativistic motion giving rise to a curved hyperbolic  geometry. 
Lastly, since it is impossible to embed non-Euclidean manifolds into Euclidean space without a big distortion, alternative geometries are well motivated extensions of NE and should, in general, be pursued. 

With the same metric space and EMD distribution as in Fig.~\ref{fig:simulatedjetEMDdist} and the same training set, we learn the function in Eq.~\ref{eq:hyperbolicembedding}, the embedding into the Poincaré disk($\mathcal{B}^2$) denoted as

\begin{equation}
\begin{aligned}
\label{eq:hyperbolicembedding}
\phi_{\theta, \mathrm{Transformer}}: (\mathcal{X}_{\mathrm{jets}} \subset \mathbb{R}^{48}, d_\mathrm{EMD} ) \rightarrow (\mathcal{B}^2, d_p)~,
\end{aligned}
\end{equation}

where the metric distance on the disk $d_p$ is given by Eq~\ref{eq:poincaredist}.

\begin{equation}
\begin{aligned}
\label{eq:poincaredist}
d_p (x, y) = \arcosh \left(1 + 2 \cdot \frac{\|x-y\|^2}{(1-\|x\|^2)(1-\|y\|^2)}\right)
\end{aligned}
\end{equation}

\begin{figure}[htbp!]
\centering
\includegraphics[width=.45\linewidth]{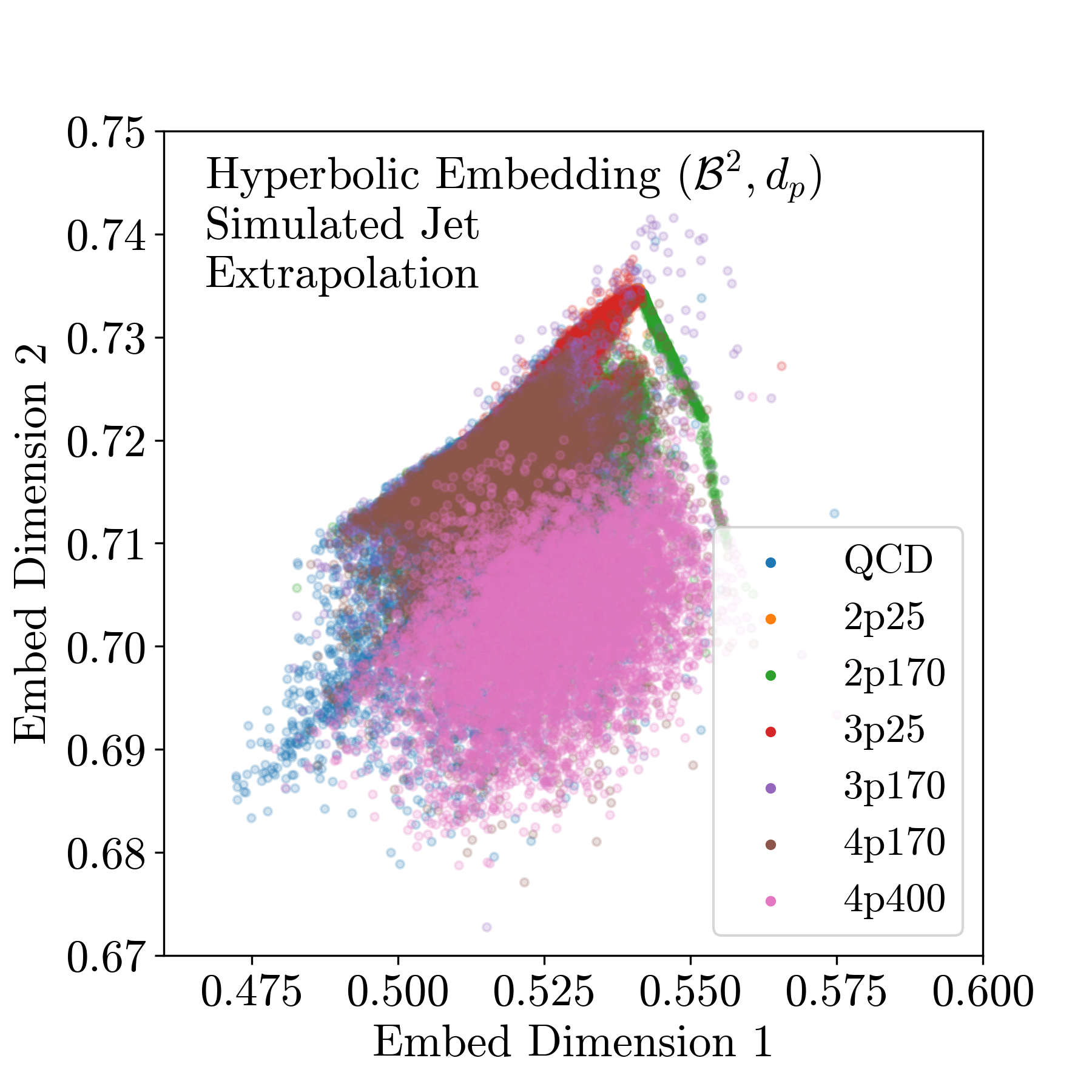}
\includegraphics[width=.445\linewidth]{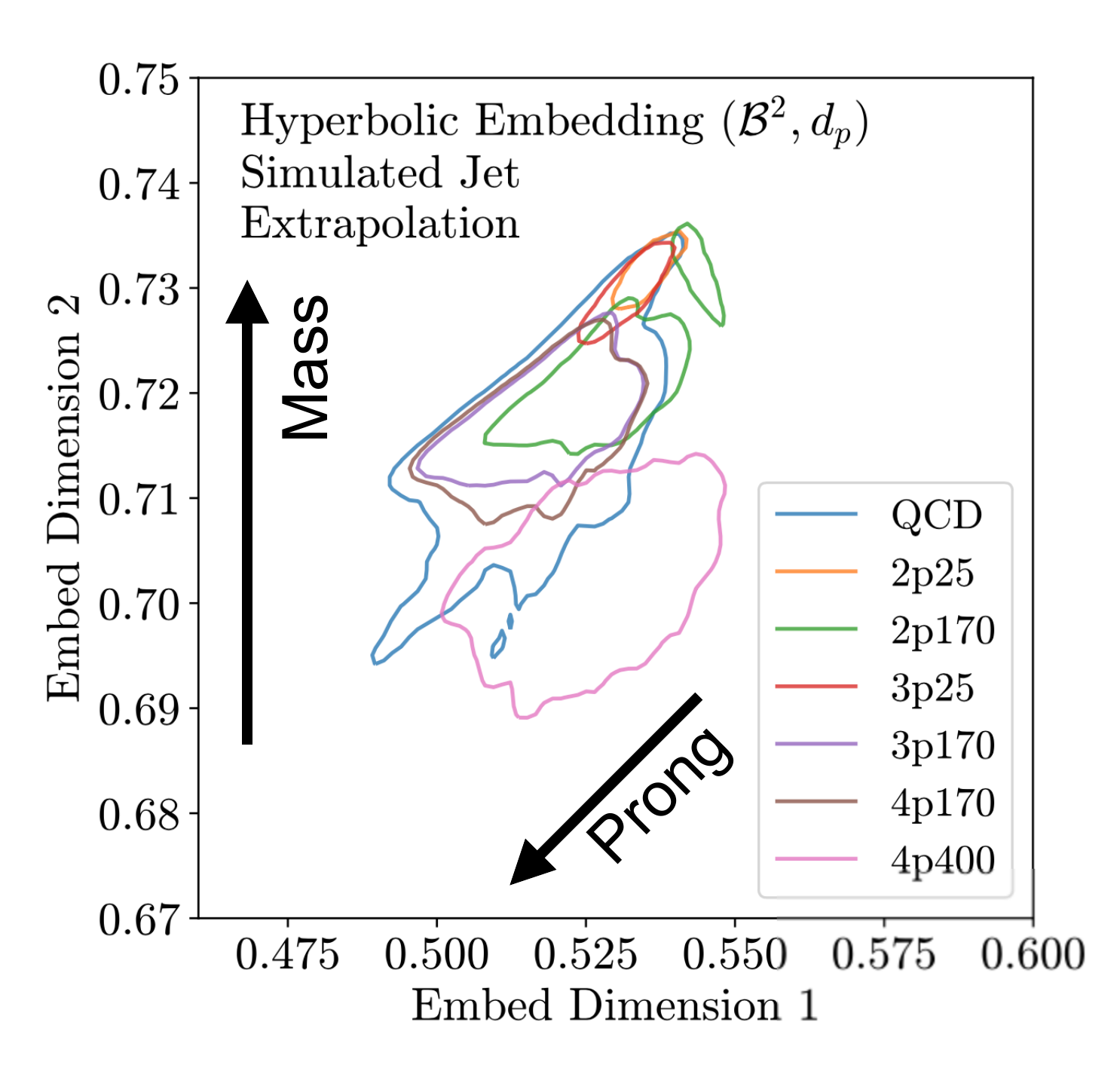}
\caption{ (Left) Scatterplot of Hyperbolic embedding of simulated jets into Poincaré disks $(\mathcal{B}^2, d_p)$  (Right) The same embedding smoothed with kernel density estimator, with contour lines corresponding to cdf value 0.5 and 0.8. Labels indicate the pronginess and the mass of the jet. For instance, 2p25 indicates 2-prong jets with mass 25$\GeV$, generated from $X\rightarrow Y Y^{\prime}$ model.}
\label{fig:hyperbolic_simulated_extrapolate}
\end{figure}

As in the Euclidean embedding cases in Section~\ref{sec:SimulatedJetsEuclidean}, we present the results in two different cases on the same extrapolation and interpolation prediction datasets. The result of the embedding into Poincaré disks for the extrapolation dataset is shown in Fig.~\ref{fig:hyperbolic_simulated_extrapolate}. There is a clear trend towards heavier objects as one goes downwards along the y-axis.  Additionally, we observe a trend toward more prongs as one moves downwards and to the left. As a consequence, we observe a strong grouping based on the mass of the jets.  The $25\GeV$ mass jets are grouped together and also $170\GeV$ jets get mapped to the same regions. Finally, The $400\GeV$ jets form a cluster of their own. This grouping based on mass seems to be stronger compared to the Euclidean embedding case. 

\begin{figure}[htbp!]
\centering
\includegraphics[width=.99\linewidth]{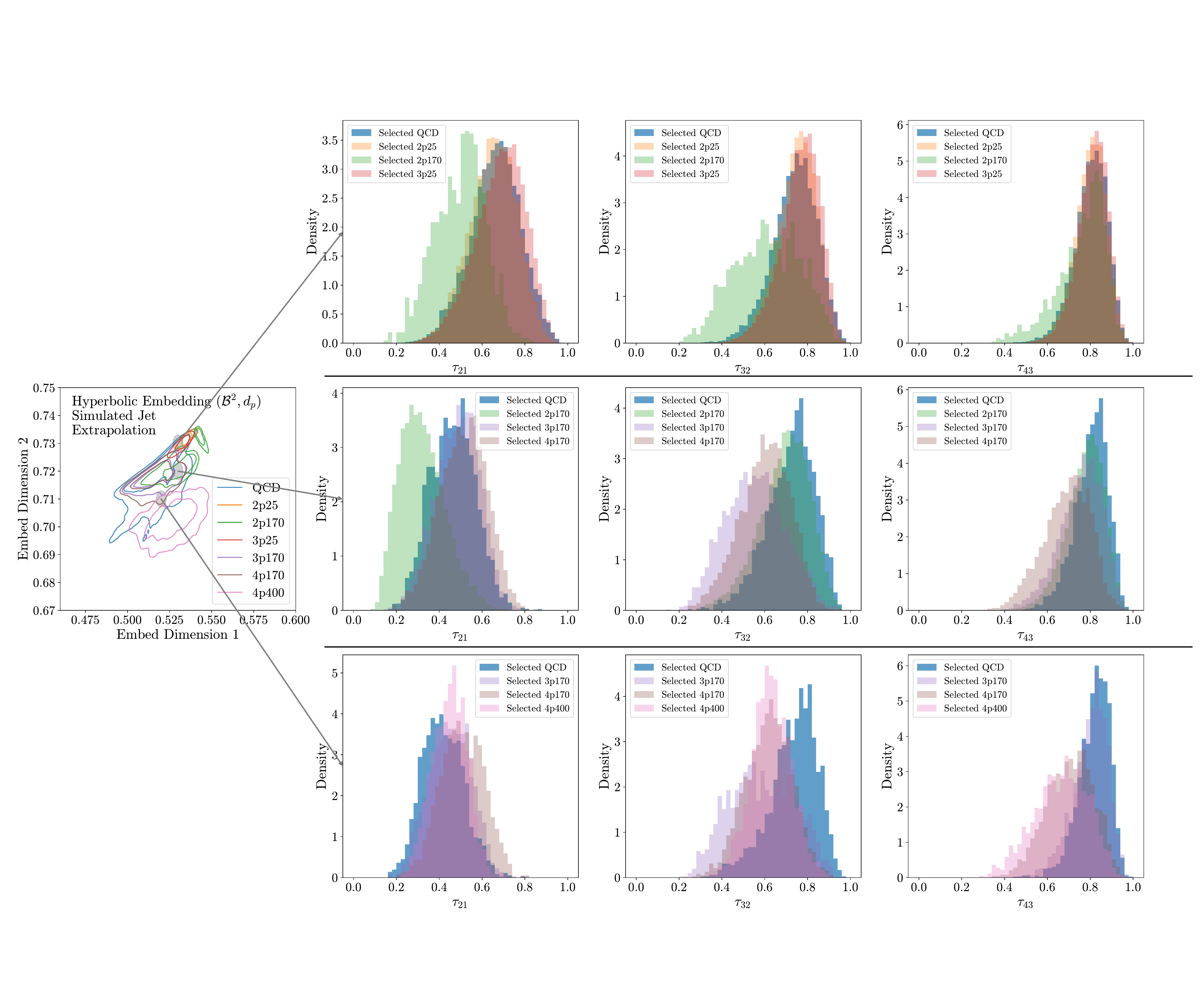}
\caption{ (Left) The embedding (Right) The same embedding smoothed with kernel density estimator, with contour lines corresponding to cdf value 0.5. 
}
\label{fig:embeddingplot_hyper_whatislearned_extrapolate}
\end{figure}

To further see whether the latent structure is learned by the embedding, Fig.~\ref{fig:embeddingplot_hyper_whatislearned_extrapolate} shows $n-$subjettiness variables of jets that get mapped to different regions of the embedded space. We can see that the distributions of $n$-subjettiness are highly correlated within the local regions of the space, even stronger than Euclidean embedding, and we conclude that interpretability is better for embedding into hyperbolic spaces compared to Euclidean spaces. 



\begin{figure}[htbp!]
\centering
\includegraphics[width=.45\linewidth]{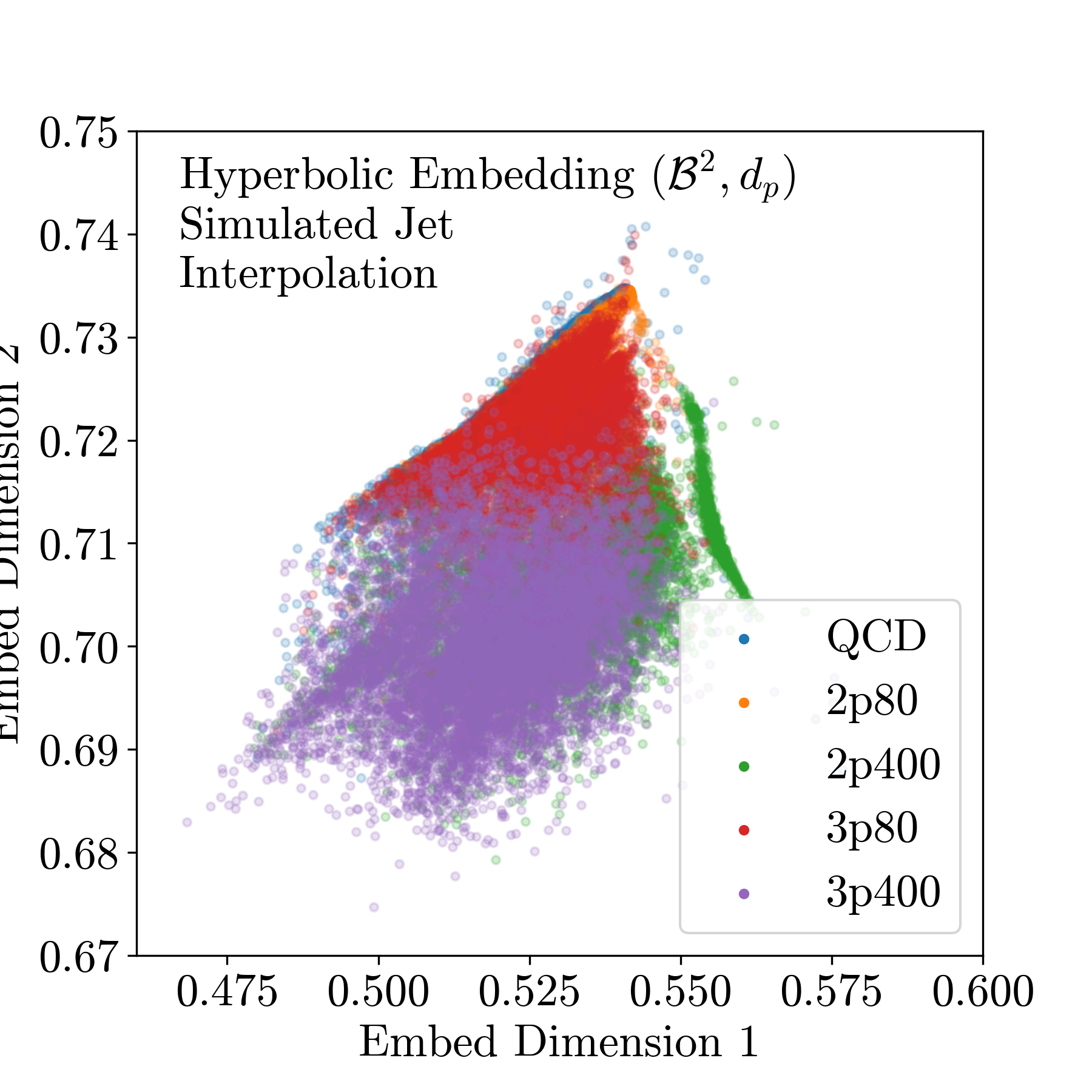}
\includegraphics[width=.45\linewidth]{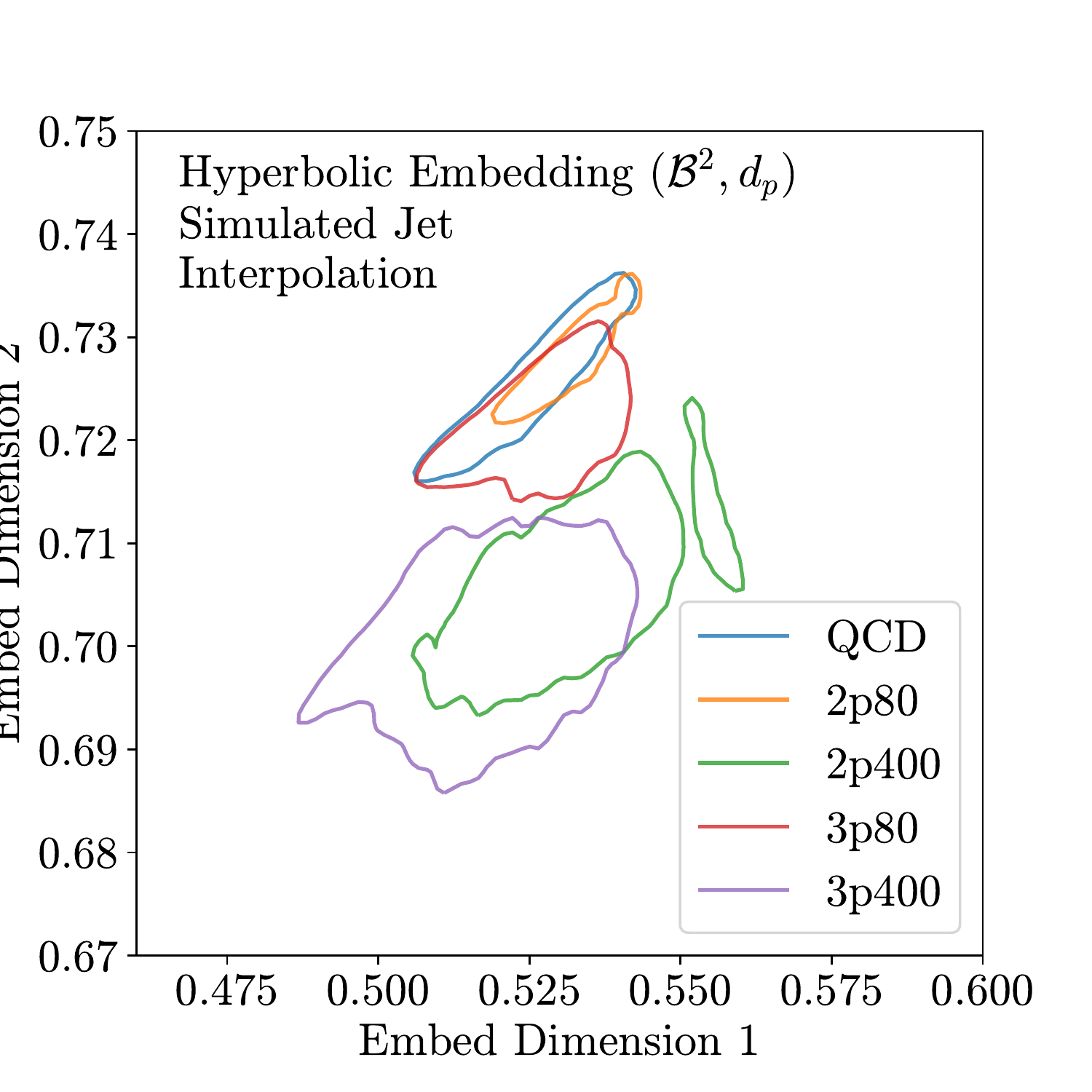}
\caption{ (Left) Scatterplot of Hyperbolic embedding of simulated jets into Poincaré disks $(\mathcal{B}^2, d_p)$  (Right) The same embedding smoothed with kernel density estimator, with contour lines corresponding to cdf value 0.5 and 0.8. Labels indicate the pronginess and the mass of the jet. For instance, 2p25 indicates 2-prong jets with mass 25$\GeV$, generated from $X\rightarrow Y Y^{\prime}$ model.}
\label{fig:hyperbolic_simulated_interpolate}
\end{figure}

The result of the embedding into Poincaré disks for the interpolation dataset is shown in Fig.~\ref{fig:hyperbolic_simulated_interpolate};  the behavior is similar to that of the extrapolation dataset. Overall, we observe there is a strong grouping of objects within this space. This implies the NE has ``self-organized'' the dataset along the EMD criterion, yielding a physically interpretable space consistent with that of EMD.

As we can see from the good separation between different types of events for both Euclidean and Hyperbolic embeddings, we see that these NE can be used to flag interesting anomalies by looking for events within this space. However, a related application that we can also perform is to quantify the effectiveness of anomaly detection algorithms in detecting interesting regions of phase space. This exciting application is covered in \ref{sec:anomalyquantification}. 

\subsection{Empirical Estimation of Distortion}
\label{subsec:distortion}

In addition to a physical clustering of events, we can also look to see how well the embedded space preserves the embedded metric within the space ~\cite{VakandaraDistortion}.
With the given embedding $\phi: (\mathcal{X}, d_\mathcal{X})\rightarrow (\mathcal{Y}, d_\mathcal{Y})$, we define the distortion as the ratio of measured EMD after NE compared to the true EMD, given by

\begin{equation}
\begin{aligned}
\label{eq:pairwiseratio}
\rho_\phi (u, v) = \frac{d_{\mathcal{Y}}(\phi(u), \phi(v))}{d_{\mathcal{X}}(u,v)}
\end{aligned}
\end{equation}
The distortion measures how far the new distances $d_{\mathcal{Y}}(\phi(u), \phi(v))$ between the embedded points deviate from the original distances $d_{\mathcal{X}}(u,v)$ for an arbitrary pair of points $(u, v) \in \mathcal{X}$.

To quantify the level of distortion, we condense the distortion response and variation to two numbers the mean, $\mu$, and the standard deviation, $\sigma$, of the distortion.  Where the variation for any embedding $\phi$ can be written in terms of the normalized ratio of the distances, $\Tilde{\rho_\phi}(u, v)$, given by

\begin{equation}
\begin{aligned}
\label{eq:normratio}
\Tilde{\rho_\phi}(u, v) = \frac{M \rho_\phi (u, v)}{\sum_{i=1}^{M} \rho_\phi (u_i, v_i)},
\end{aligned}
\end{equation}
where the summation is done for all pairs in the test dataset and $M$ denoting the total number of pairs. The distortion variation $\sigma$-distortion is defined as, letting $\Pi = P \times P$, where $P$ is a distribution over $\mathcal{X}$,

\begin{equation}
\begin{aligned}
\label{eq:sigdistortion}
\sigma{\rm-distortion} = \mathbb{E}_{\Pi} (\Tilde{\rho_\phi}(u, v) - 1)^2~.
\end{aligned}
\end{equation}

When $P$ is a uniform probability distribution over $\mathcal{X}$, then $\sigma$-distortion measures the variance of the distribution of the normalized ratio of distances, $\Tilde{\rho_\phi}(u, v)$. 
For the NE, we aim for an embedding with low distortion and a small $\sigma$-distortion. 

\begin{figure}[htbp]
\centering
\includegraphics[width=.48\linewidth]{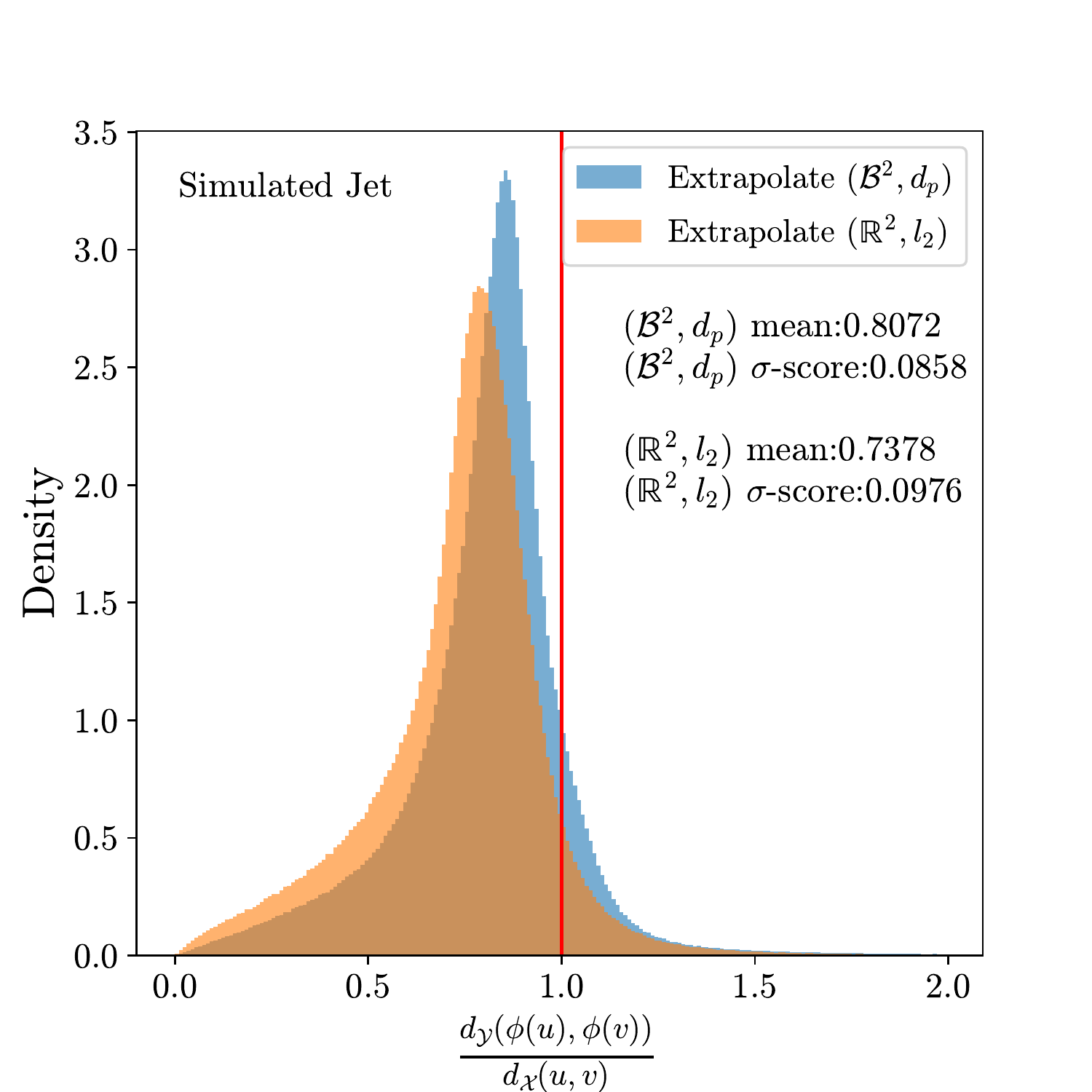}
\includegraphics[width=.48\linewidth]{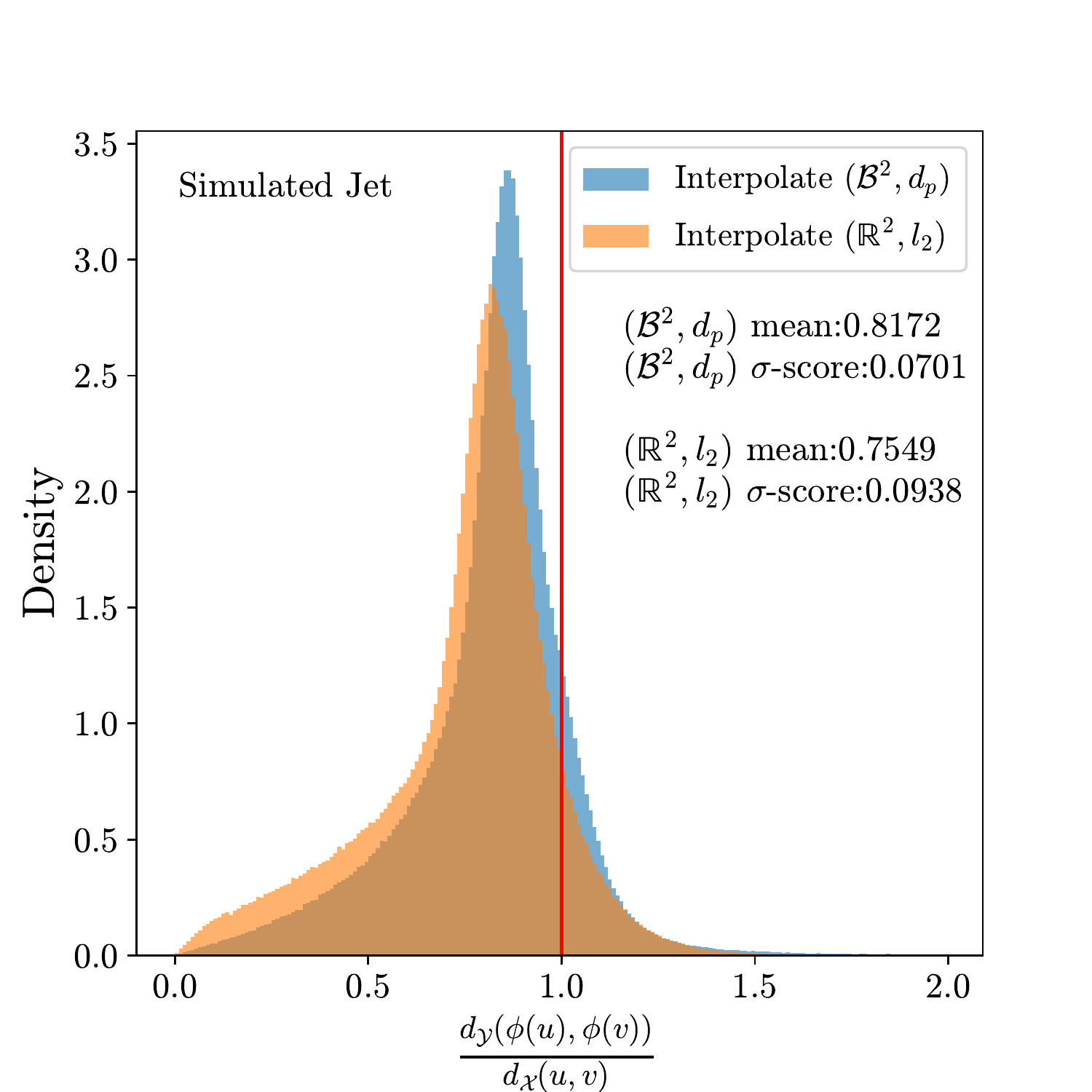}
\caption{ (Left) Histogram of pairwise ratios on the extrapolation set. (Right) Histogram of pairwise ratios on the interpolation set.
}
\label{fig:distortion_measure}
\end{figure}

To test the embedding, we compute the distortion on both the extrapolation and interpolation test datasets. The result is shown in Fig.~\ref{fig:distortion_measure}. Firstly, we verify that low distortion and low $\sigma$-score embedding is achievable in very low dimensions, in two-dimensional Euclidean and Hyperbolic spaces. Indeed, we see a sharp distribution that peaks near 1, the ideal value.  Comparing the performance of embeddings on the interpolation and extrapolation datasets, we see that the performance of the two datasets is very similar. We can conclude that the embedding is surprisingly good and has good extrapolation capabilities. 

\begin{figure}[htbp]
\centering
\includegraphics[width=.32\linewidth]{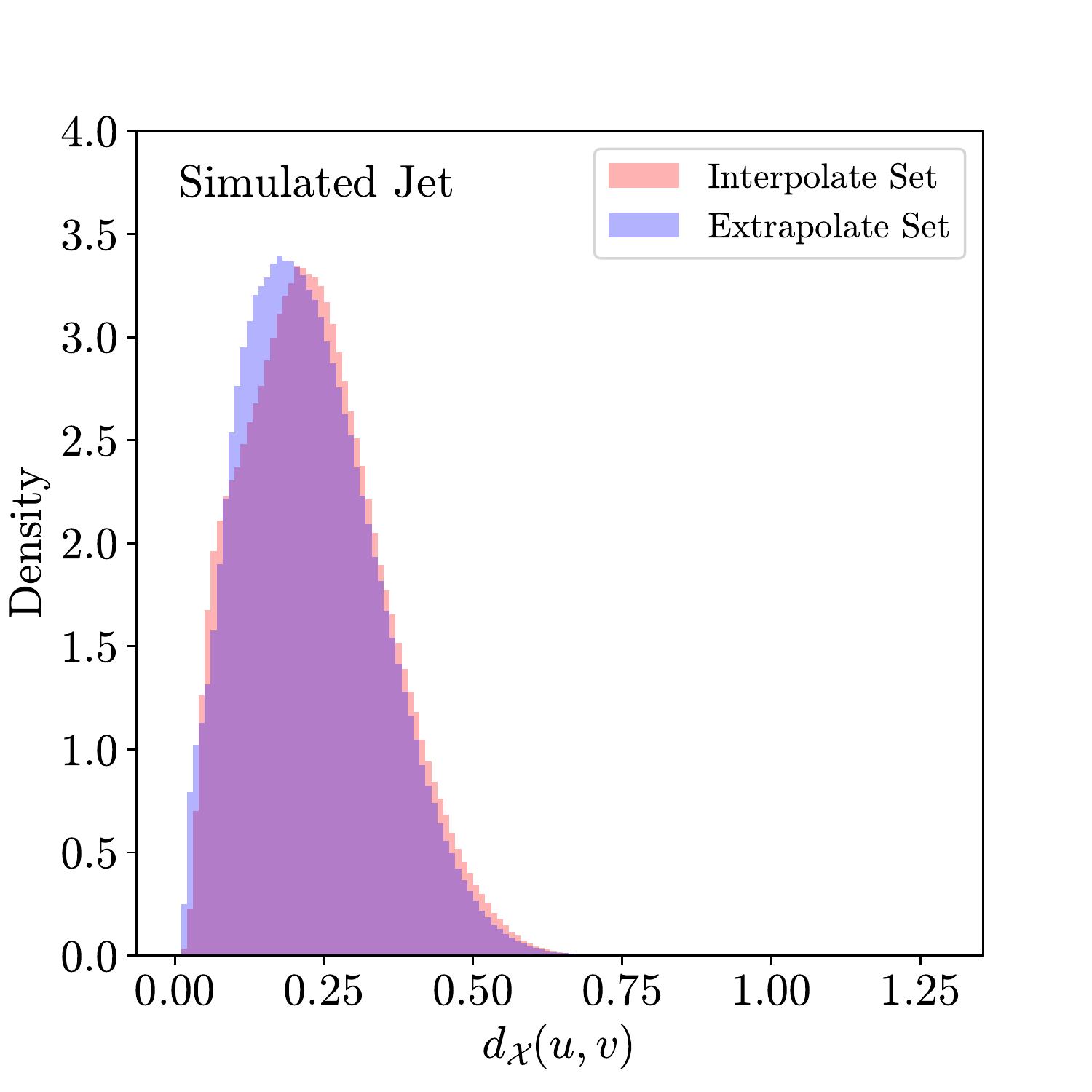}
\includegraphics[width=.32\linewidth]{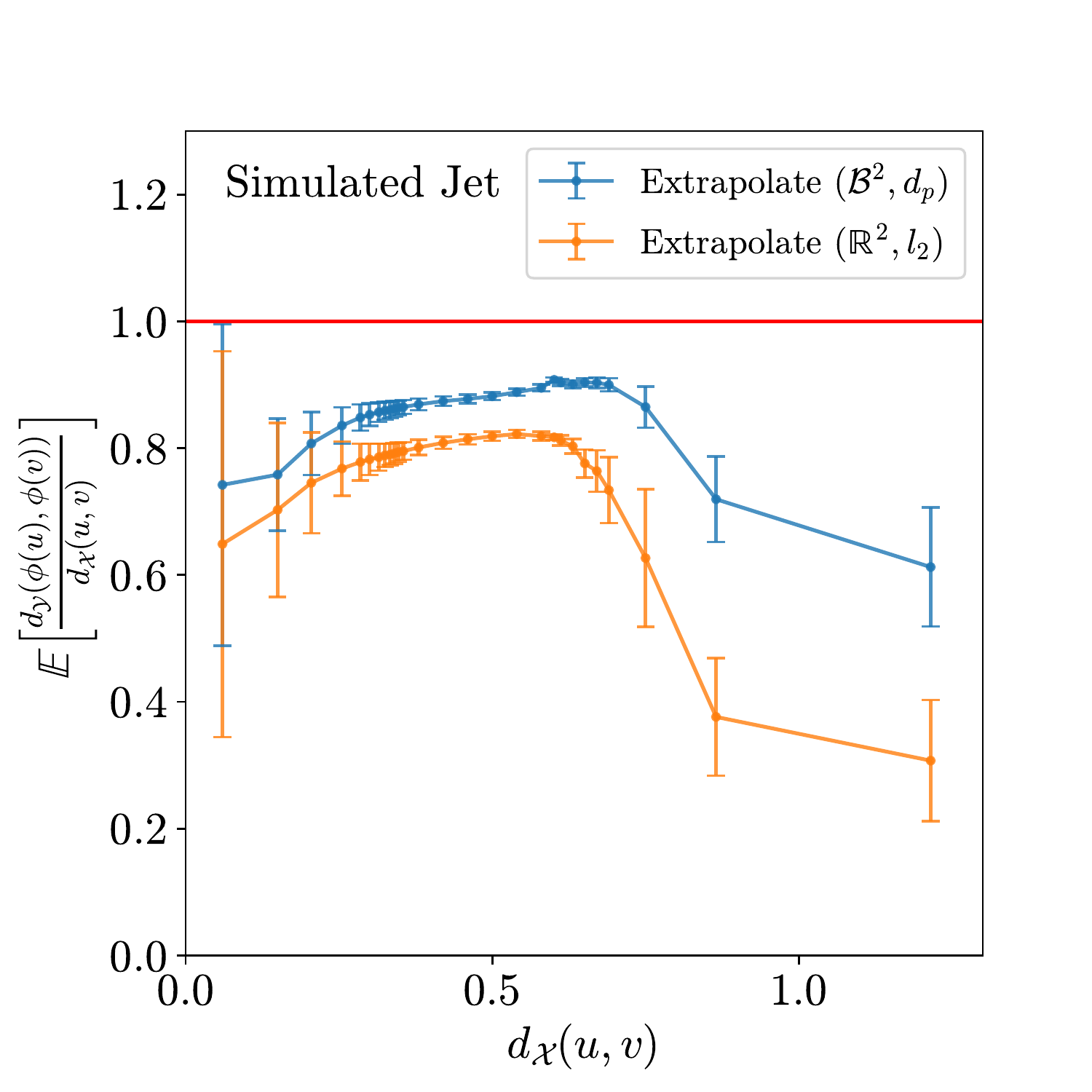}
\includegraphics[width=.32\linewidth]{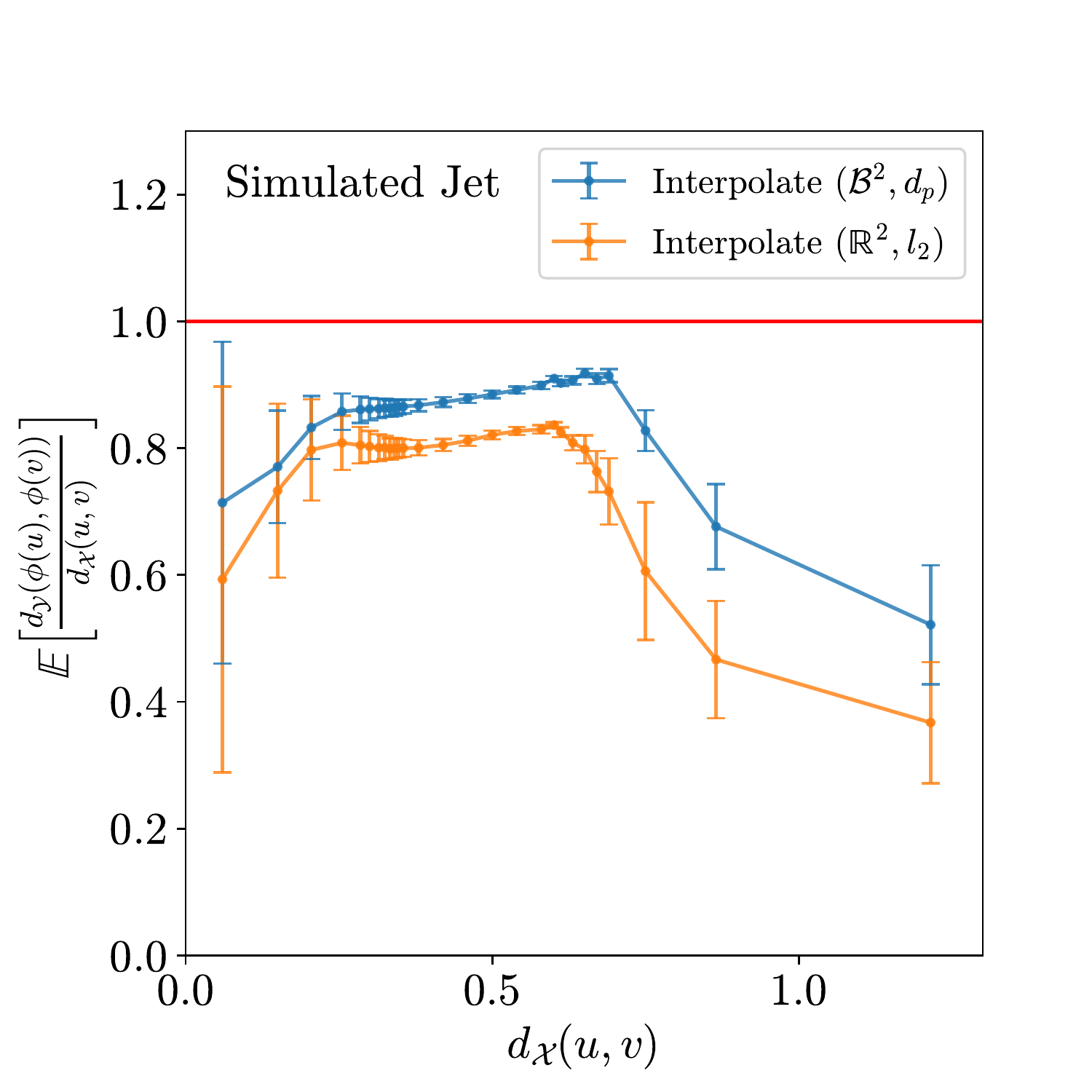}
\caption{ (Left) Histogram of original metric (EMD) distribution for interpolation and extrapolation set. (Middle) The pairwise ratios plotted as a function of the original metric distance, for the extrapolation set. The points are the mean of pairwise ratios in each bin of original metric distance, and the error bars are the $\sigma$-distortion in each bin. (Right) The pairwise ratios plotted as a function of the original metric distance, for the interpolation set. 
}
\label{fig:distortion_correlation}
\end{figure}

Fig.~\ref{fig:distortion_correlation} shows pairwise ratios and $\sigma$-distortion as a function of original metric distance. The performance is the worst (pairwise ratios deviate furthest from the ideal value 1 and $\sigma$-distortion is the largest) near extreme original metric distances 0 and 1.2, where there is less training data available. However, even with significantly less training data available, the learned neural embedding functions perform reasonably well with pairwise ratios on the same scale as the best-achieved values.

We also see that Hyperbolic embedding performs better than Euclidean on both the interpolation and extrapolation datasets. For both datasets, the Hyperbolic embedding gets the mean of pairwise ratios closer to one and achieves a lower $\sigma$-distortion. The improvement is even more striking for extreme values of the original metric distances. Looking at Fig.~\ref{fig:distortion_correlation}, we see that for both interpolation and extrapolation cases, near the largest values of the original metric distances Hyperbolic embedding significantly outperforms Euclidean embedding.

\begin{figure}[htbp]
\centering
\includegraphics[width=.24\linewidth]{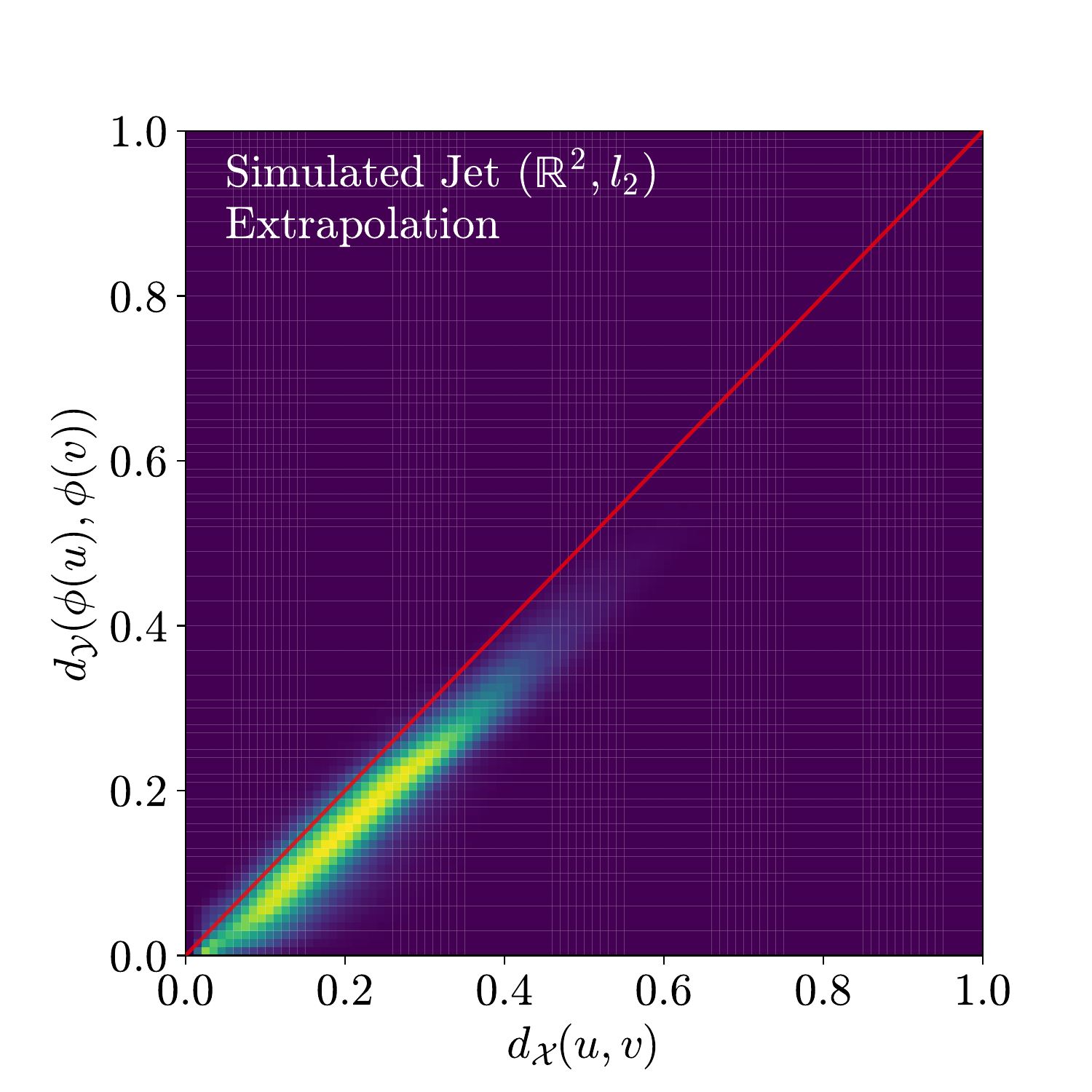}
\includegraphics[width=.24\linewidth]{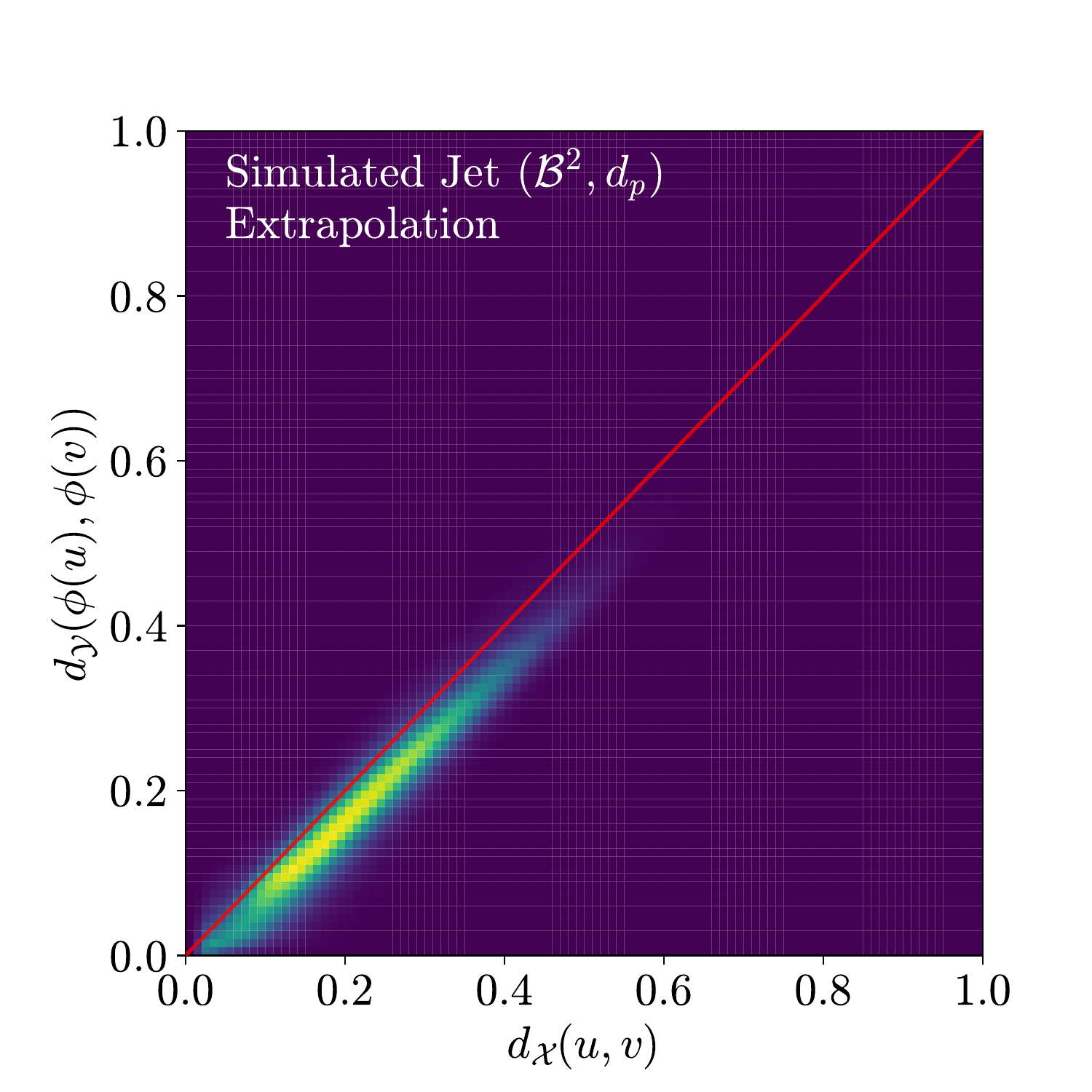}
\includegraphics[width=.24\linewidth]{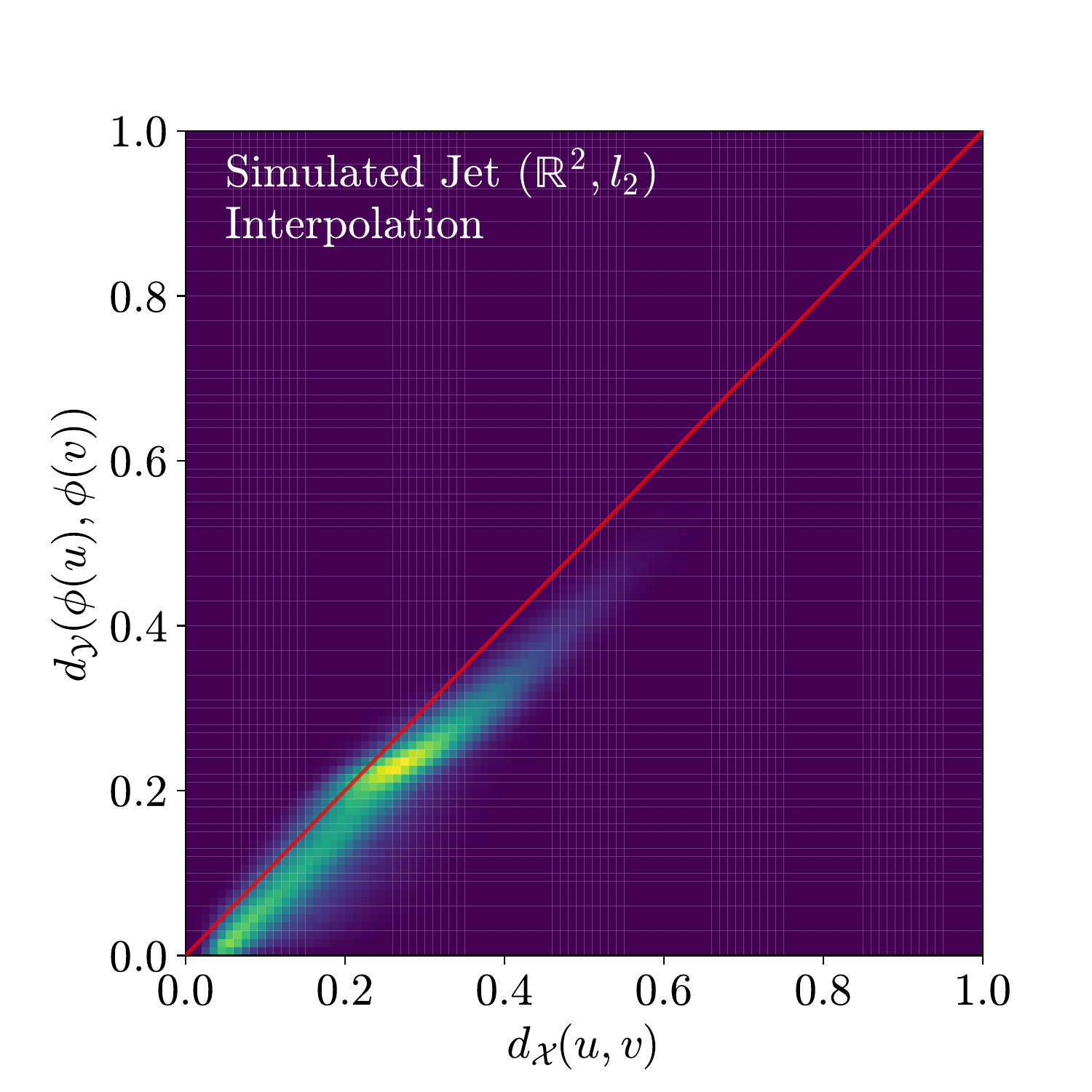}
\includegraphics[width=.24\linewidth]{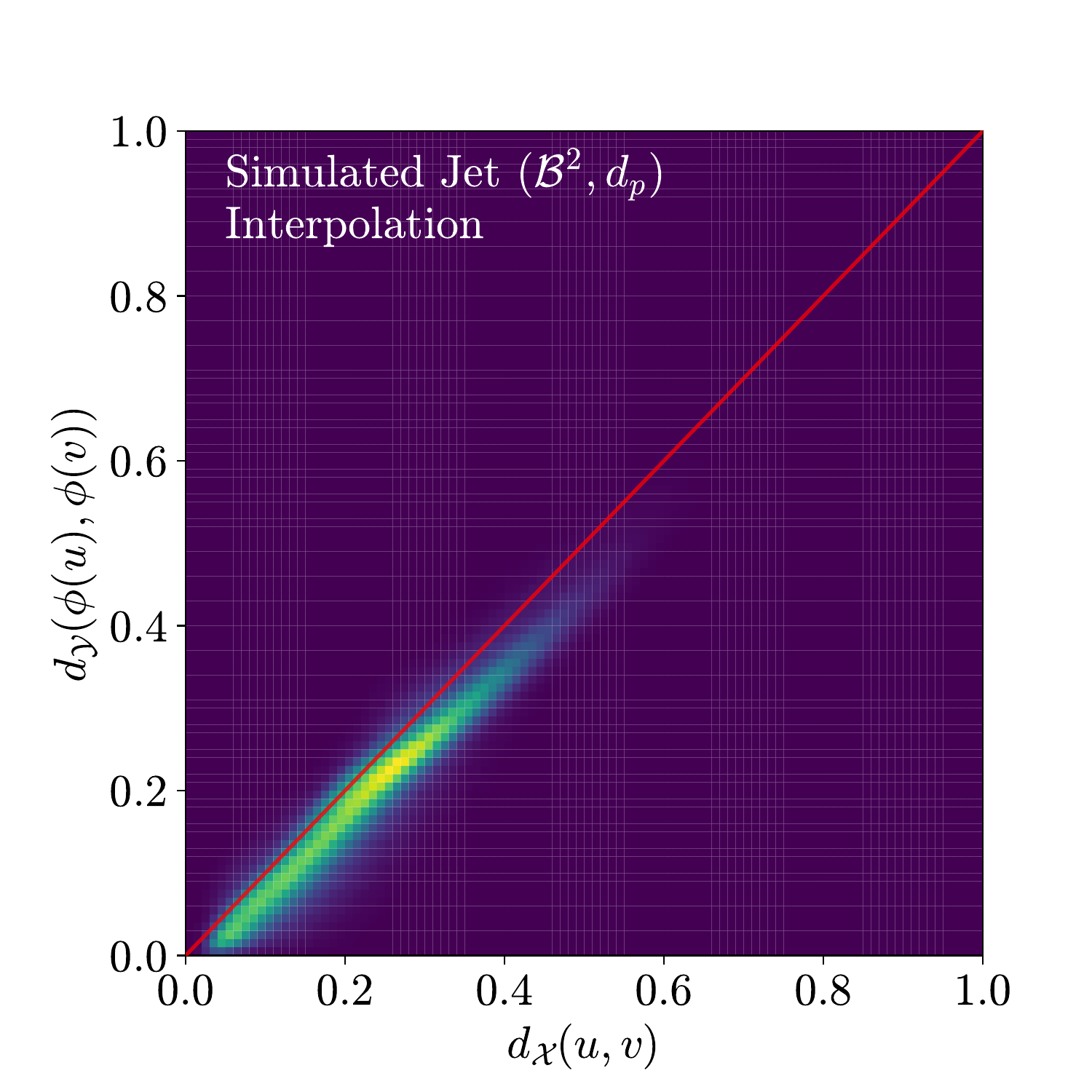}
\caption{ (Left) Correlation between the distances in the original metric space and the distances in the embedded space, for extrapolation dataset into Euclidean embedded space. (Middle Left) The same plot for extrapolation dataset into Hyperbolic embedded space. (Middle Right) The same plot for interpolation dataset into Euclidean embedded space. (Right) The same plot for interpolation dataset into Hyperbolic embedded space.
}
\label{fig:distortion_2dcorrelation}
\end{figure}

Lastly, in Fig.~\ref{fig:distortion_2dcorrelation} we visualize the relation between the distances in the original metric space and the distances in the embedded space. We see that they are very highly correlated and almost fall in $y=x$ line, as we expect, for all cases.

Overall, considering that we are working in two dimensions, a higher dimensional embedding would likely perform even better. In higher dimensional embedded spaces, Hyperbolic has the potential to further outperform Euclidean embedding due to inherent geometry.

\subsection{Comparison with Other Manifold Learning Methods}

There are many well-studied methods to embed data into Euclidean spaces. In addition to NE, t-SNE and UMAP are alternative approaches capable of embedding the original data manifold into the lower-dimensional Euclidean space. Some exploration has been done to use these embeddings on jets.  
In particular, t-SNE has been used for embedding jets in previous works \cite{Komiske:2019jim}.

Both t-SNE and UMAP are fundamentally different methods from NE and suffer from limitations in their applicability. 
First, most out-of-the-box embedding methods such as t-SNE~\cite{tsne} and UMAP~\cite{umap} deal with embedding a fixed number of points into the space by training and embedding on the same datasets.  This means that we don't have the ability to perform parallel evaluation, which limits the scalability when compared with NE.  
Secondly, since t-SNE and UMAP learn the specific relationship in a given dataset yielding an unphysical metric. The mapping cannot be applied to alternative datasets. NE avoids this problem and we have already demonstrated robust performance on a 4-pronged dataset not used within the training.

Thirdly, t-SNE focuses on preserving the local structure of a dataset at the cost of inducing severe global distortions. As a result, the Euclidean distance in t-SNE space is hard to interpret since t-SNE does not preserve distance and global structure.

\begin{figure}[htbp]
\centering
\includegraphics[width=.32\linewidth]{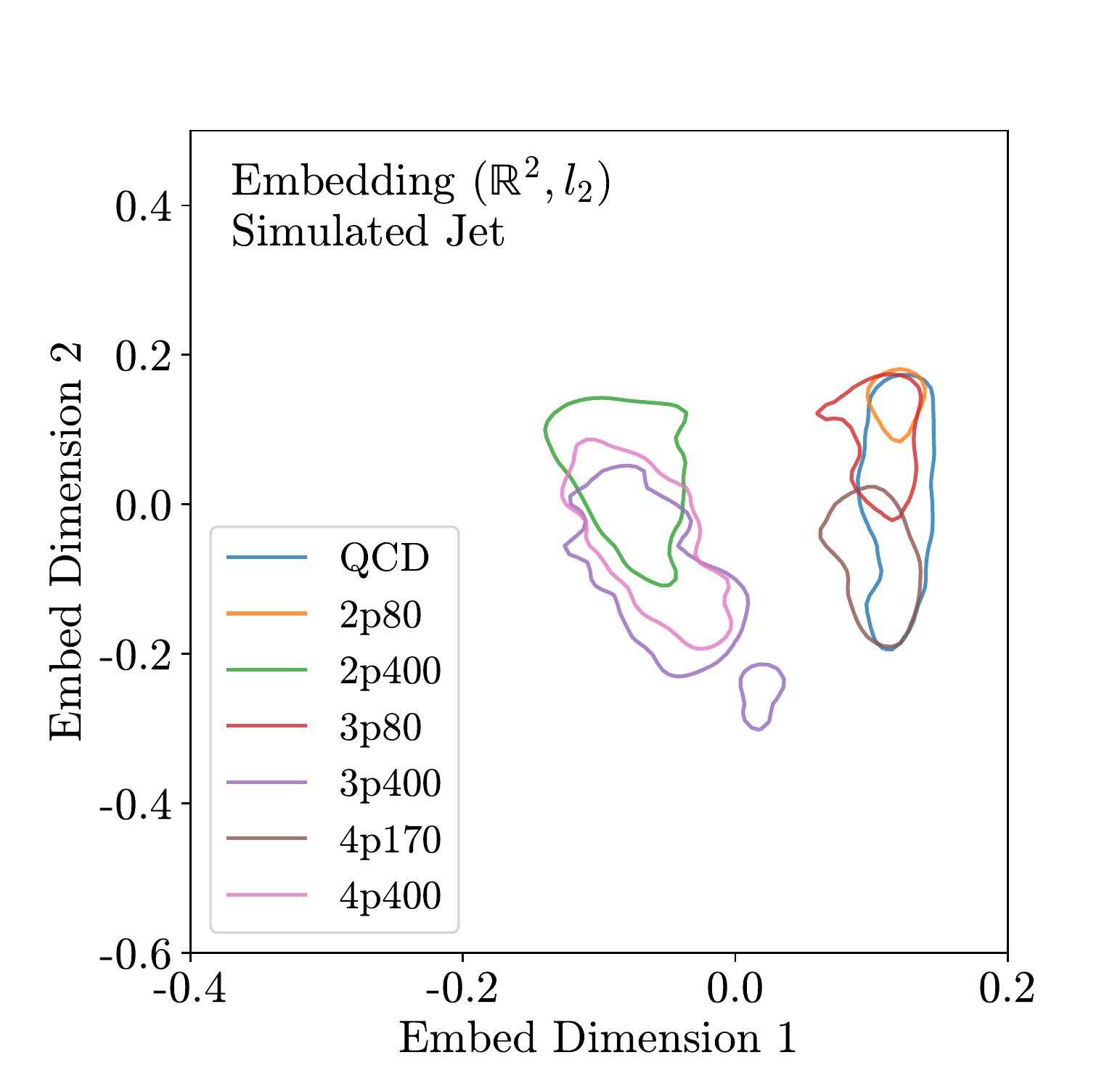}
\includegraphics[width=.32\linewidth]{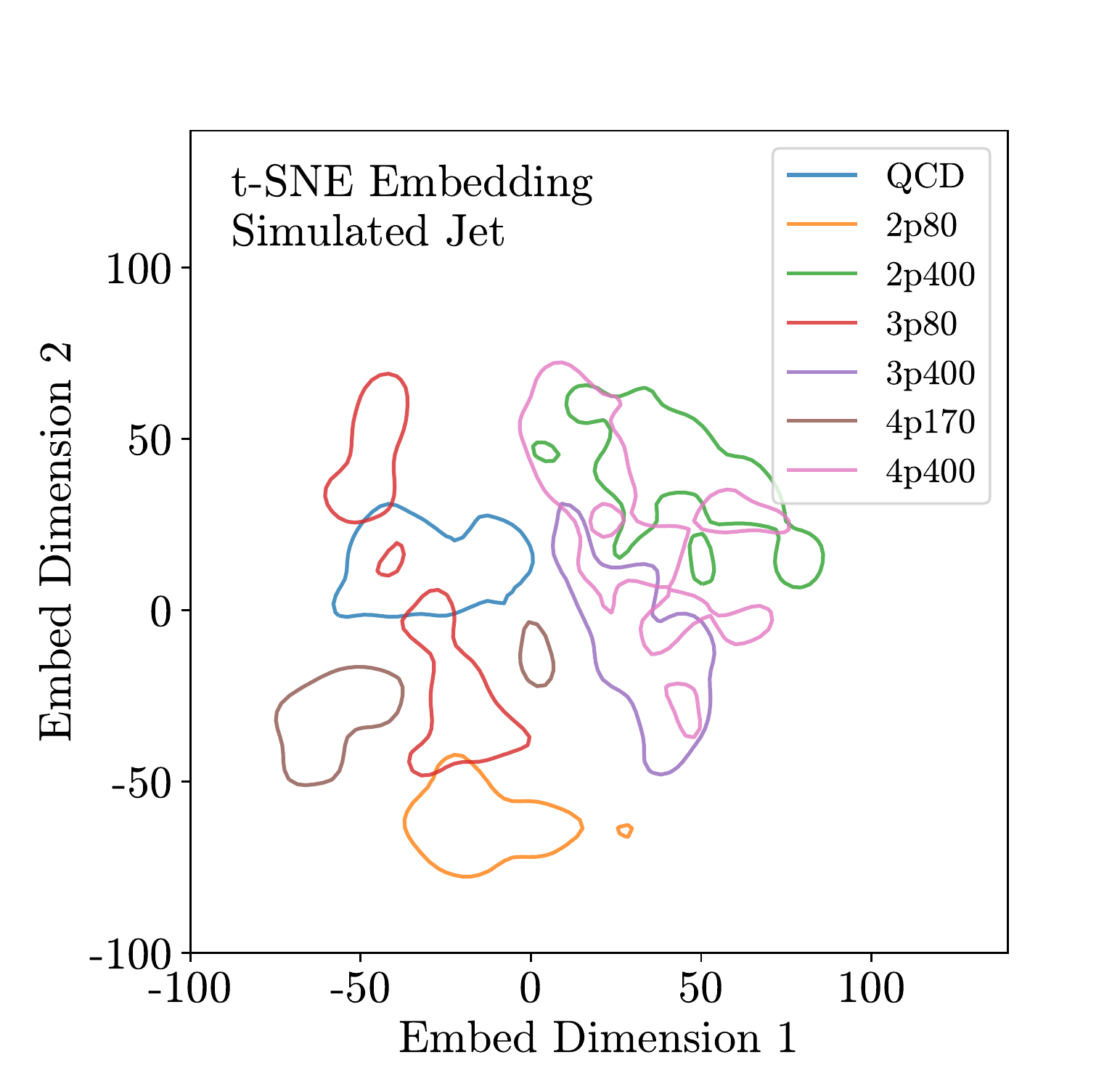}
\includegraphics[width=.32\linewidth]{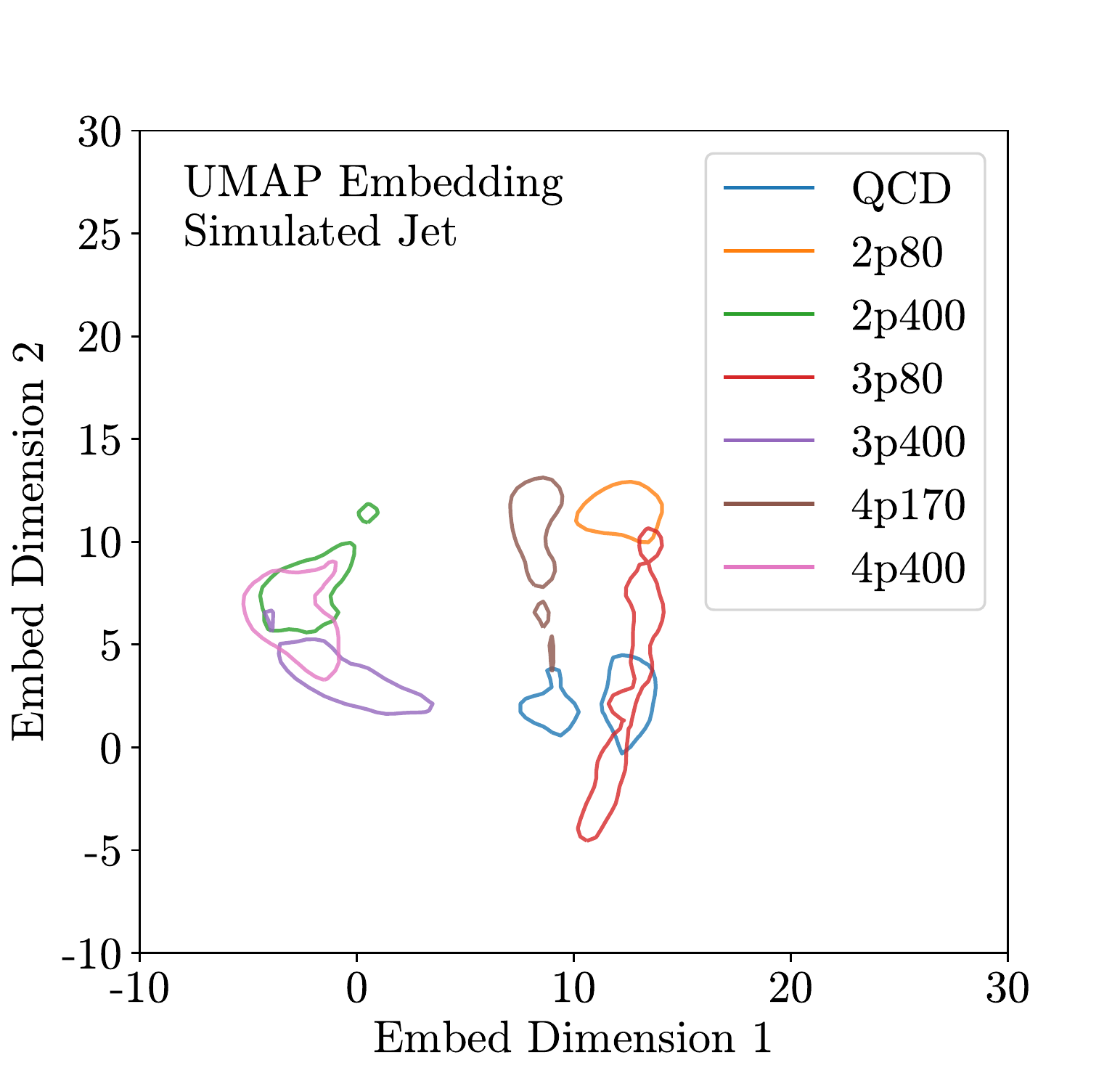}
\caption{ (Left) The neural embedding we propose smoothed with kernel density estimator, with with contour lines corresponding to the CDF value 0.4. (Middle) t-SNE result on the same dataset (Right) UMAP result on the same dataset}
\label{fig:methodcomparison}
\end{figure}

Both methods are tested with our setup in Fig.~\ref{fig:methodcomparison} along with the neural embedding method we propose on a small subset of the jet interpolation dataset. Although there is some benefit, low distortion for EMD is not guaranteed at all for UMAP and t-SNE limiting our ability physically interpret the space. We also argue that by looking at Fig.~\ref{fig:methodcomparison}, the neural embedding method offers the best interpretability since it shows a characteristic ordering of mass and pronginess. 


\section{Anomaly Quantification}
\label{sec:anomalyquantification}
Following the construction of the embedded space, we can perform a variety of explorations to understand what has been learned in the space. In this section, we look at how we can use the NE space to define a diversity metric for the scope of signatures that anomaly detection has identified.   

When trying to quantify the effectiveness of an anomaly detection algorithm, the conventional metrics fall short when the target is unknown. More specifically, in model agnostic searches at colliders, we can’t a priori know the beyond the standard model (BSM) physics signal. The conventional metrics of evaluation, such as comparing ROC curves and the significance of the extracted signal, often do not tell the full story since the performance of the algorithm depends critically on the chosen evaluation dataset. Moreover, the significance of a single dataset does not  characterize the ability of the anomaly detection algorithm to find unexpected signals.
Good performance of the algorithm on one test dataset does not guarantee performance on some very different collections of events. Currently, there is no clear way to handle this notion of “wideness” of the search capability. 



With the embedded space, we propose a new metric to indicate the coverage of phase space of a single algorithm.  Since the embedded space compresses high dimensional objects into a low dimensional space of physical features, we can utilize the notion of the volume in the embedded space as a way to define algorithm coverage. When the embedded space is a Euclidean space, the volume is straightforward to calculate, and in two-dimensional Euclidean space, this equates to the area within the embedded space. 


To understand how area coverage characterizes the wideness of an anomaly search, we introduce the idea of area adjusted ROC curve. To compute the area adjusted ROC curve, we first prepare a signal evaluation ensemble that consists of a wide variety of event topologies so that the phase space over which we wish to compare the two algorithms is broadly covered. 

With the evaluation ensemble and the background dataset, we then evaluate how the chosen algorithm covers the embedded space of the signal ensemble when compared with the background. For each point on the regular ROC curve, we map the selected signal points that pass the threshold for the true positive rate (TPR) of our chosen algorithm to the embedded space and calculate the ratio of the total embedded space area covered by the selected points; this yields the Area TPR defined in Eq.~\ref{eq:areatpr}.

\begin{equation}
\label{eq:areatpr}
    \text{Area TPR} = \epsilon_{\text{sig, area}} = \frac{\text{Selected Signal Area}}{\text{Total Signal Ensemble Area}} 
\end{equation}

Similarly, we map the selected QCD background points to the embedded space and calculate the ratio of selected background points to the total QCD area. This procedure defines the Area FPR defined in Eq.~\ref{eq:areafpr}, and it tells us the efficiency of the area of QCD rejection, defined as

\begin{equation}
\label{eq:areafpr}
    \text{Area FPR} = 1-\epsilon_{\text{bkg, area}} = \frac{\text{Selected Background Area}}{\text{Total QCD Area}} 
\end{equation}

With Area TPR and Area FPR, we can construct area adjusted ROC curve. The roc curve is adjusted based on whether the algorithm casts a wide net or only looks at the narrow region of the phase space. 

\begin{figure}[htbp]
\centering
\includegraphics[width=.99\linewidth]{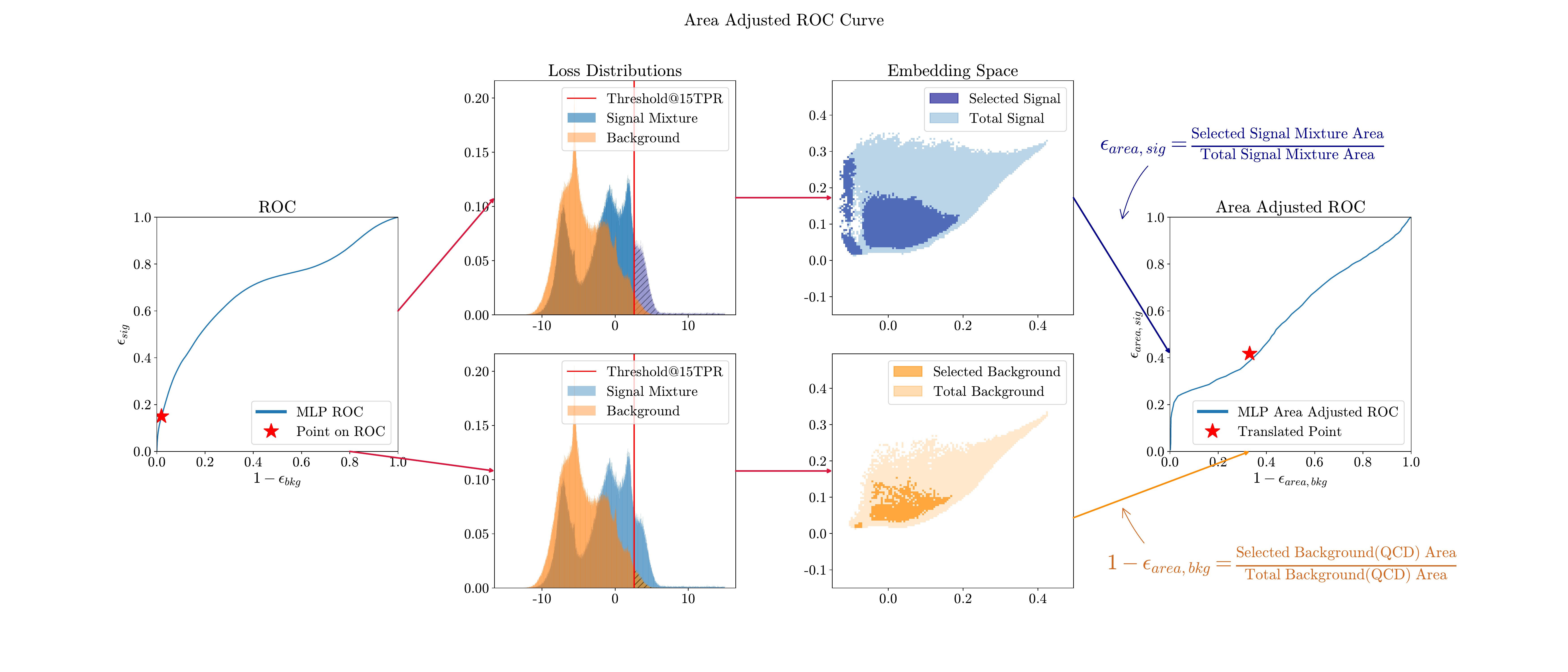}
\caption{ The process of calculating area adjusted ROC curve for anomaly detection algorithm trained with MLP architecture on QCD vs. 2-prong $170\GeV$ jet task. 
}
\label{fig:areaadjusted_roc_MLP}
\end{figure}

We diagrammatically show how this adjusted ROC curve is made in practice in Fig.~\ref{fig:areaadjusted_roc_MLP}, for a supervised learning algorithm constructed from an MLP network training QCD vs 2-prong jet with a secondary mass of $170\GeV$. 

In Fig.~\ref{fig:areaadjusted_roc_MLP}, we see how a point(a red star point) in the normal ROC curve gets translated to a point on area adjusted ROC curve. For a given TPR, we compute the area the selected points cover compared to the full area coverage of our ensemble set. Similarly, for the background, we compute the fraction of the area of the embedded space volume that gets rejected by the chosen algorithm. 

The area ratio is calculated for the Area TPR and Area FPR by dividing the embedded space into grids and counting the number of bins that has data points above a certain threshold, which we call a threshold parameter. By default, the regions with counts more than 3 are considered, but the ROC curve is stable regarding the choice of the threshold parameter. The stability of the area-adjusted ROC curve regarding the choice of this threshold parameter is discussed in Sec.\ref{sec:stabilityareaadjusted}, Fig.~\ref{fig:areaadjusted_roc_threshold}. 

This can be understood as a measure of the total phase space of events, represented in the form of the NE. The area coverage is the portion of the total phase space volume that the algorithm covers. 

\begin{figure}[htbp]
\centering
\includegraphics[width=.99\linewidth]{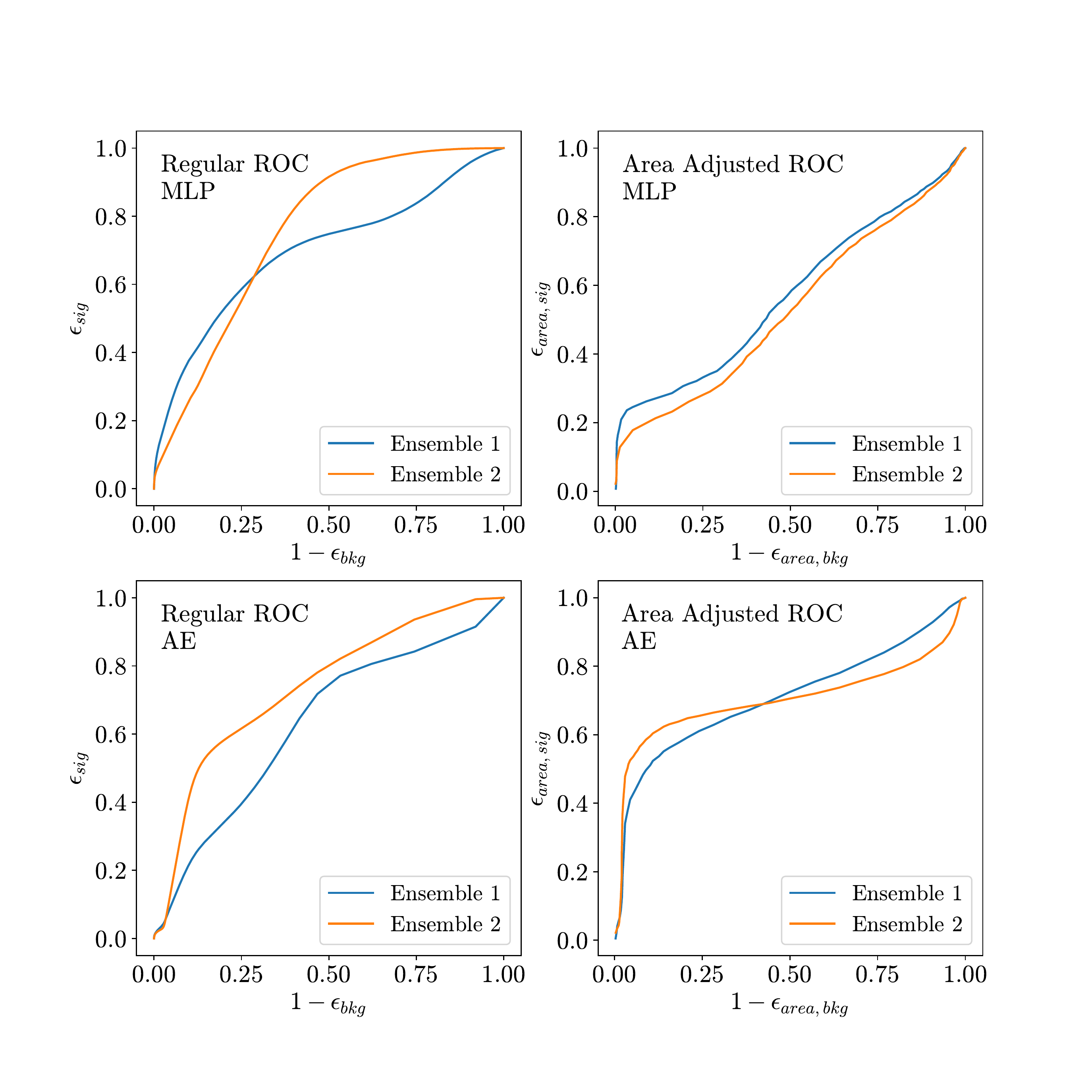}
\caption{ Comparison of stability of regular ROC curves and area adjusted ROC curves for two different anomaly detection algorithms, MLP and AE, on two different ensemble datasets. We see that area adjusted ROC curves are more stable against changing test data ensemble. (Upper Left) Regular ROC curves for MLP (Upper Right) Area adjusted ROC curves for MLP (Lower Left) Regular ROC curves for AE (Lower Right) Area adjusted ROC curves for AE
}
\label{fig:areaadjustedrocensemblestability}
\end{figure}

Careful consideration of these ROC curves should be taken into account since there is a dependence of area adjusted ROC curve on the selection of this evaluation dataset. To make this approach as general as possible for anomaly detection, we choose a dataset where the area it spans in the embedded space is as wide as possible. Later we show in Fig.~\ref{fig:areaadjustedrocensemblestability} that as long as the evaluation ensemble covers the search space well, the dependence on this set of an ensemble is very small. 

\begin{figure}[htbp]
\centering
\includegraphics[width=.99\linewidth]{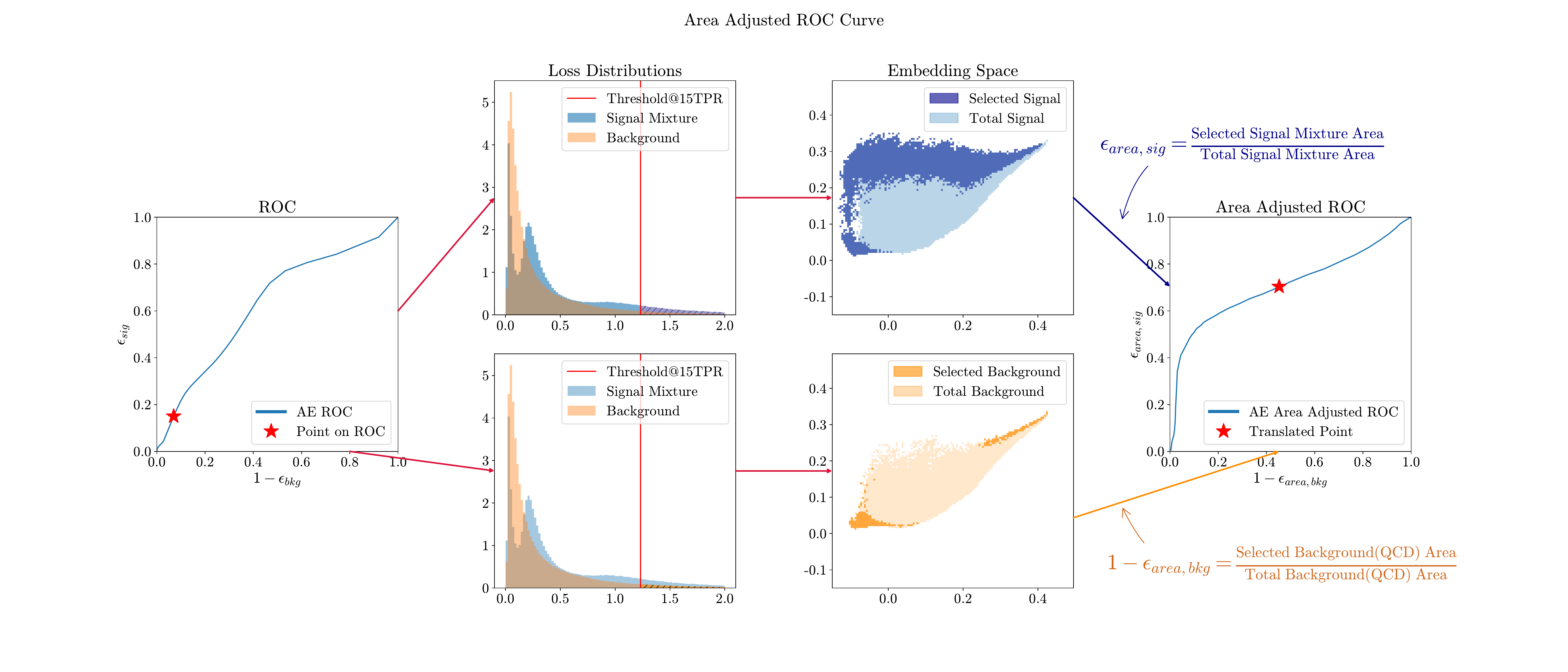}
\caption{ The process of calculating area adjusted ROC curve for anomaly detection algorithm trained with autoencoder on QCD. 
}
\label{fig:areaadjusted_roc_AE}
\end{figure}

To see the usefulness of this anomaly quantification method, we compare two different anomaly detection algorithms, a fully supervised algorithm (MLP) trained to do QCD vs 2-prong jet with mass $170\GeV$, and an unsupervised algorithm comprising of an autoencoder trained on just QCD background. Fig.~\ref{fig:areaadjusted_roc_AE} and Fig.~\ref{fig:areaadjusted_roc_MLP} shows the construction of area adjusted ROC curve for these two different algorithms.

\begin{figure}[htbp]
\centering
\includegraphics[width=.95\linewidth]{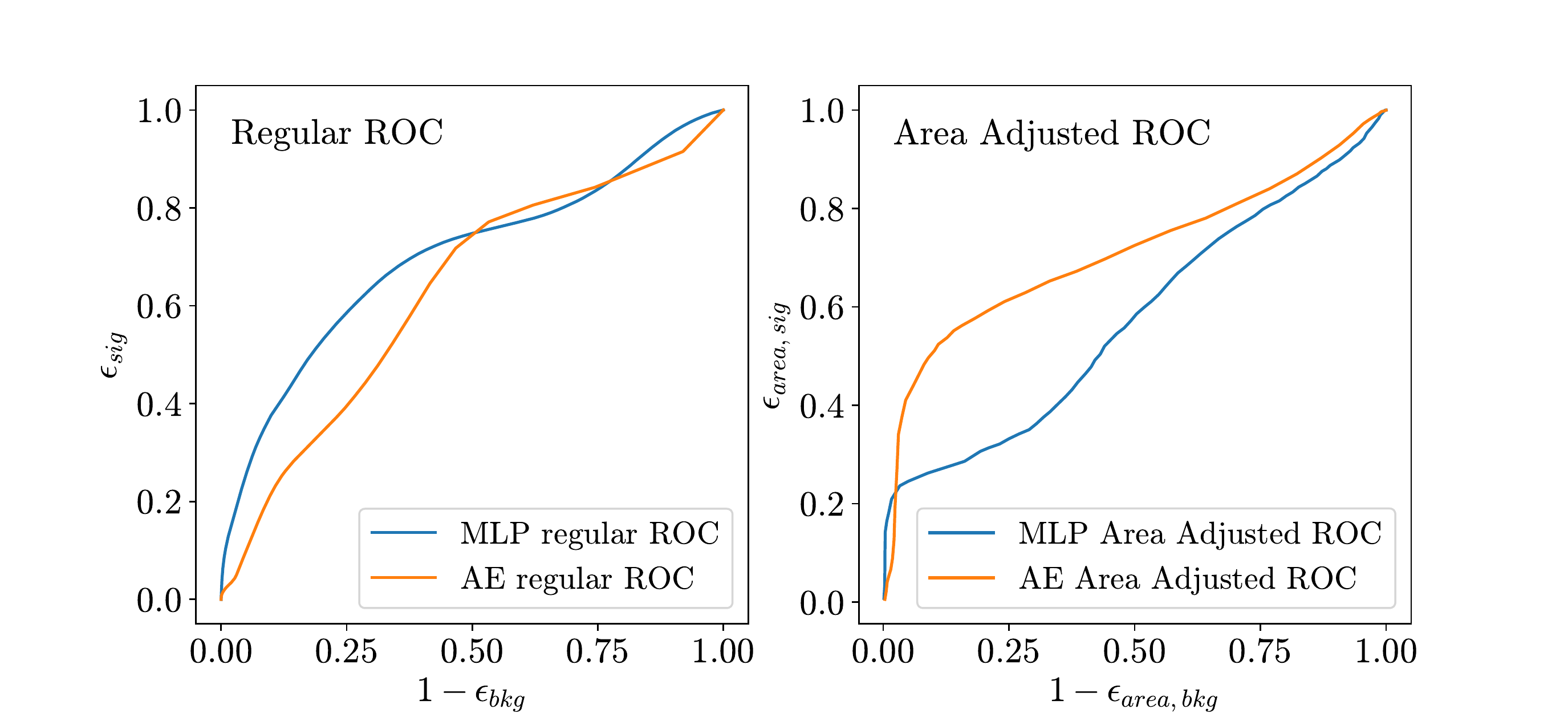}
\caption{ (Left) Comparison of regular ROC curves for two different algorithms, MLP and AE. (Right) Comparison of area adjusted roc curves for the two algorithms, MLP and AE.  
}
\label{fig:areadjusted_roc_compare}
\end{figure}

Fig.~\ref{fig:areadjusted_roc_compare} compares the normal ROC and area-adjusted ROC directly for these two algorithms. It aligns with our intuition that since MLP is hyper-optimized to do well on one specific dataset, it has lower search capability and focuses on the narrow region of the phase space.
Therefore, we see that even though the MLP algorithm seems to be doing fine on the regular ROC curve, the area adjusted ROC curve reveals that the AE algorithm is more efficient at searching the wider area in the embedded space, which, here, acts as a proxy for the phase space.   

Another benefit of this procedure is that we can visualize which region of embedded space each algorithm searches and compare this between different methods. By comparing the colored area of the selected points in the embedded space in Fig.~\ref{fig:areaadjusted_roc_AE} and Fig.~\ref{fig:areaadjusted_roc_MLP}, we can observe that for the same TPR working point, two algorithms presented in Fig.~\ref{fig:areaadjusted_roc_AE} and Fig.~\ref{fig:areaadjusted_roc_MLP} choose complementary regions of the embedded space. 

Finally, we show in Fig.~\ref{fig:areaadjustedrocensemblestability} that area adjusted ROC curves reduce the dependence on the test dataset we choose. Regular ROC curves can vary wildly depending on what test dataset we choose, as can be seen in the upper left and lower left plots comparing ROC curves on two different signal ensemble datasets. However, for area adjusted ROC curves in the upper right and lower right panels, we observe a significantly smaller variation across test datasets.

By adjusting the area with an ensemble dataset, we effectively reduce the sample dependence of the evaluation at the same time.




\section{Conclusion}
\label{sec:conclusion}
This paper introduces a method of embedding the physics data manifold with a metric structure into different lower dimensional spaces with simpler metrics. We argue that neural embedding can achieve many things in physics data analyses. In particular, neural embedding is capable of condensing complex high dimensional data into physically meaningful spaces. Furthermore, we explore various types of embedding spaces covering both Euclidean and Hyperbolic embedding.

Using collider physics simulated events of hadronically decaying objects, we demonstrate a neural embedding algorithm that embeds jets into a space where the energy mover's distance is preserved. Using a hierarchical set of progressively more realistic simulations, we find that our neural embedding can preserve the core physical features and self-organize jet datasets into their respective decay types. Furthermore, we find that a Hyperbolic embedding space improves the overall physics interpretation compared to a Euclidean embedding. 

We further demonstrate our neural embedding can be used to provide a solution to the complex problem of quantifying the performance of different model agnostic search algorithms. With the notion of volume in lower dimensional Euclidean spaces, we introduced volume-adjusted roc curves, which aim to quantify the true search breadth of a given algorithm. We find that once we apply the volume-adjusted ROC curve, an autoencoder outperforms supervised learning in its ability to search across the whole manifold of physics events. 

Additionally, we note that the optimal transport computation between sets of jets can be very time consuming. By constructing a neural embedding with the energy mover's distance metric, we avoid the need to recompute optimal transport, allowing for a significantly faster calculation that is embarrassingly parallel. 

We conclude that given a complex physics manifold with a metric structure, it can be highly beneficial to embed it into different spaces to extract meaningful information. With cheap computational cost and good capability to learn the latent structure, embedding has the potential to find use cases in a lot of collider physics scenarios. 

We also conclude that a high quality embedding of QCD physics is realizable. From this embedding, we can start to tackle the challenging problem of quantifying anomaly detection algorithms, which is a big obstacle that needs to be solved if we plan to move towards model agnostic searches.

To our knowledge, this is the first attempt at providing a viable solution to quantifying the true search capability of model agnostic search algorithms, which is a crucial step that needs to be solved to pursue model agnostic searches seriously. Embedding also provides an exciting alternative way to build a more straightforward space for various tasks without relying on latent variable or probabilistic modelings such as VAEs and flow models. Embedding really allows physicists to get the most out of the metric space properties. To our knowledge,  embedding the QCD physics manifold into different manifolds with desirable geometric properties such as the Poincaré ball has not been attempted before. 

This paper realizes a neural embedding using hadron collider events. Despite only exploring a few avenues within our embedded space, we are able to perform quantifications that were previously difficult, if not impossible. As a result, we believe embedding will be an invaluable tool in the physics data analysis pipeline. Furthermore, we believe neural embedding will find use in solving a wide variety of practical problems, such as data compression, anomaly detection, quantification of anomaly detection, organizing physics datasets, and many more.

\appendix
\section{Example of Jets}
\label{sec:example_jets}
Some of the examples of jets visualized by plotting each constituent in the $\eta-\phi$ plane with circles of sizes proportional to its $\pt$ is shown. 

\subsection{Simple Toy Jet}
\label{subsec:simpletoyjet_examples}

\begin{figure}[!htbp]
\centering
\includegraphics[width=.48\linewidth]{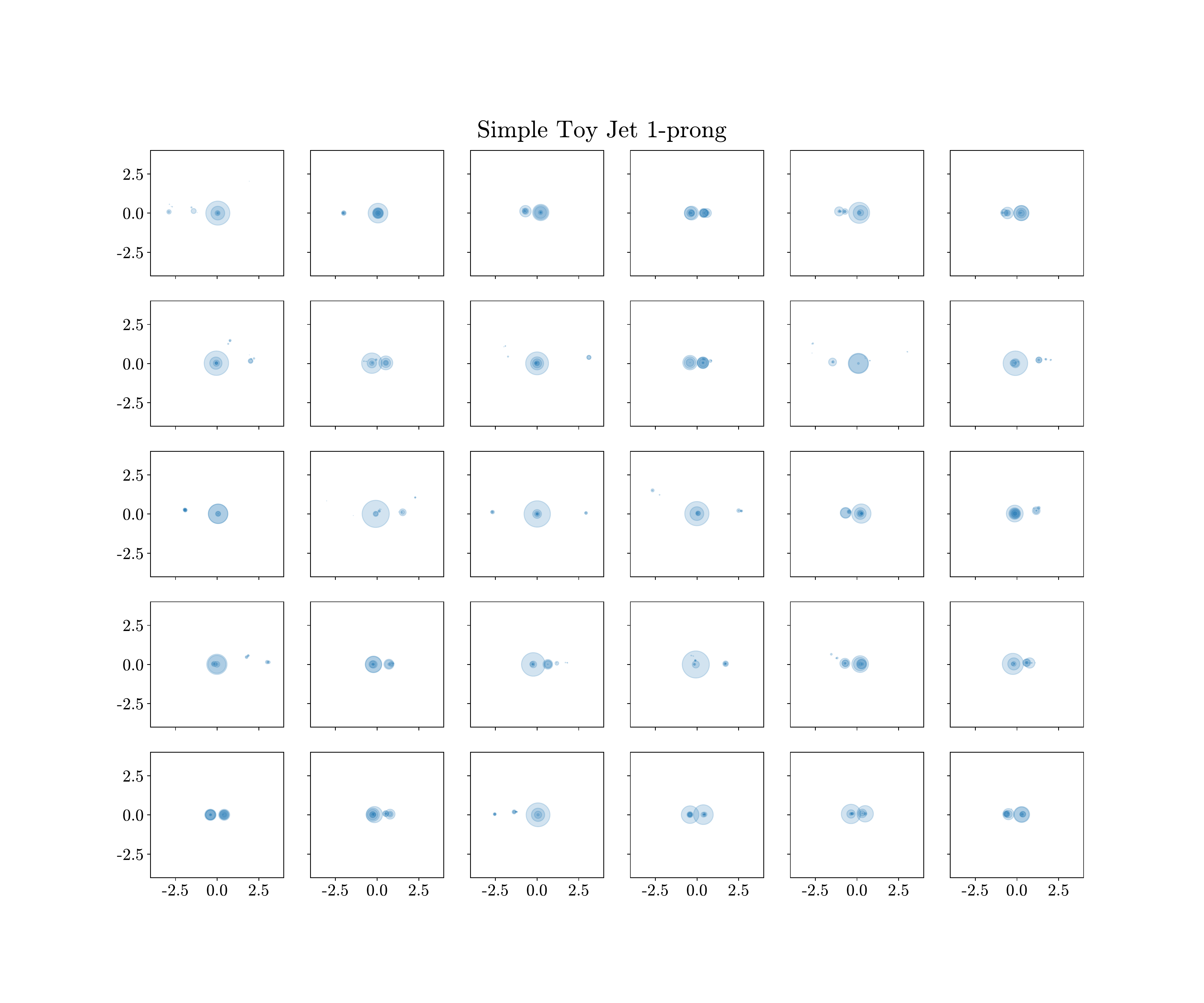}
\includegraphics[width=.48\linewidth]{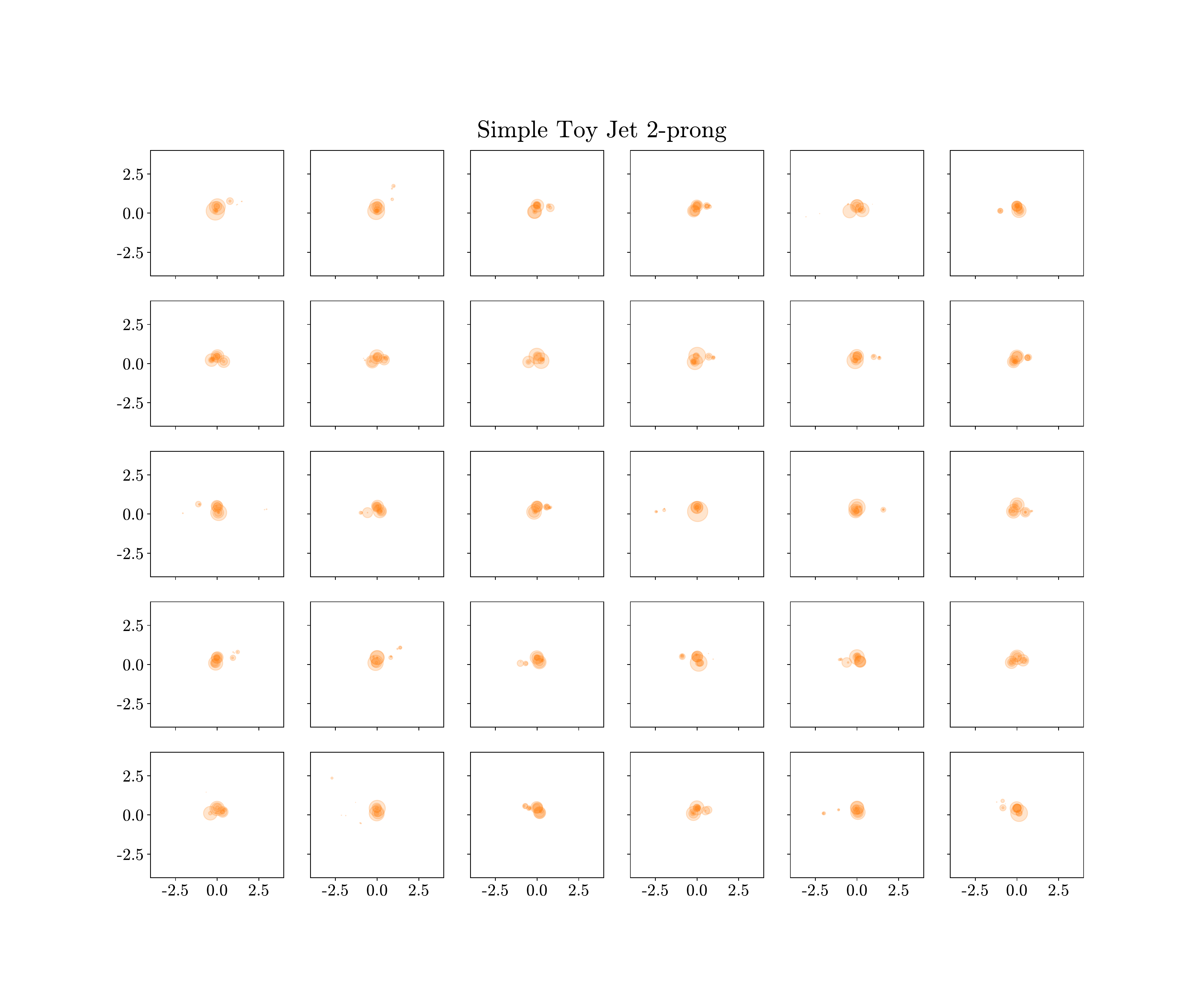}
\caption{ Samples of simple toy jet, (Left) 1-prong(QCD) (Right) 2-prong
}
\label{fig:samplesimpletoyjet1p}
\end{figure}

\begin{figure}[!htbp]
\centering
\includegraphics[width=.48\linewidth]{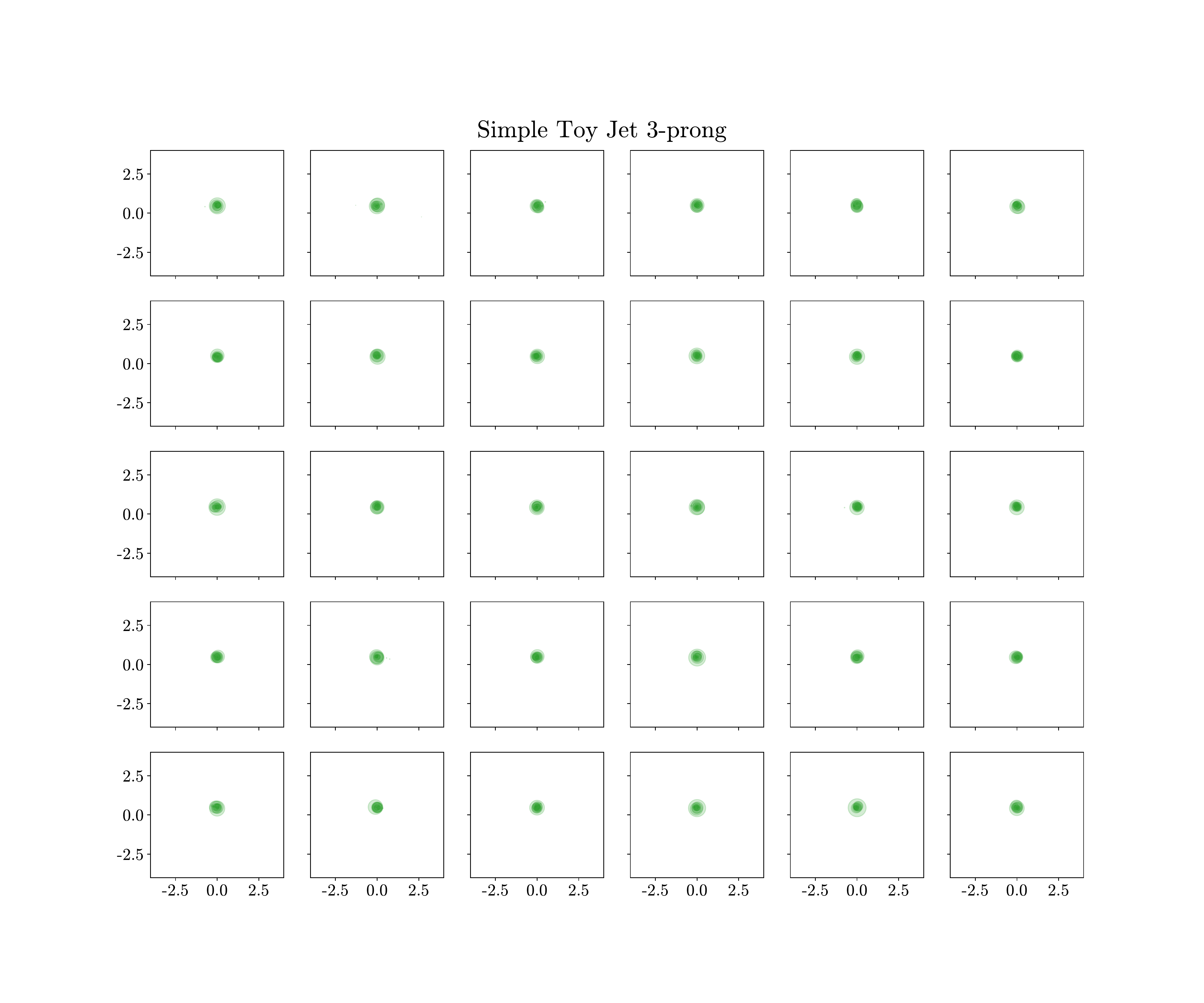}
\includegraphics[width=.48\linewidth]{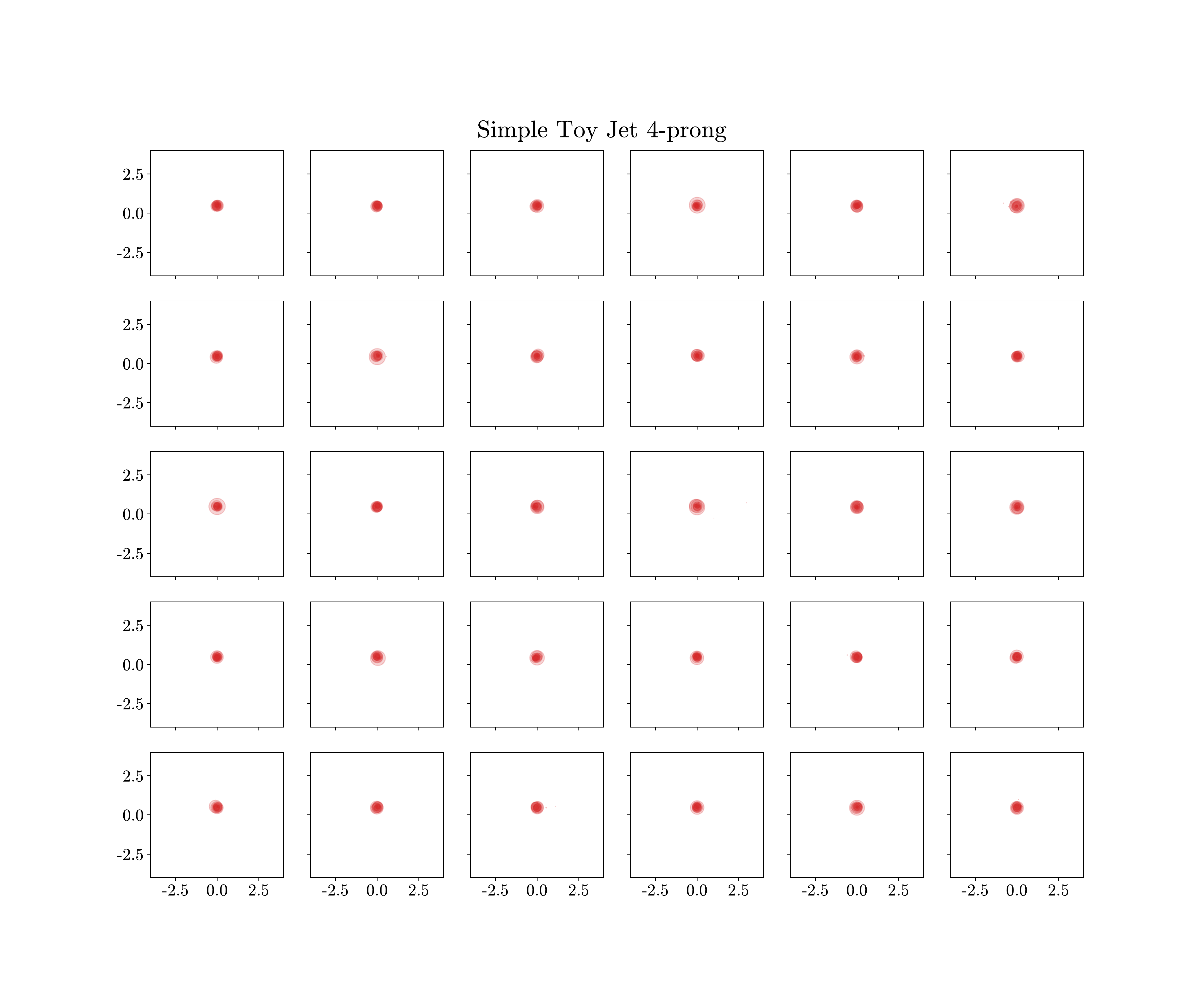}
\caption{ Samples of simple toy jet, (Left) 3-prong (Right) 4-prong
}
\label{fig:samplesimpletoyjet3p}
\end{figure}

In Fig.~\ref{fig:samplesimpletoyjet1p} and Fig.~\ref{fig:samplesimpletoyjet3p} we show some examples of simple toy jets.

\subsection{Realistic Toy Jet}
\label{subsec:realistictoyjet_examples}

\begin{figure}[!htbp]
\centering
\includegraphics[width=.48\linewidth]{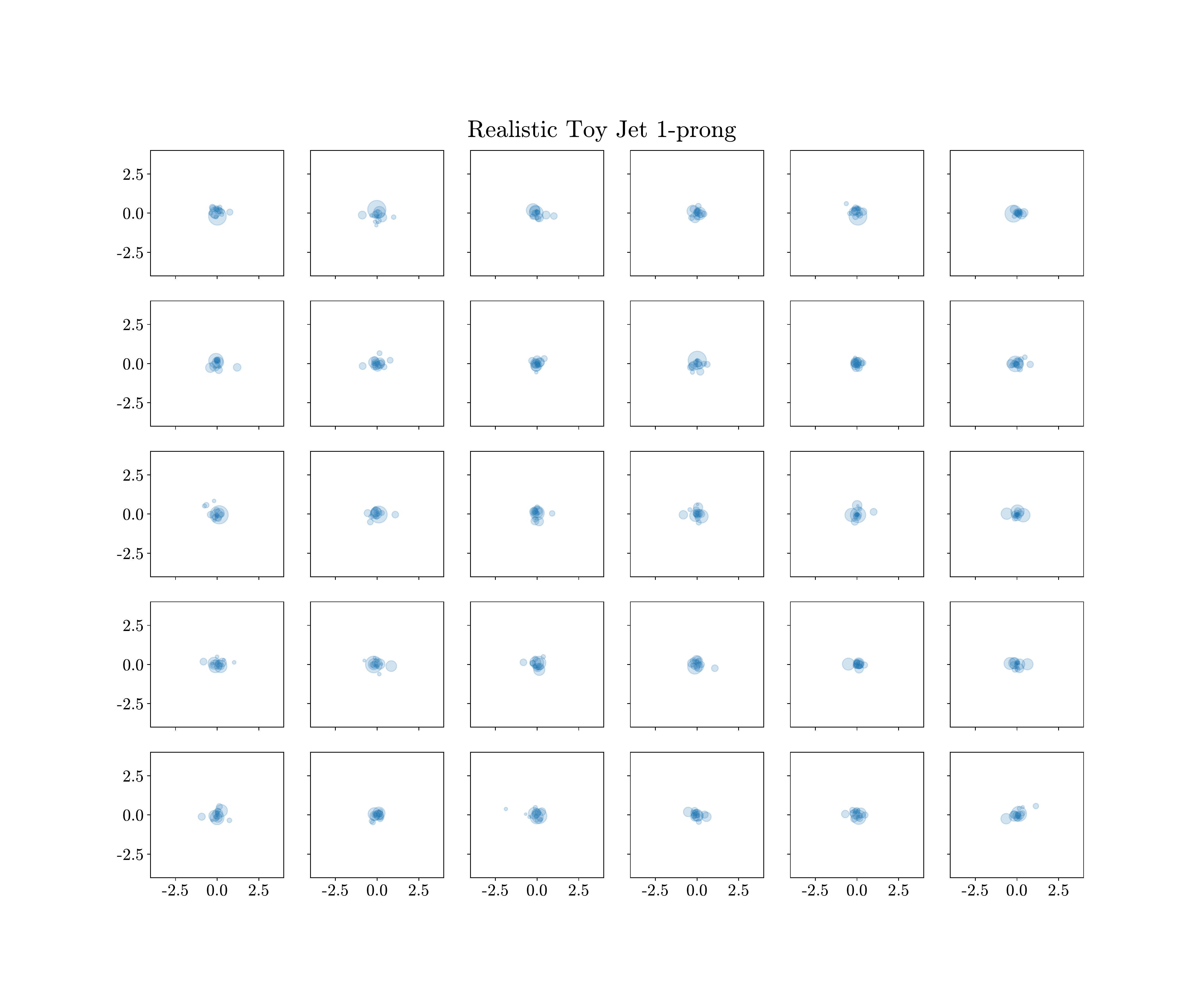}
\includegraphics[width=.48\linewidth]{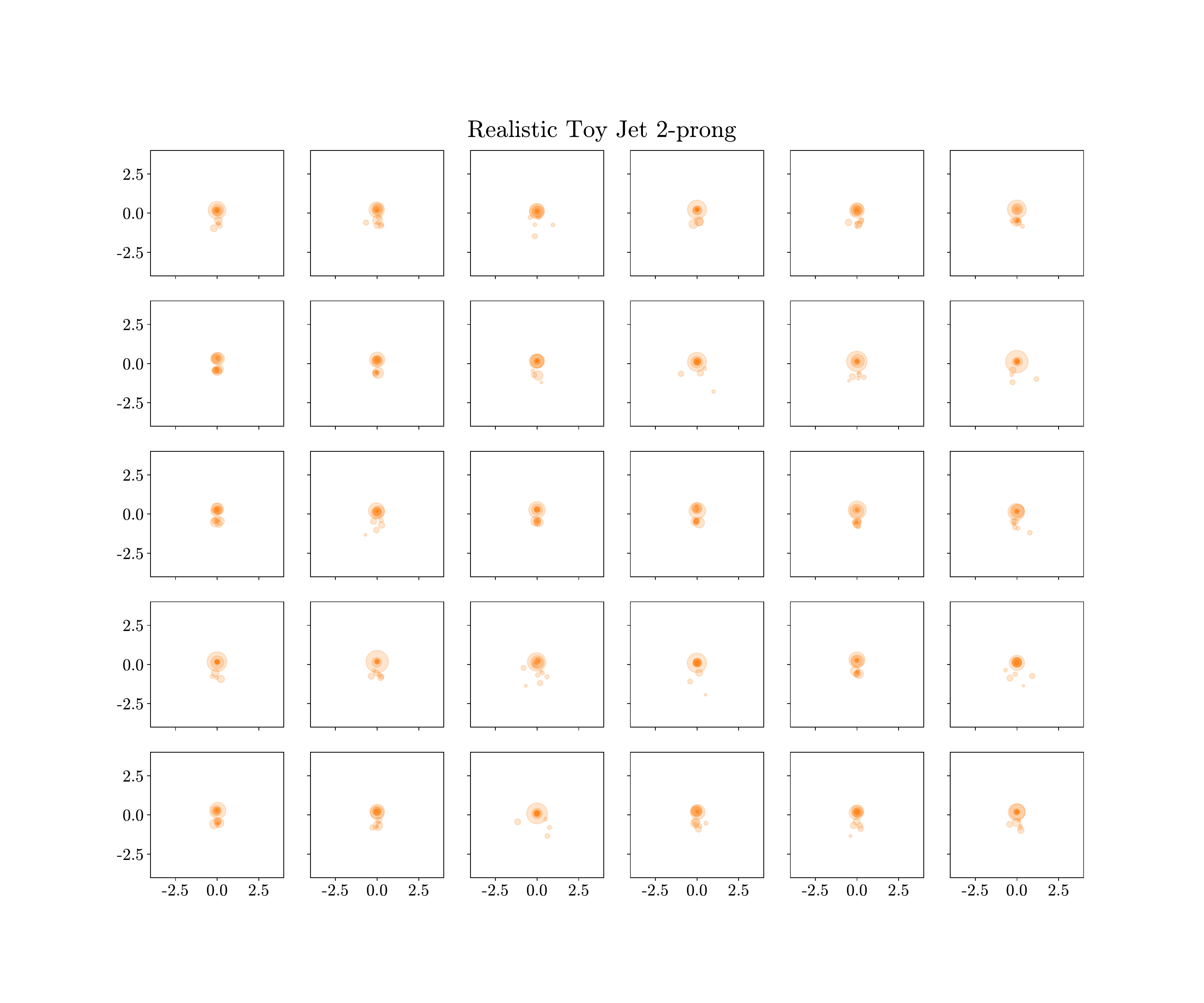}
\caption{ Samples of realistic toy jet, (Left) 1-prong(QCD) (Right) 2-prong
}
\label{fig:samplerealistictoyjet_1p}
\end{figure}

\begin{figure}[!htbp]
\centering
\includegraphics[width=.48\linewidth]{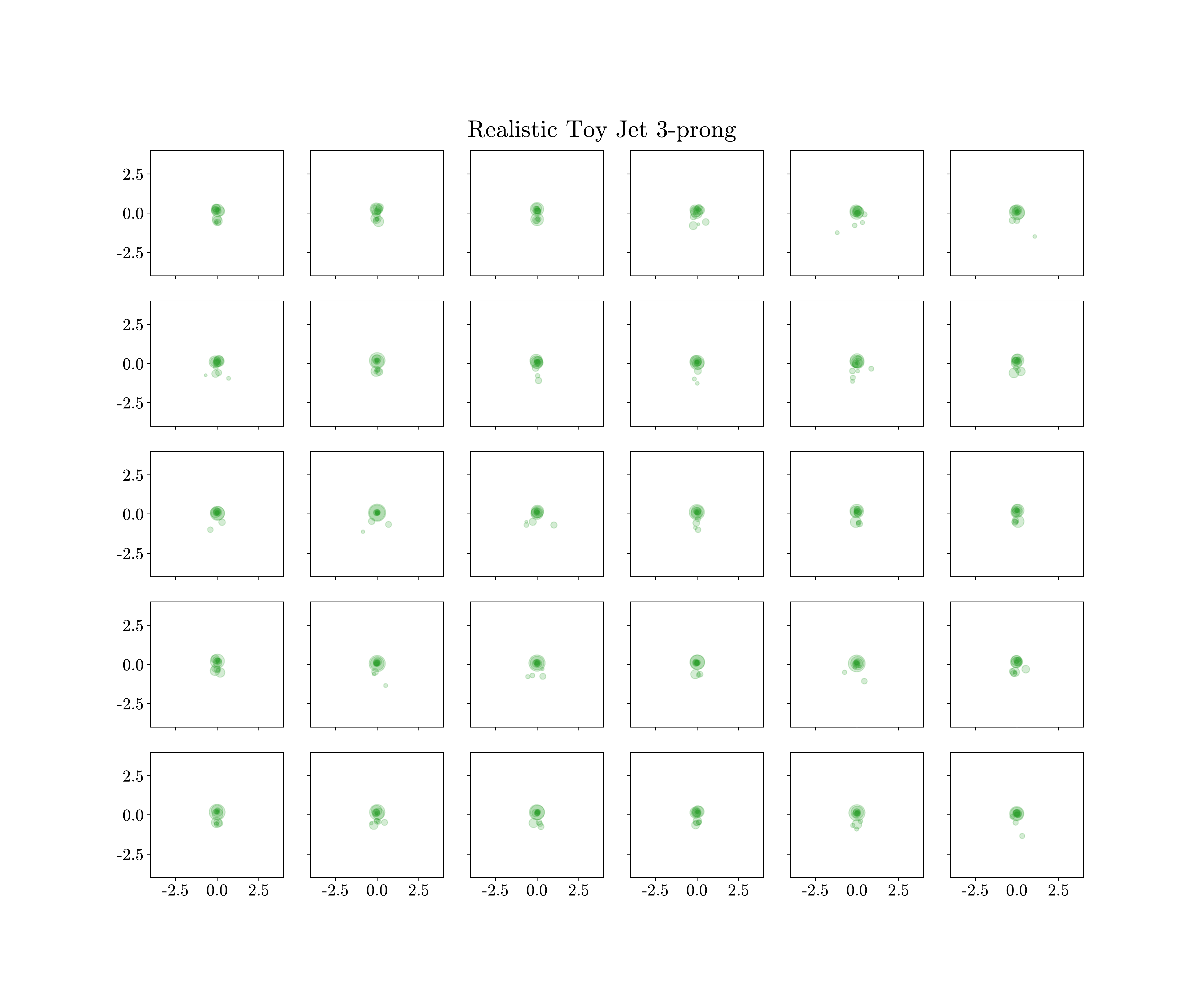}
\includegraphics[width=.48\linewidth]{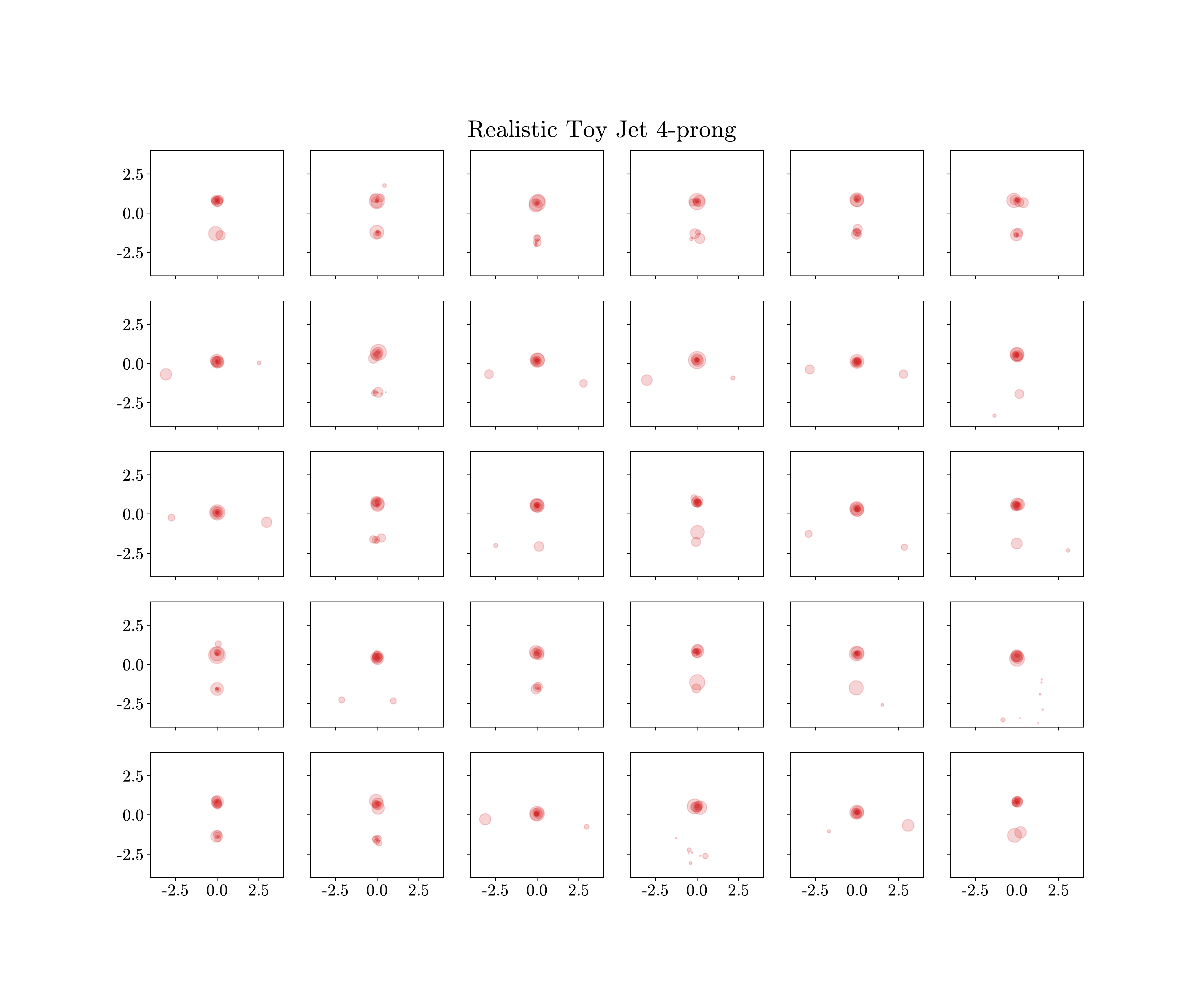}
\caption{ Samples of realistic toy jet, (Left) 3-prong (Right) 4-prong
}
\label{fig:samplerealistictoyjet_3p}
\end{figure}

In Fig.~\ref{fig:samplerealistictoyjet_1p} and Fig.~\ref{fig:samplerealistictoyjet_3p} we show some examples of realistic toy jets.

\subsection{Simulated Jet}
\label{subsec:simulatedjet_examples}

\begin{figure}[!htbp]
\centering
\includegraphics[width=.48\linewidth]{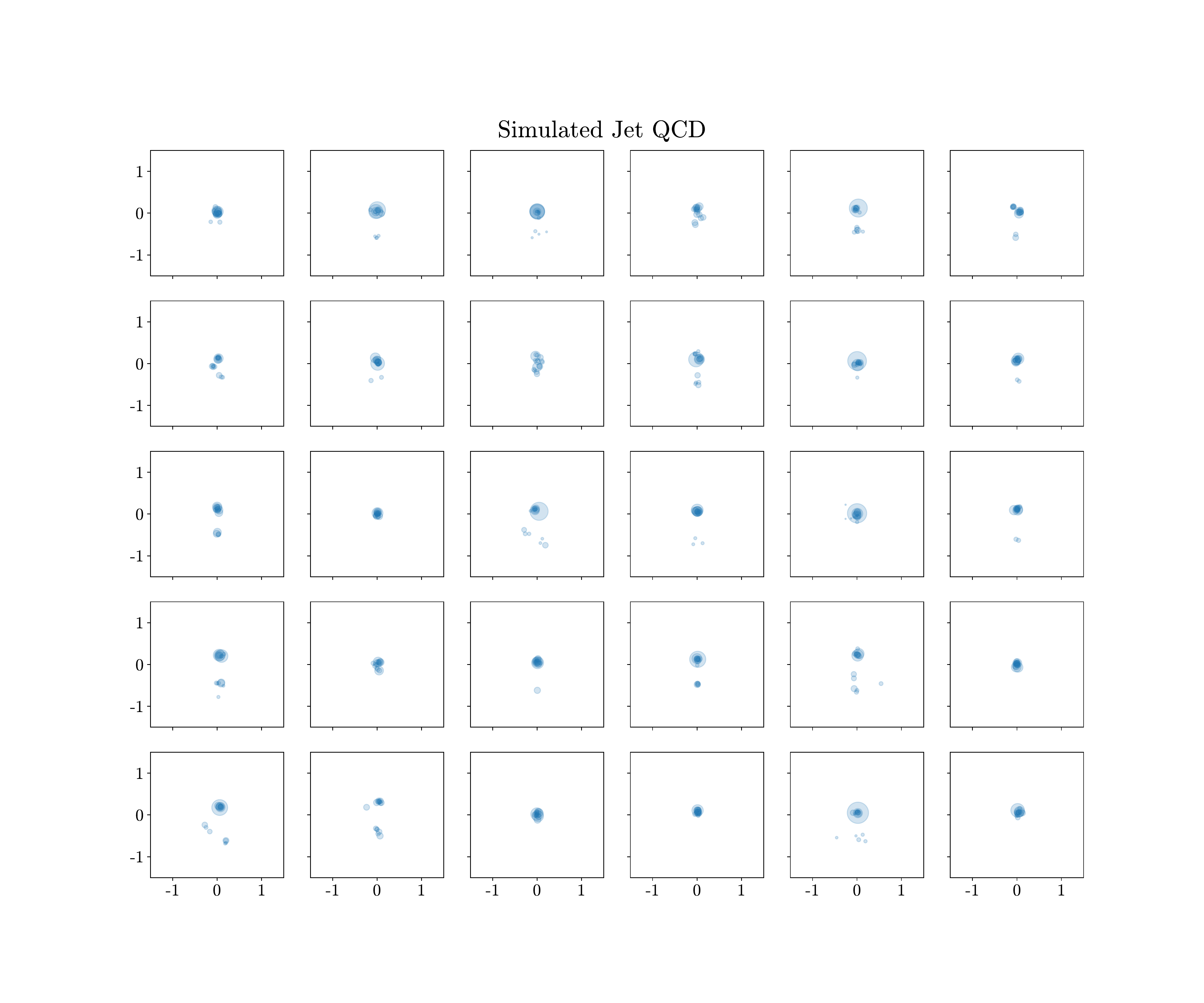}
\includegraphics[width=.48\linewidth]{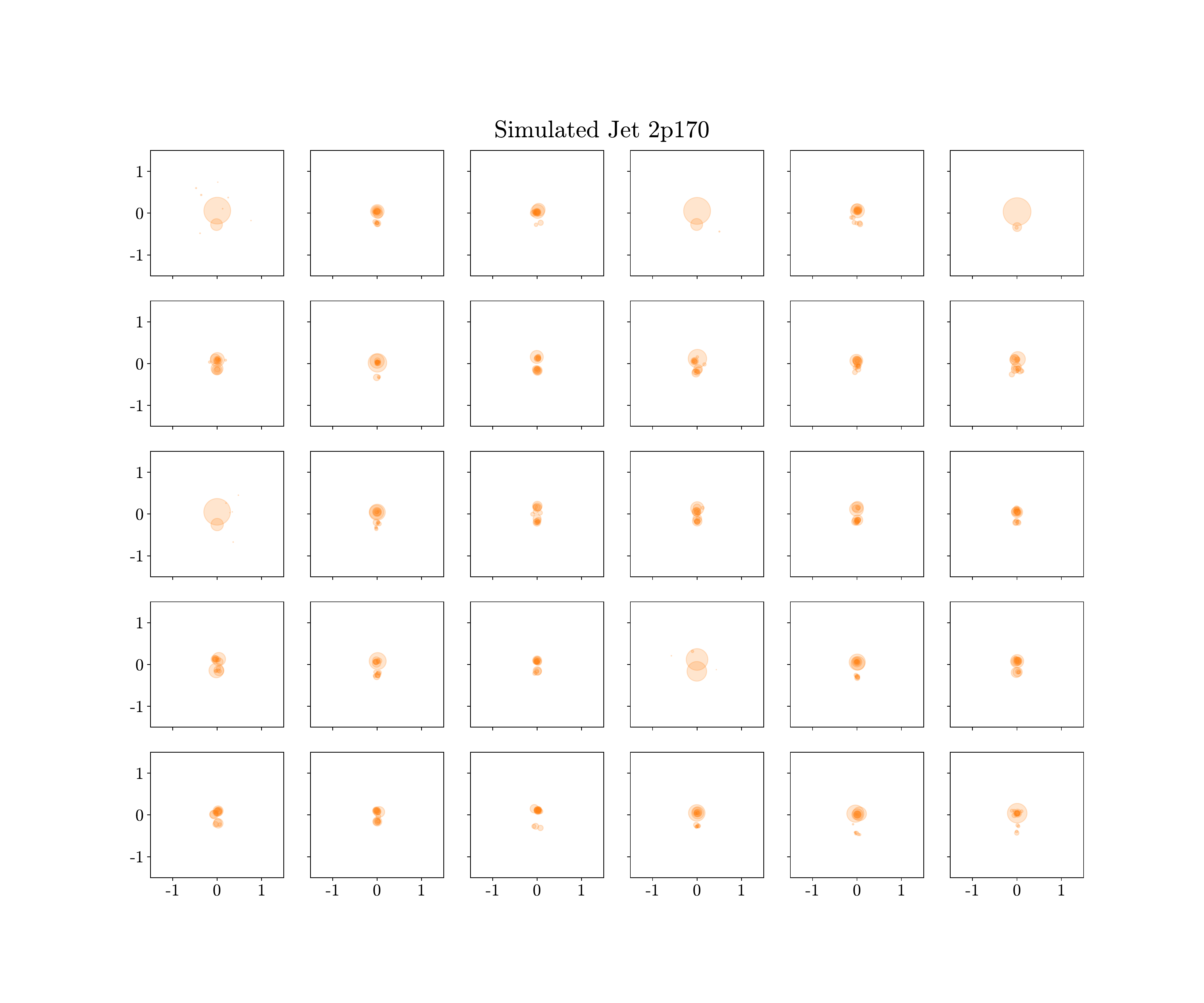}
\caption{ Samples of simulated jet, (Left) QCD (Right) 2-prong 170 $\GeV$
}
\label{fig:samplesimulatedjet_1p}
\end{figure}

\begin{figure}[!htbp]
\centering
\includegraphics[width=.48\linewidth]{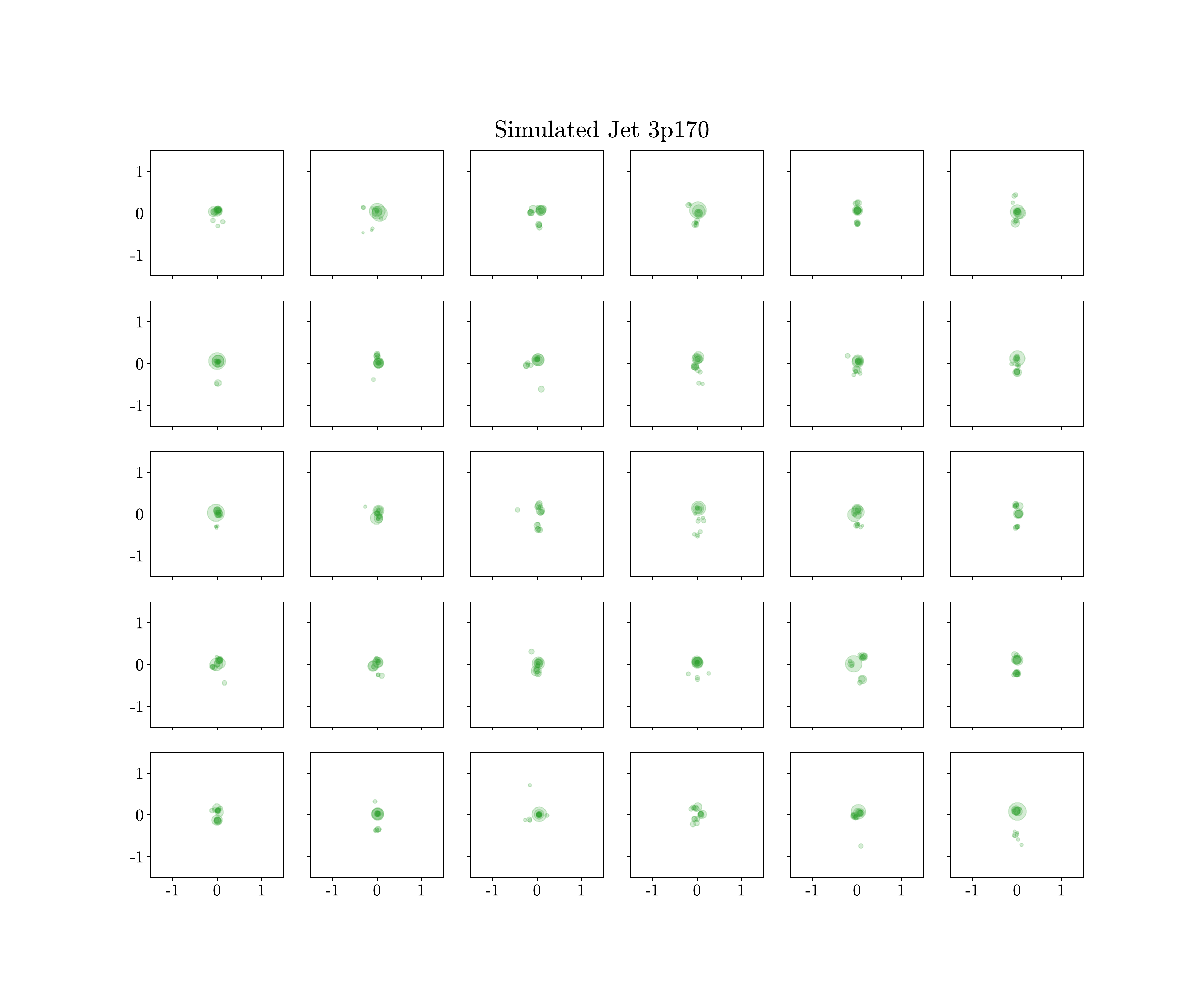}
\includegraphics[width=.48\linewidth]{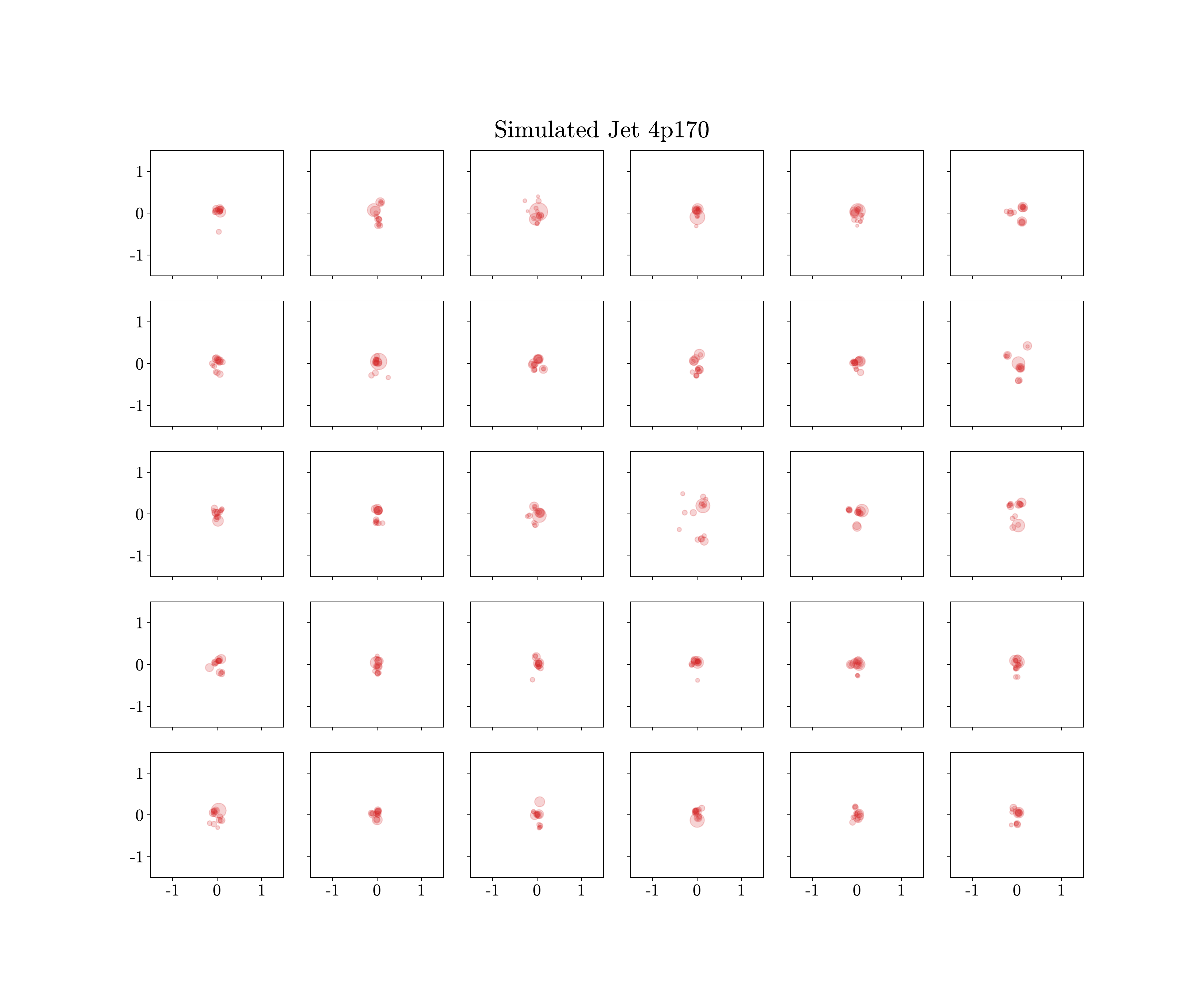}
\caption{ Samples of simulated jet, (Left) 3-prong 170 $\GeV$
(Right) 4-prong 170 $\GeV$}
\label{fig:samplesimulatedjet_3p}
\end{figure}

In Fig.~\ref{fig:samplesimulatedjet_1p} and Fig.~\ref{fig:samplesimulatedjet_3p} we show some examples of simulated jets.

\section{Stability of Area Adjusted ROC Curve}
\label{sec:stabilityareaadjusted}
\begin{figure}[htbp]
\centering
\includegraphics[width=.6\linewidth]{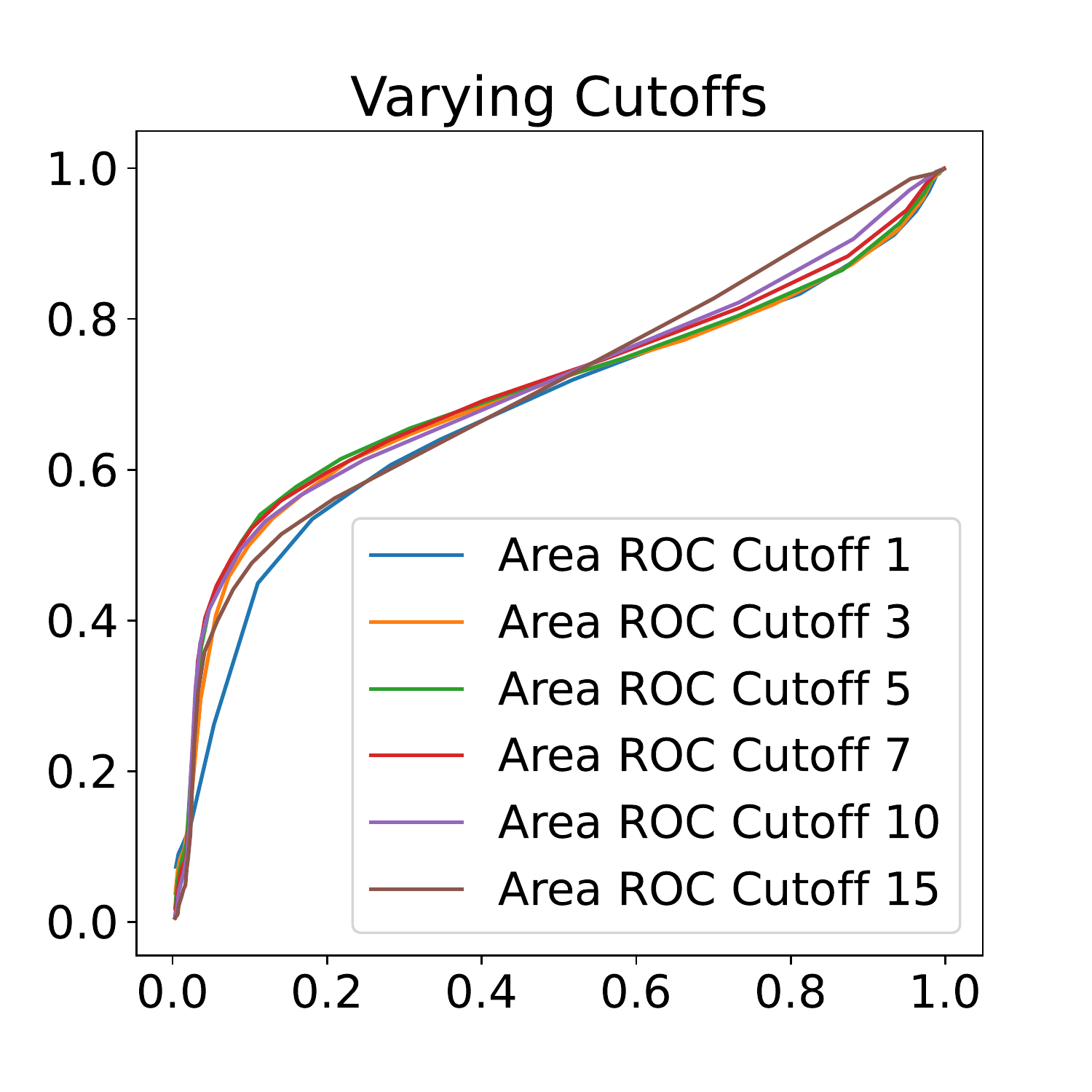}
\caption{ The area adjusted roc curve calculated for different thresholds
}
\label{fig:areaadjusted_roc_threshold}
\end{figure}

We  study the effect of binning by varying the threshold parameter when calculating the area ROC curve. Fig.~\ref{fig:areaadjusted_roc_threshold} shows the area adjusted ROC curve with varying threshold parameter for minimum required number of events in each 2D bin. We see that the new ROC curve is robust against choosing binning and thresholding of this minimum number of events. The cutoff was varied up to 1 percent of the evaluation set, and within that range of the cutoff the area adjusted ROC curve is stable.  

\section{Neural Network Architecture Details}
\label{sec:nndetails}
\subsection{CNN}

The CNN architecture used in \ref{sec:mnist} is made of 3 2-D convolution layers with kernel size 5, with max pooling and ReLU activation. The linear layers have 1000, 400, 200 neurons with leaky ReLU activation, with batch normalization \cite{batchnorm}, and dropout \cite{dropout}. There are 1M total number of parameters for this model. 
\subsection{Transformers}

For the transformer architecture used in \ref{sec:simpletoyjets}, \ref{sec:realistictoyjets},\ref{sec:SimulatedJetsEuclidean} \ref{sec:simulatedjethyperbolic}, architecture search was performed for each setting to achieve the minimum distortion. Since the feature embedding dimension was always fixed to 32, the positional encoding in Eq.~\ref{eq:posenc} was used. 

\begin{equation}
\label{eq:posenc}
f(t)^{(i)} := \begin{cases} \sin{(\omega_t \cdot t)} & \text{if } i=2k \\
\cos{(\omega_t \cdot t)}& \text{if } i=2k+1 \end{cases} \\
\quad \text{where } \omega = \frac{1}{10000^{2k/32}}
\end{equation}

For transformers, dropout \cite{dropout}  was used for regularization. The details of architectures for each cases is summarized in the Table~\ref{tab:nndetails}. 

\begin{table}[h]
\centering
\caption{Summary of network architectures for jet data } \label{tab:nndetails}

\vspace{3pt}

\small
\begin{tabular}{cccccc}
\toprule
\textbf{Section} & \textbf{Dataset} & \textbf{ Attention Heads} & \textbf{Linear Layers} & \textbf{Dropout Prob.} & \textbf{Params} \\ \midrule
Section \ref{sec:simpletoyjets} & Simple Toy Jets \ref{subsec:simpletoyjet}  & 4  & 1200,450,30 &0.25 &1.6M \\
Section \ref{sec:realistictoyjets} & Realistic Toy Jets \ref{subsec:realistictoyjet}  & 4  & 1000,400,20 & 0.2 & 1.3M  \\
Section \ref{sec:SimulatedJetsEuclidean} & Simulated Jets \ref{sec:SimulatedJetsEuclidean}  & 4  & 1000,400,20 & 0.2 & 1.3M\\
Section \ref{sec:simulatedjethyperbolic} & Simulated Jets \ref{sec:SimulatedJetsEuclidean}  & 4  & 1000,500,20 & 0.25  &1.4M\\
\bottomrule
\end{tabular}
\end{table}



\clearpage

\acknowledgments

We thank Jesse Thaler, Matthew Schwartz, and Javier Duarte for useful discussions and comments. Additionally, we thank the discussion group with Katherine Fraser, Samuel Homiller, Rashmish K. Mishra, and Patrick McCormack where the idea for this paper originated. P.H. acknowledges support by DOE grant de-sc0021943 and NSF CSSI award \#1934700. SEP acknowledges support by DOE grant DE-SC0021225, and the Institute for Fundamental Interactions and Artificial Intelligence (NSF Award  \#2019786). We thank B. Wyslouch, J. Formaggio, and P. Fisher for providing office space on the 5th floor of MIT building 24. 




\bibliographystyle{JHEP}
\bibliography{references}

\end{document}